\documentclass[preprintnumbers, floatfix, letterpaper, twocolumn,aps,prd,nofootinbib,longbibliography]{revtex4-1}
\usepackage{graphicx}
\usepackage{epstopdf}
\usepackage{latexsym}
\usepackage{amssymb}
\usepackage{amsmath}
\usepackage{color}
\usepackage{mathrsfs}
\usepackage{xparse}
\usepackage[center]{subfigure}

\begin{document}

%%%%%%%%%%%%%%%%%%%%%%%%%%%%%%%%%%%%%%%%%%%%%%%%%%%%%%%%%%%%%%%
  \renewcommand\arraystretch{2}
 \newcommand{\bq}{\begin{equation}}
 \newcommand{\eq}{\end{equation}}
 \newcommand{\bqn}{\begin{eqnarray}}
 \newcommand{\eqn}{\end{eqnarray}}
 \newcommand{\nb}{\nonumber}
 \newcommand{\lb}{\label}
 \newcommand{\cb}{  }
    \newcommand{\cc}{\color{cyan}}
        \newcommand{\cm}{\color{magenta}}
\newcommand{\rc}{\rho^{\scriptscriptstyle{\mathrm{I}}}_c}
\newcommand{\rd}{\rho^{\scriptscriptstyle{\mathrm{II}}}_c} 
\NewDocumentCommand{\evalat}{sO{\big}mm}{
 \IfBooleanTF{#1}
   {\mleft. #3 \mright|_{#4}}
   {#3#2|_{#4}}%
}

\title{Uniform Asymptotic Approximation Method with P\"{o}schl-Teller Potential}
 
\author{Rui Pan$^{a}$}
\email{rui$\_$pan1@baylor.edu}

\author{John Joseph Marchetta$^{b}$}
\email{john$\_$marchetta1@baylor.edu}

\author{Jamal Saeed$^{a}$}
\email{jamal$\_$saeed1@baylor.edu}

\author{Gerald Cleaver$^{b}$}
\email{Gerald$\_$Cleaver@baylor.edu}

\author{Bao-Fei Li$^{c, d}$} 
\email{libaofei@zjut.edu.cn}

\author{Anzhong Wang$^{a}$ \footnote{The corresponding author}}
 \email{anzhong$\_$wang@baylor.edu; the corresponding author}
 
 \author{Tao Zhu$^{c, d}$}
 \email{zhut05@zjut.edu.cn}
 
\affiliation{ $^{a}$ GCAP-CASPER, Physics Department, Baylor University, Waco, TX 76798-7316, USA\\
$^{b}$ EUCOS-CASPER, Physics Department, Baylor University, Waco, TX 76798-7316, USA\\
$^{c}$ Institute for Advanced Physics $\&$ Mathematics, Zhejiang University of Technology, Hangzhou, 310032, China\\
$^{d}$ United Center for Gravitational Wave Physics (UCGWP), Zhejiang University of Technology, Hangzhou, 310032, China}

\date{\today}

\begin{abstract}
In this paper, we study  analytical approximate solutions of the second-order homogeneous differential equations with the existence of only 
two turning points (but without poles), by using the uniform asymptotic approximation (UAA) method. To be more concrete, we consider the 
P\"{o}schl-Teller (PT) potential, for which analytical solutions are known. Depending on the values of the parameters involved in the PT potential, 
we find that the upper bounds of the errors of the approximate solutions in general are  $\lesssim 0.15\% \sim 10\%$, to the first-order 
approximation of the UAA method. The approximations can be easily extended to high-order, with which the errors are expected to be much 
smaller. Such obtained analytical solutions can be used to study cosmological perturbations in the framework of quantum cosmology, as well 
as quasi-normal modes of black holes.

\end{abstract}


\maketitle

\section{Introduction}
\renewcommand{\theequation}{1.\arabic{equation}}
\setcounter{equation}{0}

 A century after the first claim by Einstein that general relativity (GR)  needs to be quantized, the unification of Quantum Mechanics and  GR still remains  an open question,
despite  enormous efforts \cite{QGs}.  Such a theory is  necessary not only for conceptual reasons but also for the understanding of fundamental issues,  such as the big bang 
and black hole    singularities.   Various theories  have been proposed and among them, string/M-Theory and Loop Quantum Gravity  (LQG) have been extensively investigated 
 \cite{string,LQG}. {  {Differences between the two approaches are described in \cite{TTLQGS,RHLQGS}.}} 
  
LQG was initially based  on a canonical approach to quantum gravity (QG) introduced earlier by Dirac, Bergmann, Wheeler,  and DeWitt \cite{WDW}.   
However, instead of using metrics as the quantized objects \cite{WDW}, LQG is formulated in terms of {   {densitized triads and connections, and is a  non-perturbative and background-independent}} 
quantization of GR  \cite{Ashtekar86}. The gravitational  sector is described by the  {SU}(2)-valued  Ashtekar connection and its associated conjugate momentum, the densitized 
triad, from which  one  defines the holonomy of Ashtekar's connection and the flux of the densitized triad. Then, one can construct  the full kinematical Hilbert space   in a rigorous 
and well-defined way  \cite{LQG}. An open  question  of LQG is its semiclassical limit, that is, are there solutions of LQG that closely approximate those of GR in the semiclassical limit? 

Although the above  question still remains  open, concrete examples can be found in the context of loop quantum cosmology 
(LQC) (For recent reviews of LQC, see \cite{LQC_rew1,LQC_rew2,LQC_rew3,LQC_rew4,LQC_rew5,LQC_rew6,LQC_rew7,LQC_rew8,LQC_rew9,LQC_rew10} and references therein). 
Physical implications of LQC have  also been  studied using {\em the effective descriptions} of the quantum spacetimes derived from coherent states \cite{Taveras08}, 
whose validity has been verified  numerically for  various spacetimes \cite{AG15,numlsu-1}, especially for states sharply peaked on classical trajectories at late times \cite{KKL20}.  
The effective dynamics  provide a definitive answer on the  resolution of the big bang singularity \cite{generic,SVV06,ZL07,AS10,AS11,CZ15},  
replaced by a quantum bounce when the energy density of matter reaches a maximum value determined purely by
the underlying quantum geometry.

 To connect LQC with observations, cosmological perturbations in LQC have been also investigated intensively in the past decade, and a variety of different approaches to extend LQC to include 
 cosmological perturbations have been developed. These include the dressed metric  \cite{Agullo:2012sh,Agullo:2012fc,Agullo:2013ai}, hybrid 
 \cite{Fernandez-Mendez:2012poe,Fernandez-Mendez:2014raa,Martinez:2016hmn,ElizagaNavascues:2020uyf}, 
 deformed algebra \cite{Bojowald:2007hv,Bojowald:2007cd,Bojowald:2008gz, Bojowald:2008jv} and  separate universe \cite{Wilson-Ewing:2012dcf, Wilson-Ewing:2015sfx} approaches.
 For a brief review on each of these approaches, we refer readers to \cite{LQC_rew9}.
 
 One of the major challenges in the studies of cosmological perturbations in LQC is how to solve  {for} the mode functions $\mu_k$  {from the modified Mukhanov-Sasaki equation}. So far, 
 {   {it has mainly been done  {numerically}}}  \cite{LQC_rew1,LQC_rew2,LQC_rew3,LQC_rew4,LQC_rew5,LQC_rew6,LQC_rew7,LQC_rew8,LQC_rew9,LQC_rew10}. However, this is often required to be conducted  
 with high-performance computational resources \cite{AM15}, which are not accessible to general audience. 
 
 In the past decade, we have systematically developed the uniform asymptotic approximation (UAA) method initially proposed by Olver \cite{Olver97,Olver56,Olver75}, and applied it successfully to various circumstances \cite{P0,P1,P2,P3a,P3,P4,P5,P6,P7,P8,P8a,P10,P11,P12,P13,P14,P15, Zhu:2022dfq, Li:2022xww} \footnote{It should be noted that the first application of the UAA method to cosmological perturbations in GR was
 carried by  Habib et al, \cite{Habib:2002yi,Habib:2004kc}.}. In this paper, we shall continuously work on it by considering the case in which the effective potential has only zero points but without singularities. 
 To be more concrete, we shall consider the P\"{o}schl-Teller (PT) potential, for which analytical solutions are known \cite{dong_wave_2011}. The consideration of this potential is also motivated from the studies
 of cosmological perturbations in dressed metric and hybrid approaches \cite{ZWCKS17,WZW18}, in which it was shown explicitly that the potentials for the mode functions can be well approximated by the
 PT potential with different choices of the PT parameters.  In particular, in the dressed metric approach, the mode function  satisfies the following equation
 \cite{ZWCKS17}
  \bqn\lb{sch}
\mu_k''(\eta)+\left[k^2-\mathscr{V}(\eta)\right]\mu_k(\eta)=0,
\eqn
in which $\mathscr{V}(\eta)$ serves as an effective potential. During the bouncing phase  it is given by
\bqn\lb{potential}
\mathscr{V}_{\text{dressed}}(\eta)\equiv \frac{\gamma_\text{B} m_{\text{Pl}}^2(3- \gamma_\text{B} t^2/t_{\text{Pl}}^2)}{9(1+\gamma_\text{B} t^2 /t_{\text{Pl}}^2)^{5/3}},
\eqn
  where $ \gamma_\text{B}$ is a constant introduced in \cite{ZWCKS17}, and $m_{\text{Pl}}$ and $t_{\text{Pl}}$ are respectively, the Planck mass and time. This potential can be well approximated
  by a PT potential
\bqn\lb{PT}
\mathscr{V}_{\text{PT}}(\eta) = \frac{\mathscr{V}_0}{\cosh^2{\alpha (\eta-\eta_\text{B})}},
\eqn
with 
\bqn
\lb{cc}
\mathscr{V}_0=\frac{ \gamma_\text{B} m_{\text{Pl}}^2}{3}  = \frac{\alpha^2}{6}.
\eqn
Here $\eta$ is the conformal time related to the cosmic time $t$ by $d\eta = dt/a(t)$. On the other hand, in the hybrid approach, the effective potential during the bouncing phase is given by
\bqn\lb{mass_dressed}
\mathscr{V}_{\text{Hybrid}}(\eta) = - \frac{\gamma_{\rm B} m_{\rm Pl}^2 (1- \gamma_{\rm B} t^2/t^2_{\rm Pl})}{9 (1+\gamma_{\rm B} t^2/t^2_{\rm Pl})^{5/3}},
\eqn
which can be also modeled by the PT potential (\ref{PT}) but now with  \cite{WZW18}
 \bqn
V_0= \frac{m_{\rm Pl}^2 \gamma_B}{9}, \quad \alpha^2=\frac{2}{3} m_{\rm Pl}^2 \gamma_B.
\eqn
 For more details, we refer readers to \cite{ZWCKS17,WZW18}.

The rest of the paper is organized as follows: In Sec. II we provide a brief review of the UAA method with two turning points, and show that the first-order approximate solution will be described by 
the parabolic cylindrical functions.  In Sec. III {   {we construct the explicit approximate analytical solutions with the PT potential}}, and find that the parameter space can be divided into three different cases: 
A) $k^2 \gg \beta^2$, B) $k^2 \simeq \beta^2$, and C) $k^2 \ll \beta^2$, where $k$ and $\beta$ are real constants. 
After working out the error control function $\mathscr{T}$ [cf. Appendix C] in each case, we are able to determine the parameter $q_0$,
 introduced in the process  of the UAA method in order to minimize the errors. Then, we show the upper bounds of errors of our approximate solutions with respect to the exact one, given in Appendix B. 
 In particular, in Case A), the upper bounds are $\lesssim  0.15\%$, while in Case B) they are no larger than $10\%$. In Case C), the errors are also very small, except the minimal points [cf. Fig. 10], at
 which the approximate solutions deviate significantly from the analytical one. The causes of such large errors are not known, and still under our investigations. In each of these three cases, we also develop our numerical
 codes, and find that the numerical solutions trace the exact one very well, and the upper bounds of errors are always less than $10^{-4}\%$ in each of the three cases. 
 The paper is ended in Sec. IV, in which our main
 conclusions are summarized. There are also three appendices, A, B, and C, in which some mathematical formulas are presented.

%%%%%%%%%%%%%%%%%%%%%%%%%%%%%%%%%%%%%%%%%%%%%%%%%%%%%%%%%%
\section{The Uniform Asymptotic Approximation Method}
\renewcommand{\theequation}{2.\arabic{equation}}
\setcounter{equation}{0}
%%%%%%%%%%%%%%%%%%%%%%%%%%%%%%%
%%%%%%%%%%%%%%%%%%%%%%%%%%%%%%

Let us start with the following second-order differential equation
\bqn
\lb{eq2.1}
\frac{d^2\mu_k(y)}{dy^2}=f(y)\mu_k(y).
\eqn
It should be noted that all second-order linear homogeneous ordinary differential equations (ODEs) can be written in the above form by properly choosing the variable
$y$ and $\mu_k(y)$. Instead of working with the above form, we introduce two functions $g(y)$ and $q(y)$, so that the function $f(y)$ takes the form
 \cite{Olver97}
\bqn
\lb{eq2.1a}
f(y) = \lambda^2 g (y)+q(y), 
\eqn
where $\lambda$ is a  large positive dimensionless constant and serves as a bookmark,  so we can expand $\mu_k(y)$ as
\bq
\lb{eq2.1b}
\mu_k(y)  = \sum_{n=0}^{\infty}{\frac{\mu^{(n)}_k(y)}{\lambda^n}}.
\eq
After all the calculations are done, one can always set $\lambda=1$ by simply absorbing the factor $\lambda^{-n}$ into $\mu^{(n)}_k(y)$. It should be noted that there exist
cases in which the above expansion does not converge, and in these cases we shall expand  $\mu_k(y) $ only to finite terms, say, ${\cal{N}}$, so that
$\mu_k(y)$ is well approximated by the sum of these ${\cal{N}}$ terms. On the other hand, 
the main reason to introduce two functions $g(y)$ and $q(y)$, instead of only  $f(y)$,  is to minimize errors by properly choosing $g(y)$ and $q(y)$.

  In general, the function $g(y)$ has singularities and/or zeros in the interval of our interest. We  call the zeros  and singularities of $g(y)$ as {\em turning points} 
  and {\em poles}, respectively. The {\em uniform asymptotic approximate} (UAA) solutions of $\mu_k(y)$ depend on the properties of  $g(y)$  around their poles and turning points  \cite{Olver56,Olver97,Olver75}. The cases in which $g(y)$ 
  has both poles and turning points were studied in detail in \cite{P2,P3,P8}, so in this paper we shall focus ourselves on the cases where singularities are absent and  only turning points exist. As to be shown below, the treatments of these cases will be different from the ones considered in  \cite{P2,P3,P8}. In particular, in our previous studies the function $q(y)$ was uniquely determined by requiring that {\em the error control function be finite and minimized  at the poles}, while in the current cases no such poles exist. So,  to fix $q(y)$, other analyses  of the  error control function must be carried out.

  \begin{figure}
\includegraphics[width=7cm]{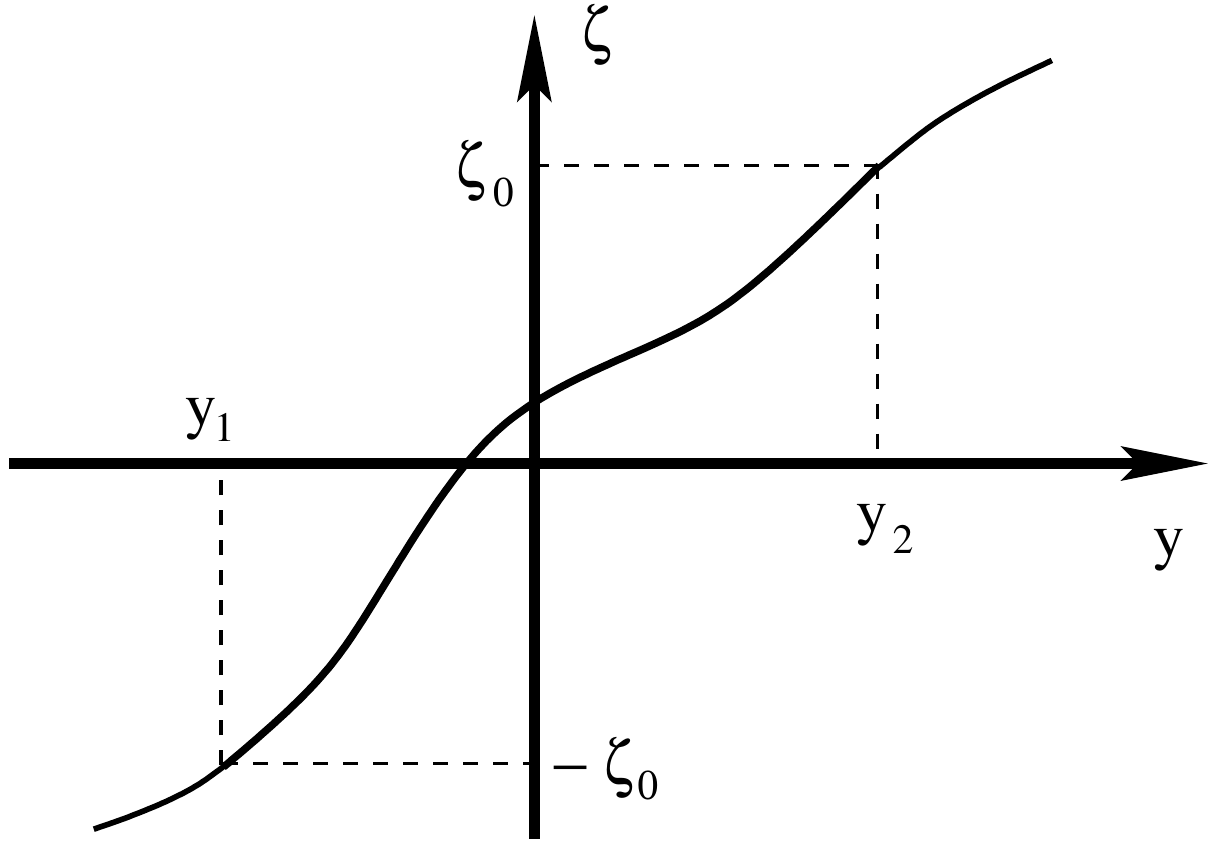}
\caption{The function $\zeta(y)$ vs $y$, which is assumed to be always an increasing function of $y$.
} 
\label{fig2}
\end{figure}

  \subsection{The UAA Method}
  
  The  UAA method includes three major steps:  (i) the Liouville transformations; (ii) the minimization of the error control function; and  (iii) the choice of the function $y(\zeta)$, where $\zeta$ is a new variable.  In the following, we shall consider each of them separately.

\subsubsection{The Liouville Transformations}

The  Liouville transformations consist of introducing a new variable $\zeta(y)$, which is assumed that {\em the inverse $y = y(\zeta)$ always exists and is thrice-differentiable}. Without loss of the generality,  we also assume that  $y(\zeta)$ is a monotonically increasing function [cf. Fig. \ref{fig2}]. Then, in terms of $U(\zeta)$,  {which  is }defined by
\bq
\lb{eq2.2}
U(\zeta) \equiv \dot y ^{-1/2} \mu_k,  
\eq
Eq.(\ref{eq2.1}) takes the form,
\bq
\lb{eq2.3}
\frac{d^2U(\zeta)}{d\zeta^2} = \left[\lambda^2 \dot y^2 g + \psi(\zeta)\right] U(\zeta),  
\eq  
where
\bqn
\lb{eq2.4}
&& \dot y \equiv \frac{dy(\zeta)}{d\zeta} > 0, \quad \zeta'(y) \equiv \frac{d\zeta(y)}{dy}=  \frac{1}{\dot y},
\eqn
and 
\bqn
 \lb{eq2.5}
 \psi(\zeta) &\equiv& \dot y^2 q + \dot y^{1/2}\frac{d^2}{d\zeta^2}\left(\dot y^{-1/2}\right)\nb\\
&=& \dot y^2 q - \dot y^{3/2}\frac{d^2}{dy^2}\left(\dot y^{1/2}\right) \equiv \psi(y).
\eqn

It should be noted that Eqs.(\ref{eq2.1}) and (\ref{eq2.3}) are completely equivalent, and so far no approximations are taken. However, the advantage of the form of Eq.(\ref{eq2.3}) is that,
by properly choosing $q(y)$,
the term $\left|\psi(\zeta)\right|$ can be much smaller than   $\left|\lambda^2 \dot y^2 g\right|$, that is, 
\bqn
 \lb{eq2.5b}
\left|\frac{\psi}{\lambda^2\dot y^2 g}\right| \ll 1,
\eqn
{   {so that  the exact solution of Eq.(\ref{eq2.1}) can be well approximated}}
by the first-order solution of Eq.(\ref{eq2.3})  with $\psi(\zeta)=0$.  This immediately  {raises} the question: how to choose $q(y)$ so that the condition (\ref{eq2.5b}) holds. To explain this in detail, let us  {move onto} the next subsection.

\subsubsection{Minimization of Errors}

To minimize the errors, let us  first introduce {\em the error control function} \cite{Olver56,Olver97,Olver75,P2,P3,P8}
\bq
\lb{eq2.6}
\mathscr{T}(\zeta)   \equiv - \int\frac{\psi(\zeta)}{{\left|\dot y^2 g\right|^{1/2}} } d\zeta.
\eq
Then, introducing the free parameters $a_n$ and $b_n$ into the functions $g(y)$ and $q(y)$, so we have   
\bq
\lb{eq2.6a}
 g(y) = g\left(y, a_n\right), \quad  q(y) = q\left(y, b_n\right), 
 \eq
 where $n = 1, 2, ..., N$, with $N$ being an integer.  It is clear that for such chosen $g(y)$ and $q(y)$, the error control function $\mathscr{T}(\zeta)$ will also depend on $a_n$ and $b_n$. To minimize the errors,
 one way is to minimize the error control function by properly choosing $a_n$ and $b_n$, so that
 \bqn
\lb{eq2.6b}
 \frac{\partial \mathscr{T}\left(\zeta, a_n, b_n\right)}{\partial a_n}  = 0, \;\;\; && \frac{\partial \mathscr{T}\left(\zeta, a_n, b_n\right)}{\partial b_n}  = 0,\nb\\
 &&  ~~~~~(n = 1, 2, ..., N).
  \eqn

 \subsubsection{Choice of  $y(\zeta)$}

On the other hand, the errors also depend on the choice of $y(\zeta)$, which  in turn sensitively depends on the properties of the functions $g(y)$ and $q(y)$ near their poles and turning points.
In addition, it must be chosen so that  the resulting  equation of the first-order approximation  (obtained by seting $\psi(\zeta) = 0$) can be solved explicitly (in terms of known functions). Considering all the above, it has been
found that $y(\zeta)$ can be chosen as \cite{Olver56,Olver97,Olver75,P2,P3,P8}
 \bqn
\lb{eq2.6c}
\dot y^2 g = \begin{cases}
\text{sgn}(g), & \text{zero turning point}, \cr
\zeta, & \text{one turning point}, \cr
\zeta_0^2 - \zeta^2, &\text{two turning point}, \cr
\end{cases} ~~~~
  \eqn
in the cases with zero, one and two turning points, respectively. Here $\text{sgn}(g) = 1$ for $g > 0$ and $\text{sgn}(g) = -1$ for $g < 0$.

 In the rest of this paper, we shall consider only the cases with
 two turning points.

\subsection{UAA Method for Two Turning Points}

 For the cases with two turning points, we can always write $g(y)$ as 
\bqn
\lb{eq2.7}
 g (y)=p(y)(y-y_1)(y-y_2),
\eqn
where  $y_1$ and $y_2$ are the two  turning points, and $p(y)$ is a function of $y$ with $p(y_i) \not= 0, \; (i = 1, 2)$. 
 In general, according to the properties of $y_1$ and $y_2$, we can divide all the cases  into three different subclasses:
 
 \begin{enumerate}
 
 \item  $y_1$ and $y_2$ are two distinct real roots of $g(y) = 0$; 
  
\item  $y_1=y_2$, a double real root of $g(y) = 0$; and 

\item $y_1$ and $y_2$ are two complex roots of  $g(y) = 0$. Since $g(y)$ is real,  in this case these two roots must be complex conjugate,
$y_1 = y_2^*$. 
 
 \end{enumerate}

To apply  the UAA method to 
Eq.(\ref{eq2.4}), we assume that the following conditions are satisfied \cite{P2,P3,P8}:

\begin{itemize}

\item  When far away from any of the two turning points,   we require
\bq
\lb{CD1}
\left|\frac{q(y)}{ g(y)}\right| \ll 1.
\eq
 \item When  near any of these two points,  we require
 \bq
 \lb{CD2}
 \left|\frac{q(y)(y-y_i)}{ g(y)}\right| \ll 1, \; (i = 1, 2),
 \eq
 provided that the two turning points are far away from each other, that is, when $ \left|y_1 - y_2\right| \gg 1$. 
\item If the two turning points are close to each other, $\left|y_1 - y_2\right| \simeq 0$,  then near these points we require
\bq
\lb{CD3}
\left|\frac{q(y)(y-y_1)(y-y_2)}{ g(y)}\right| \ll 1.
\eq
  
\end{itemize}

\begin{figure}
\includegraphics[width=7cm]{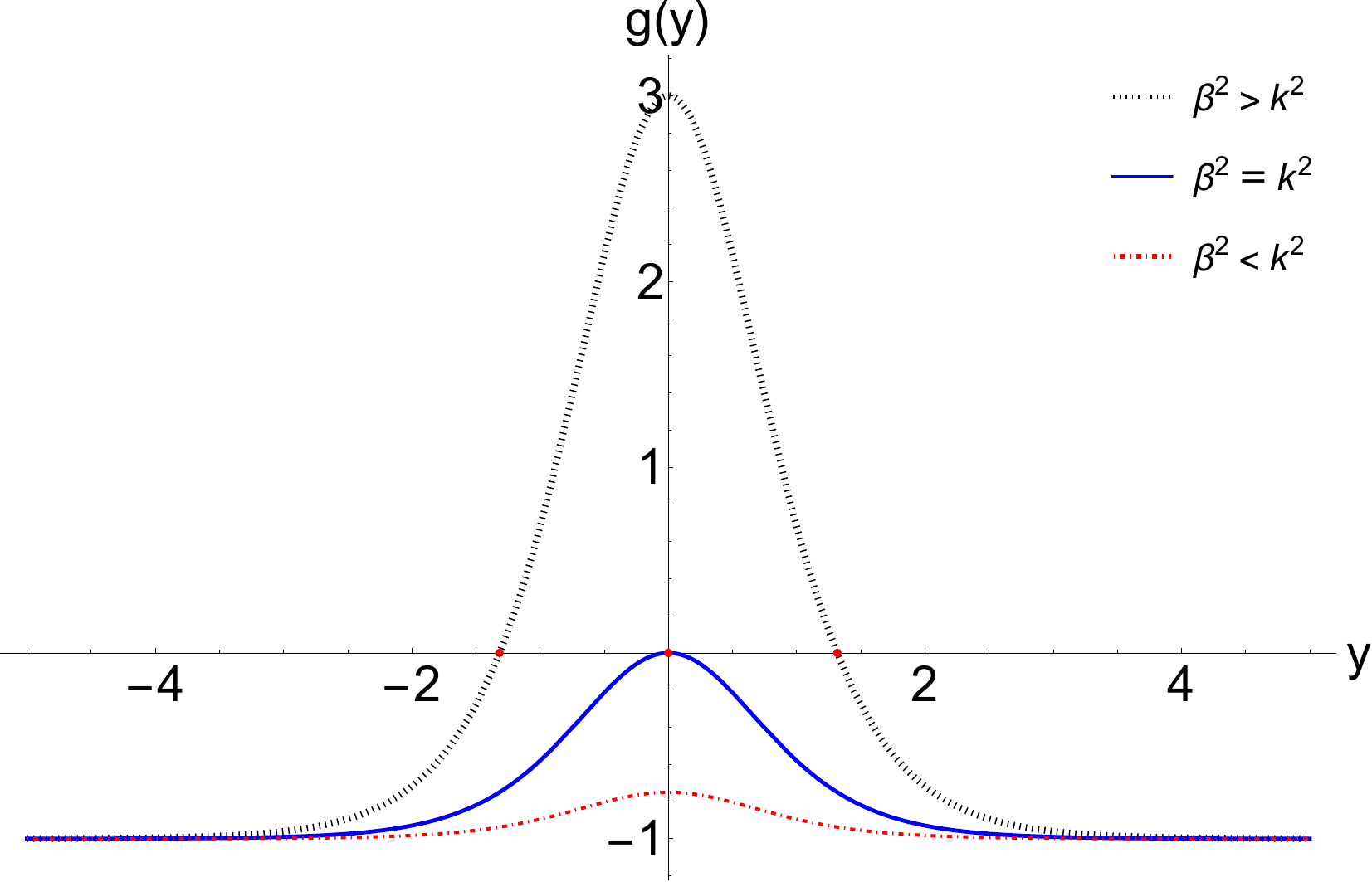}
\caption{The function ${g}(y)$ defined by  {Eq.} (\ref{3.1b})  for different choices of $k$ and $\beta$.  In particular, the dotted black line denotes the case 
$k^2 < \beta^2$, and the  solid blue line denotes the case $k^2 = \beta^2$, while  the dash-dotted red line denotes the case $k^2 > \beta^2$.
} 
\label{fig1}
\end{figure}

It should be noted that, when $\left|y_2 - y_1\right| \gg 1$, the two turning points are far away, and each of them can be treated 
as an isolated single turning point  \cite{Olver56,Olver97}.  In addition, without loss of generality, 
we assume that $g(y) < 0$ for $y > y_2$ or $y < y_1$, when
 $y_1$ and $y_2$ are real.  When $y_2$ and $y_1$ are complex conjugate, we assume that $g(y) < 0$ [cf. Fig. \ref{fig1}]. 
 Then, in this case  we  adopt a method to treat all these three classes listed above together \cite{Olver75,P2,P3,P8}.  In particular, 
we choose   $\dot y^2 g$  
as 
\bqn
\lb{xi0A}
\dot y^2 g = \zeta_0^2 - \zeta^2  \begin{cases}
> 0, & g > 0, \cr
= 0, & g = 0,\cr
< 0, & g < 0, \cr
\end{cases}
\eqn
so that $\zeta$ is an increasing function of $y$ [cf. Fig. \ref{fig2}] and
\bq
\lb{eq2.8}
\sqrt{|g(y)|} \; dy = \sqrt{\left|\zeta_0^2 - \zeta^2\right|}\; d\zeta.
\eq
When we integrate the above equation, without loss of the generality, we shall choose the integration constants  so that 
\bq
\lb{eq2.9}
\zeta(y_1)=-\zeta_0, \quad \zeta(y_2)=\zeta_0. 
\eq
Then, we find that
\bqn
\lb{eq2.9a}
\zeta_0^2 = \begin{cases}
> 0, & \text{$y_{1, 2}$ real, and $y_1 \not= y_2$},\cr
= 0, & \text{$y_{1, 2}$ real, and $y_1 = y_2$},\cr
< 0, & \text{$y_{1, 2}$ complex},\cr
\end{cases}
\eqn
 with 
\bqn
\lb{eq2.10}
\zeta_0^2&=& \pm \frac{2}{\pi}\int_{y_1}^{y_2} \sqrt{ |g (y)|}dy\nb\\
&=&   \pm \frac{2}{\pi}\int_{-\zeta_0}^{\zeta_0}\sqrt{\left|\zeta_0^2 - \zeta^2\right|} d\zeta,
\eqn
where $``+"$ corresponds to the cases that the two turning points $y_1$ and $y_2$ are both real, and $``-"$ to the cases that the two turning points $y_1$ and $y_2$ are complex conjugate.   
When  $y_1$ and $y_2$ are  complex conjugate, the integration of Eq.(\ref{eq2.10}) is along the imaginary axis \cite{Olver75}. When the two real roots are equal, we have $\zeta_0 = 0$.

To proceed further,  let us first  derive the relation between $\zeta(y)$ and $y$ by first integrating the right-hand side of Eq.(\ref{eq2.8}). To this goal, it is found easier to distinguish the case in which 
$y_1$ and $y_2$ are real from the one in which they are complex conjugate.

\subsubsection{When $y_{1, 2}$ Are Real}

Let us first consider the case when $y_1$ and $y_2$ are real. Then, when $y>y_2$, we have $\zeta(y)>\zeta_0$ [cf. Fig.\ref{fig2}]. Hence,
 from Eq. (\ref{eq2.8}) we find
\bqn
\lb{eq2.11}
&& \int_{y_2}^y\sqrt{-g(y')}dy' =  \int_{\zeta_0}^{\zeta}\sqrt{v^2-\zeta_0^2}dv\nb\\
&&\;\;\;=\frac{1}{2}\zeta\sqrt{\zeta^2-\zeta_0^2}-\frac{\zeta_0^2}{2}\ln\left(\frac{\zeta+\sqrt{\zeta^2-\zeta_0^2}}{\zeta_0}\right)\nb\\
&&\;\;\;=\frac{1}{2}\zeta\sqrt{\zeta^2-\zeta_0^2}-\frac{\zeta_0^2}{2}\operatorname{arcosh}{\left(\frac{\zeta}{\zeta_0}\right)},\;
(y \ge y_2). ~~~~~~
\eqn
When $y\le y_1$, we have $\zeta(y)\le -\zeta_0$. Then,  from Eq. (\ref{eq2.8}) we find
\bqn
&& \int^{y_1}_y\sqrt{- g (y')}dy' =   \int^{-\zeta_0}_{\zeta}\sqrt{v^2-\zeta_0^2}dv\nb\\    
&&\;\;\;=-\frac{1}{2}\zeta\sqrt{\zeta^2-\zeta_0^2}+\frac{\zeta_0^2}{2}\ln\left(\frac{-\zeta-\sqrt{\zeta^2-\zeta_0^2}}{\zeta_0}\right)\nb\\
&&\;\;\;=-\frac{1}{2}\zeta\sqrt{\zeta^2-\zeta_0^2}-\frac{\zeta_0^2}{2}\ln\left(\frac{-\zeta+\sqrt{\zeta^2-\zeta_0^2}}{\zeta_0}\right)\nb\\
&&\;\;\;=-\frac{1}{2}\zeta\sqrt{\zeta^2-\zeta_0^2}-\frac{\zeta_0^2}{2}\operatorname{arcosh}{\left(-\frac{\zeta}{\zeta_0}\right)},
(y \le y_1). ~~~~~~~
\eqn
When $y_1\leq y\leq y_2$, we have $-\zeta_0 <\zeta(y)<\zeta_0$, and
\bqn
&& \int_{y_1}^y \sqrt{ g (y')}dy' = \int_{-\zeta_0}^\zeta \sqrt{\zeta_0^2-v^2}dv = \frac{1}{2}\zeta\sqrt{\zeta_0^2-\zeta^2}\nb\\
&& ~~~~~~~~ +\frac{\zeta_0^2}{2}\arccos\left(-\frac{\zeta}{\zeta_0}\right), \; (y_1 \le y \le y_2).
\eqn

\subsubsection{When $y_{1, 2}$ Are Complex Conjugate}

Now let us turn to consider the case when $y_1$ and $y_2$ are complex. For this case $\zeta_0^2$ is always negative, $\zeta_0^2 < 0$,  thus from Eq. (\ref{eq2.2}) we find
\cite{Olver75}
\bqn
&& \int_{0}^y \sqrt{- g (y')}dy' = \int_0^\zeta \sqrt{\zeta^2-\zeta_0^2}d\zeta\nb\\
&&\;\;\;=\frac{1}{2}\zeta\sqrt{\zeta^2-\zeta_0^2}-\frac{\zeta_0^2}{2}\ln\left(\frac{\zeta+\sqrt{\zeta^2-\zeta_0^2}}{|\zeta_0|}\right).
\eqn

\subsubsection{The First-order Approximate Solutions}

With the choice of Eq.(\ref{xi0A}), we find that Eq. (\ref{eq2.4}) reduces to
\bqn
\lb{eomy1y2}
\frac{d^2U}{d\zeta^2}=\Big[\lambda^2 \left(\zeta_0^2-\zeta^2\right)+\psi(\zeta)\Big]U,
\eqn
where we assume that $\zeta \in (-\zeta_2, \zeta_2)$, with $\zeta_2$ being a real and positive constant, which can be arbitrarily large  $\zeta_2 \rightarrow \infty$.

Neglecting the $\psi(\zeta)$ term, we find that  the approximate solutions can be expressed in terms of the parabolic cylinder functions 
$W(\frac{1}{2}\lambda \zeta_0^2,\pm \sqrt{2\lambda} \zeta)$ \cite{Olver75}, and are given by
\bqn
\lb{Uf}
U(\zeta)&=& \alpha_k \Bigg\{W\left(\frac{1}{2}\lambda \zeta_0^2, \sqrt{2\lambda}\zeta \right)+\epsilon_1\Bigg\}\nb\\
&&+\beta_k \Bigg\{W\left(\frac{1}{2}\lambda \zeta_0^2, -\sqrt{2\lambda }\zeta \right)+\epsilon_2\Bigg\},
\eqn
from which we have
\bqn\lb{solutionW}
\mu_k(y)&=&\alpha_k \left(\frac{\zeta^2-\zeta_0^2}{- g (y)}\right)^{\frac{1}{4}} \left[W\left(\frac{1}{2}\lambda \zeta_0^2, \sqrt{2\lambda}\zeta \right)+\epsilon_1\right]\nb\\
&&+\beta_k \left(\frac{\zeta^2-\zeta_0^2}{- g (y)}\right)^{\frac{1}{4}} \left[W\left(\frac{1}{2}\lambda \zeta_0^2, -\sqrt{2\lambda }\zeta \right)+\epsilon_2\right],\nb\\
\eqn
where $\alpha_k$ and $\beta_k$ are two integration constants, $\epsilon_1$ and $\epsilon_2$ are  the errors of the corresponding approximate solutions, whose upper bounds are given by
Eqs.(\ref{errors}) and (\ref{ecf}) in Appendix A.

For the choice of Eq.(\ref{xi0A}), we find that  the associated error control function defined by Eq.(\ref{eq2.6}) now takes the form 
\bqn\lb{ECFa}
\mathscr{T}(\zeta)&=&-\int^{\zeta}\left\{\frac{q}{ g }-\frac{5}{16}\frac{ g '^2}{g^3}+\frac{1}{4}\frac{g''}{g^2}\right\}\sqrt{v^2-\zeta_0^2}dv\nb\\
&&+\int^{\zeta} \left\{\frac{5\zeta_0^2}{4(v^2-\zeta_0^2)^3}+\frac{3}{4(v^2-\zeta_0^2)^2}\right\}\sqrt{v^2-\zeta_0^2}dv\nb\\
&=&-\int^{y}\left\{\frac{q}{ g }-\frac{5}{16}\frac{ g '^2}{g^3}+\frac{1}{4}\frac{g''}{g^2}\right\}\sqrt{- g } dy'\nb\\
&&+\int^{\zeta} \left\{\frac{5\zeta_0^2}{4(v^2-\zeta_0^2)^{5/2}}+\frac{3}{4(v^2-\zeta_0^2)^{3/2}}\right\}dv,\nb\\
\eqn
for $ g < 0$, and
\bqn\lb{ECFb}
\mathscr{T}(\zeta)&=&\int^{\zeta}\left\{\frac{q}{ g }-\frac{5}{16}\frac{ g '^2}{g^3}+\frac{1}{4}\frac{g''}{g^2}\right\}\sqrt{\zeta_0^2 - v^2}dv\nb\\
&&-\int^{\zeta} \left\{\frac{5\zeta_0^2}{4(v^2-\zeta_0^2)^3}+\frac{3}{4(v^2-\zeta_0^2)^2}\right\}\sqrt{\zeta_0^2-v^2}dv\nb\\
&=&\int^{y}\left\{\frac{q}{ g }-\frac{5}{16}\frac{ g '^2}{g^3}+\frac{1}{4}\frac{g''}{g^2}\right\}\sqrt{g } dy'\nb\\
&&+\int^{\zeta} \left\{\frac{5\zeta_0^2}{4(\zeta_0^2-v^2)^{5/2}}-\frac{3}{4(\zeta_0^2-v^2)^{3/2}}\right\}dv,\nb\\
\eqn
for $ g > 0$.

\section{UAA Solutions with the P\"oschl-Teller Potential}
\renewcommand{\theequation}{3.\arabic{equation}} 
\setcounter{equation}{0}
To study the case in which only turning points exist, in this paper we consider the second-order differential equation (\ref{eq2.1}) with a 
P\"oschl-Teller (PT) potential \cite{ZWCKS17,WZW18}  
\bq
\lb{3.1}
\left(\lambda^2  g+q\right)=-\left(k^2-\frac{\beta_0^2}{\cosh^2(\alpha y)}\right),
\eq
as in this case exact solutions exist, where  $k$ is the comoving wavenumber, and $\beta_0$ is a real and positive constant.
The two  parameters $\beta_0$ and $\alpha$ determine the height and the spread of the PT potential, respectively. Under the rescaling $\alpha y\rightarrow y$,
 the $\alpha$ parameter can be absorbed into the wavenumber $k$ and  $\beta_0$ by  redefining $\left(k/\alpha \rightarrow k, \beta_0/\alpha \rightarrow \beta_0 \right)$. 
 As a result, there is no loss of generality to set
$\alpha=1$ from now on. Then, the exact solutions in this case exist, and are presented in Appendix B.

On the other hand, to apply the UAA method to this case, and to minimize the errors of the analytic approximate solutions,  we  tentatively choose $q$ as
\bq
\lb{3.1a}
q=\frac{q^2_0}{\cosh^2(y)},
\eq
where $q_0$ is a free parameter,  to be determined below by minimizing the error control function (\ref{eq2.6}) with the choice of $\dot{y}^2 g$ given by Eq.(\ref{xi0A}).
 Then, 
 we have
\bq
\lb{3.1b}
 g(y)=\frac{\beta^2}{\cosh^2(y)}-k^2,
\eq
where $\beta \equiv \sqrt{\beta^2_0-q^2_0}$. In this paper, without loss of generality, 
we shall choose $q_0$ so that $\beta$ is always real, that is
\bq
\lb{3.2da}
\beta^2 \equiv \beta_0^2 - q_0^2 > 0.
\eq
 Thus, from $ g(y)= 0$ we find that the two roots are given by
\bq
\lb{3.1c}
 y_i = \pm \cosh^{-1}\frac{\beta}{k} = \pm \cosh^{-1}\frac{\sqrt{\beta^2_0-q^2_0}}{k}.
\eq
It is clear that, depending on the relative magnitudes of $\beta_0$ and $k$, as well as the choices of $q_0$, two turning points can be either complex or real. 
In Fig. \ref{fig1}, we plot out the three different cases, $k^2 < \beta^2$,  $k^2 = \beta^2$, and $ k^2 > \beta^2$, from which it can be seen clearly that the two turning points
are real and different for  $k^2 < \beta^2$,  real and equal for  $k^2 = \beta^2$, and  complex conjugate for  $k^2 > \beta^2$, respectively.  
Then, from Eqs.(\ref{3.1a}) and (\ref{3.1b}), we find that 
\bq
\lb{3.2a}
\left|\frac{q(y)}{ g(y)}\right| = \left|\frac{q_0^2}{\beta^2 - k^2\cosh^2(y)}\right| \simeq q_0^2 e^{-2|y|}, 
\eq
for $|y| \gg 0$, and
\bq
\lb{3.2b}
\left|\frac{q(y)(y-y_i)}{ g(y)}\right| \simeq  \frac{q_0^2}{y + y_j}, ( i \not= j), 
\eq
for $|y| \simeq |y_i|$ and $|y_1 - y_2| \gg 1$, and  
\bq
\lb{3.2c}
\left|\frac{q(y)(y-y_1)(y-y_2)}{ g(y)}\right| \simeq  q_0^2, 
\eq
for $|y| \simeq |y_1|$ and $|y_1 -  y_2| \simeq 0$.
In the following, let us consider the three cases: (a) $k ^2 \gg \beta^2$; (b) $k^2 \simeq  \beta^2$; and (c) $\beta^2 \gg k^2$, separately.

\subsection{$k^2 \gg \beta^2$}

 In this case, we have $g(y)$ is always negative, $g(y) < 0$, so that the two turning points of $ g(y)=0$ are complex conjugate and  are given by
\bq
\lb{3a3}
y_{1} = y_2^* =  -i\cos^{-1}\left(\frac{\beta}{k}\right) \simeq -  \frac{i\pi}{2}.
\eq
 
As discussed in the last section, now $\zeta_0^2 < 0$, for which  Eq.(\ref{eomy1y2}) can be cast in the form
\bq
\lb{eom2}
\frac{d^2W(\zeta)}{d^2\zeta}=\Big\{-\lambda^2\left(\zeta^2+\hat\zeta^2_0\right)+\psi\Big\}W(\zeta),
\eq
where $\hat\zeta^2_0 \equiv - \zeta^2_0 > 0$. Note that in writing down the above equation, we had replaced   
$U$ by $W$. In addition,  the new variable $\zeta$ is related to $y$ via   
\begin{widetext}
\bqn
\lb{relation1}
 \int^y_0 \sqrt{- g(y)}dy = \int^{\zeta}_0 \sqrt{v^2 + \hat\zeta^2_0}dv
 =  \frac{1}{2}\hat\zeta^2_0\ln\left(\zeta+\sqrt{\zeta^2+\hat\zeta^2_0}\right)
+\frac{1}{2}\zeta\sqrt{\zeta^2+ \hat\zeta^2_0} -\frac{1}{2}\hat\zeta^2_0\ln \hat\zeta_0,
\eqn
\end{widetext}
from which we find that 
$\hat\zeta_0$  is given explicitly by 
\bq
\lb{zeta0}
\hat\zeta^2_0=2\left(k-\beta\right) > 0. 
\eq
Moreover, in the case of the PT potential, the integration of Eq. (\ref{relation1}) can be carried out explicitly, giving 
\begin{widetext}
\bqn
 \int^y_0dy \sqrt{- g} = \epsilon_y\sqrt{1-x^2}\sqrt{k^2-\beta^2} \times \mathrm{AppellF_1}\left(\frac{1}{2},-\frac{1}{2},1,\frac{3}{2}; \frac{1-x^2}{1-k^2/\beta^2},1-x^2\right),
\eqn
\end{widetext}
where $\epsilon_y$ denotes the sign of $y$ with  $x\equiv1/\cosh(y)$,  and $\mathrm{AppellF_1}$  is the Appell hypergeometric function. 
Ignoring the $\psi$ term in   Eq. (\ref{eom2}), we find the general solution
\begin{widetext}
\bqn
\lb{sol1}
\mu_k(y)=\left(\frac{\zeta^2+\hat\zeta^2_0}{- g(y)}\right)^{1/4}\Bigg\{\alpha_k W\left(-\frac{\hat\zeta^2_0}{2},\sqrt{2}\zeta\right)
+\beta_k W\left(-\frac{\hat\zeta^2_0}{2},-\sqrt{2}\zeta\right)\Bigg\},
\eqn
\end{widetext}
where $W$ denotes the Weber parabolic cylinder function \cite{WeberF}, and $\alpha_k$ and $\beta_k$  are two integration parameters which generally depend on the comoving wavenumber $k$.

The validity of the analytic solution (\ref{sol1}) depends on the  criteria given by Eqs.(\ref{CD1}) - (\ref{CD3}), while its accuracy can be predicted by the error control function $\mathscr{T}$.  
 In the current case, we find that   $\mathscr{T}$  of Eq.(\ref{ECFa})  can be written as a combination of three terms
as that given by Eqs.(\ref{C.3})  \footnote{In this case,
the associated error control function is $ \mathscr{V}_{\zeta_1,\zeta}(\mathscr{T})$ for any given $\zeta_1$, where $\zeta_1 \in \left(-\infty, \infty\right)$ \cite{Olver75}. 
In this paper, we choose $\zeta_1 = 0$, so the integrations will be carried out in the interval  $\zeta \in[0, \infty)$, corresponding  to $y \in [0, \infty)$. Due to the symmetry of the equation, one can easily obtain the solutions
for the region  $y \in (-\infty, 0]$ by simply replacing $y$ by $-y$ (or $\zeta$ by $-\zeta$).}, where 
\begin{widetext}
\bqn
\lb{definition1}
\mathscr{T}_1&=&\int^y_0 \frac{q}{\sqrt{- g}}dy
=\frac{q^2_0\epsilon_y}{\beta}\ln \left( \frac{\sqrt{1-x^2}\beta+\sqrt{k^2-\beta^2 x^2}}{\sqrt{k^2-\beta^2}} \right),\nb\\
\mathscr{T}_2&=&\int^y_0 \left(\frac{5 g'^2}{16 g^3}-\frac{ g^{''}}{4 g^2}\right)\sqrt{- g}dy
=- \epsilon_y \Bigg\{\frac{1}{4\beta}\ln\left (\frac{\sqrt{1-x^2}\beta+\sqrt{k^2-x^2\beta^2}}{\sqrt{k^2-\beta^2}}\right ) - \frac{\sqrt{1-x^2}A}{12(k^2-\beta^2)(k^2-\beta^2 x^2)^{3/2}}\Bigg\},\nb\\
\mathscr{T}_3&=&\int^{\zeta}_{0}\left(\frac{-5\hat\zeta^2_0}{4\left(v^2+\hat\zeta^2_0\right)^{5/2}}+\frac{3}{4\left(v^2+\hat\zeta^2_0\right)^{3/2}}\right)dv
=-\frac{\zeta\left(\zeta^2+6\hat\zeta^2_0\right)}{12\hat\zeta^2_0\left(\zeta^2+\hat\zeta^2_0\right)^{3/2}},
\eqn
\end{widetext}
 where $A$ is given by Eq.(\ref{C.5}).  
 It should be noted that  $\mathscr{T}_1$, $\mathscr{T}_2$ and $\mathscr{T}_3$ given in Eq. (\ref{definition1}) all vanish when $y=0$ (for which we have $x=1$ and $\zeta=0$), that is,
\bq
\lb{eq3.22}
\mathscr{T}(\zeta=0) = 0.
\eq
Besides, as the PT potential is an even function, the error control function is antisymmetric about the origin, namely, $\mathscr{T}(-y)=-\mathscr{T}(y)$. As a result, we will study its behavior only on the positive $y$ axis, $y \ge 0$. With the help of  Eq. (\ref{relation1}), the numeric value of the error control function at any point $y>0$ can be found from Eq.(\ref{definition1}).  In particular, for 
$ \beta/k \ll 1$, we find that
\begin{widetext}
\bqn
\lb{error3}
\mathscr{T}  = \frac{q^2_0}{k} \sqrt{1 - x^2} - \frac{\zeta\left(\zeta^2+6\hat\zeta^2_0\right)}{12\hat\zeta^2_0\left(\zeta^2+\hat\zeta^2_0\right)^{3/2}} +
{\cal{O}}\left(x^2, \frac{\beta^2}{k^3}\right)
\rightarrow \frac{1}{24k}\left[\left(24q_0^2 -1\right) - \left(\frac{\beta}{k}\right) + {\cal{O}}\left(\frac{\beta^2}{k^2}\right)\right], 
\eqn
\end{widetext}
as $x \rightarrow 0$ (or $y \rightarrow \infty$). Note that $\zeta \rightarrow \infty$ as $y \rightarrow \infty$, which can be seen clearly from Eq.(\ref{relation1}). Thus, to minimize the error control function for very large values of $y$, we must choose
\bq
\lb{q0}
q_0^2 = \frac{1}{24} \simeq 4.167\times 10^{-2}.
\eq

\begin{widetext}
 
\begin{figure}
\includegraphics[width=8.5cm]{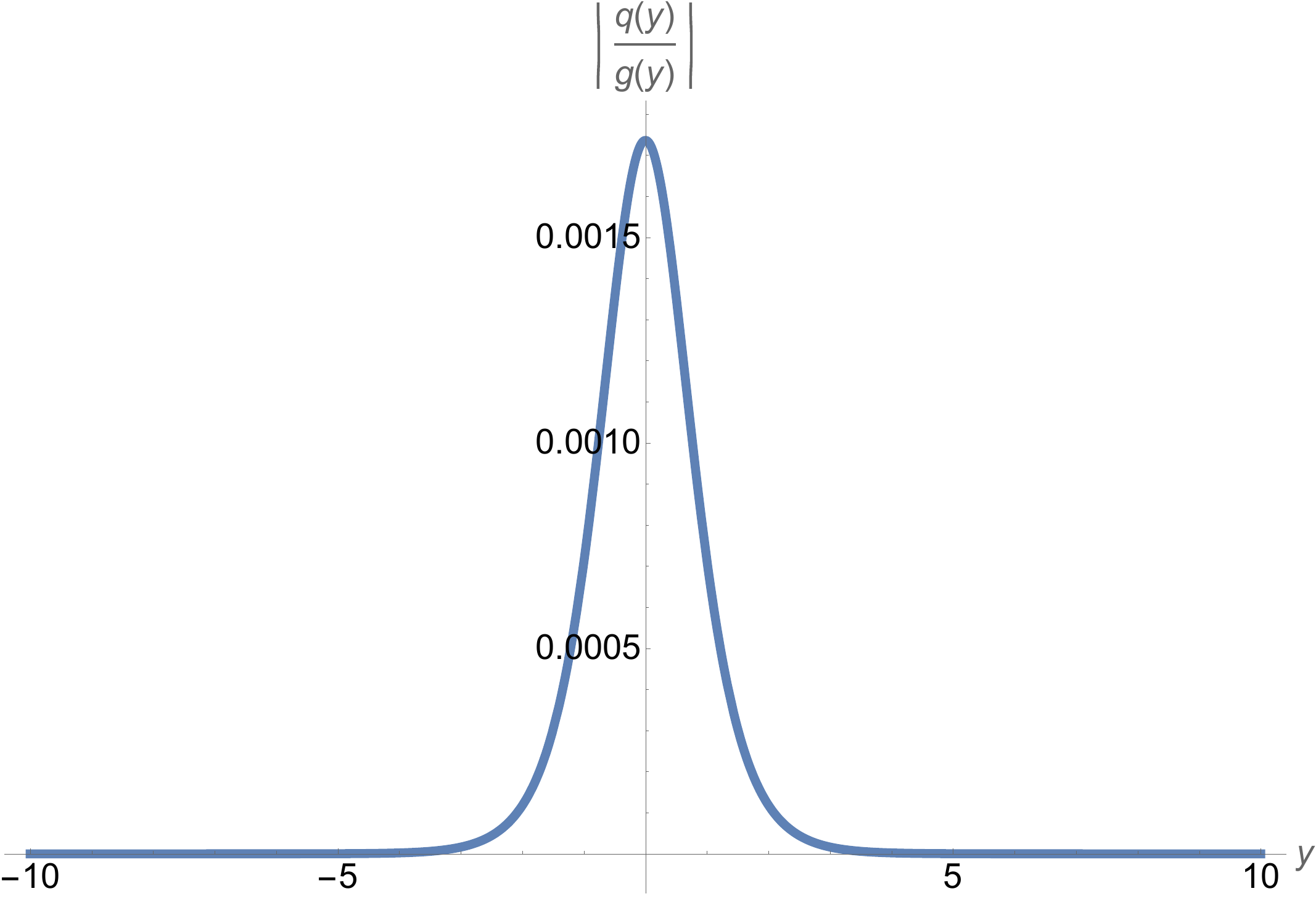}
\includegraphics[width=8.5cm]{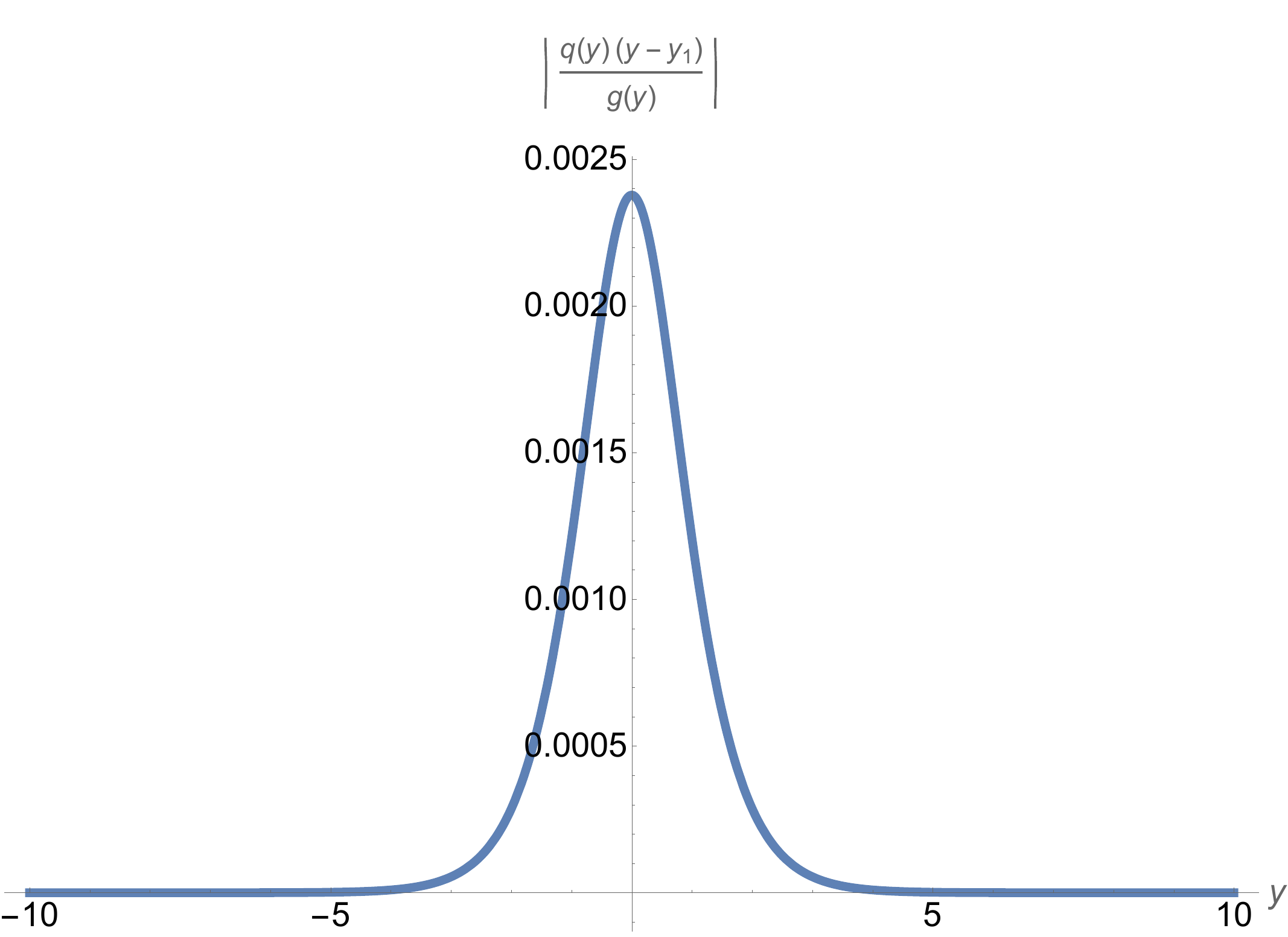}\\
\vspace{.5cm}
\includegraphics[width=8.5cm]{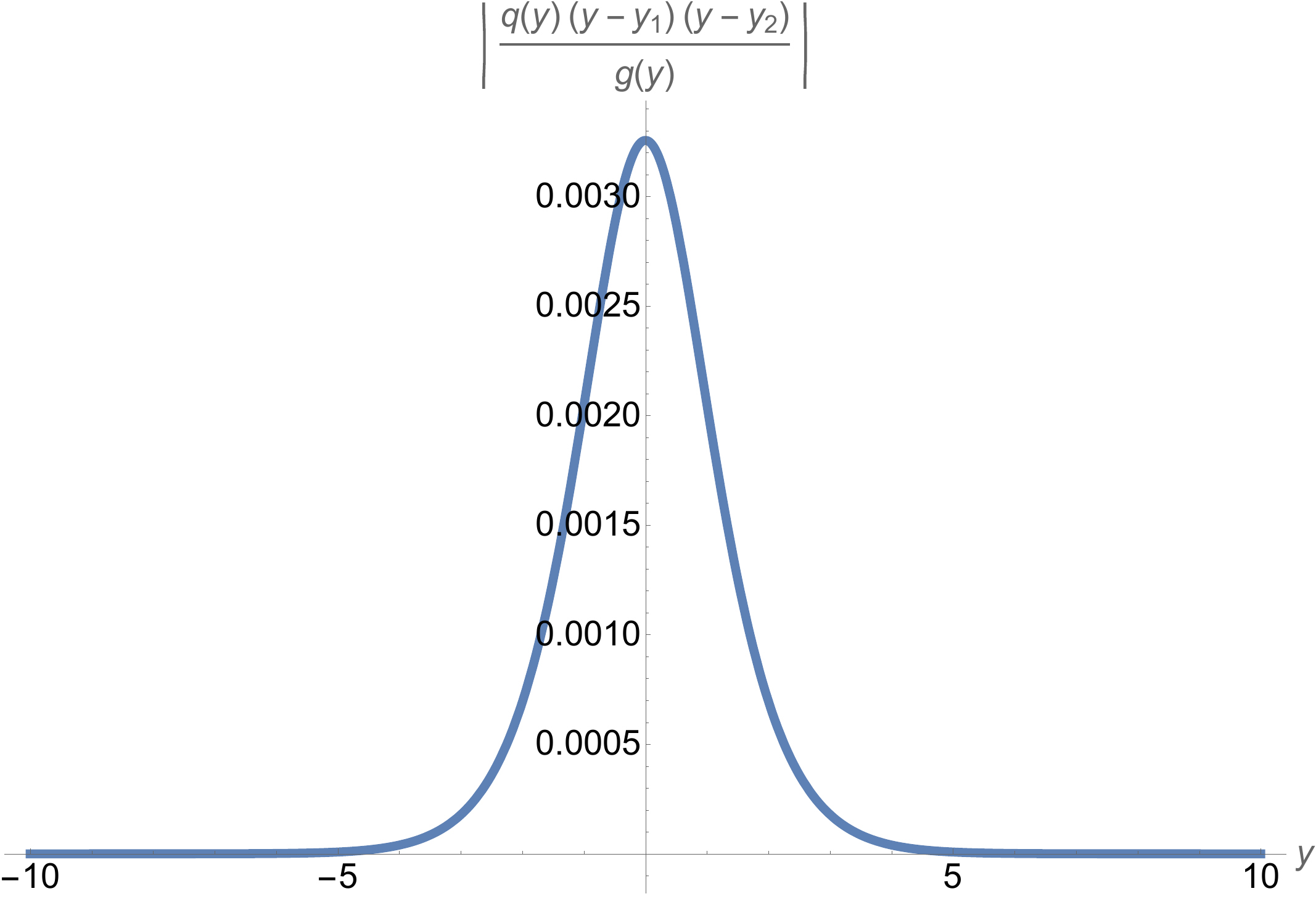}
\includegraphics[width=8.5cm]{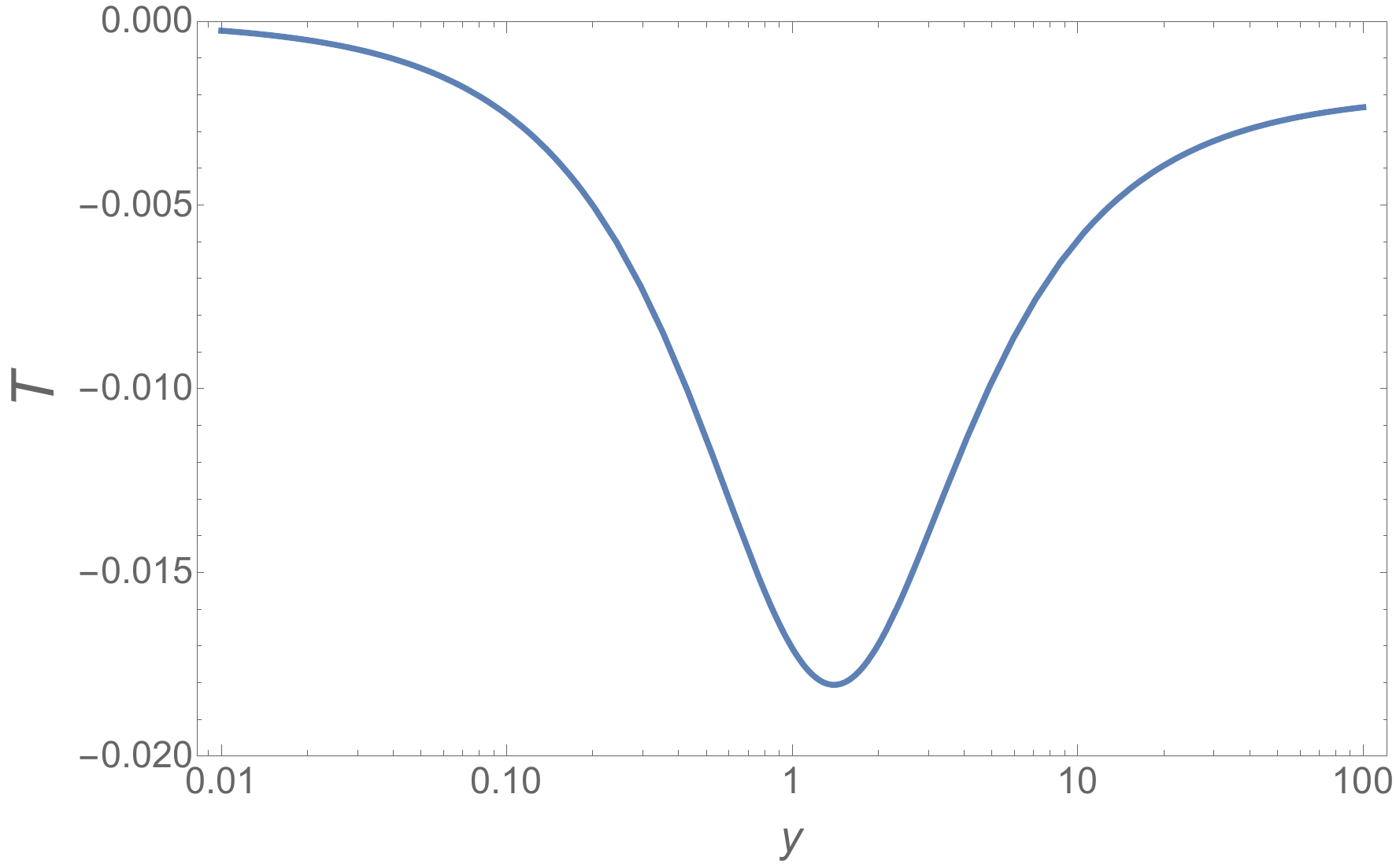}\\
\caption{Plots of the quantities $\left|{q}/{g}\right|$,  $\left|{q(y-y_1)}/{g}\right|$, $\left|{q(y-y_1)(y-y_2)}/{{  {g}}}\right|$,
 and the error control function ${\cal{T}}$  for
$k = 5.0$,  $\beta = 1.0$ and $q_0=1/\sqrt{24}$, for which we have $y_2 = y_1^* =1.36944i$.} 
\label{k5b1}
\end{figure}

 \end{widetext}

In Fig. \ref{k5b1}   
we plot the functions, $\left|{q}/{g}\right|$,  $\left|{q(y-y_1)}/{g}\right|$, and $\left|{q(y-y_1)(y-y_2)}/{{  {g}}}\right|$,  together with the error control function defined by Eqs.(\ref{C.3}) - (\ref{C.5}) for
$(k, \beta)  = (5.0, 1.0)$,  with $q_0$ being given by Eq.(\ref{q0}). (Recall  $\beta_0 \equiv \sqrt{\beta^2 + q_0^2}$).
From these figures it is clear that the conditions (\ref{CD1}) - (\ref{CD3}) are well satisfied, and the error control function remains small all the time. In particular, it decreases as  $\beta/k$ decreases.

 \begin{widetext}
 
\begin{figure}
\includegraphics[width=8.cm]{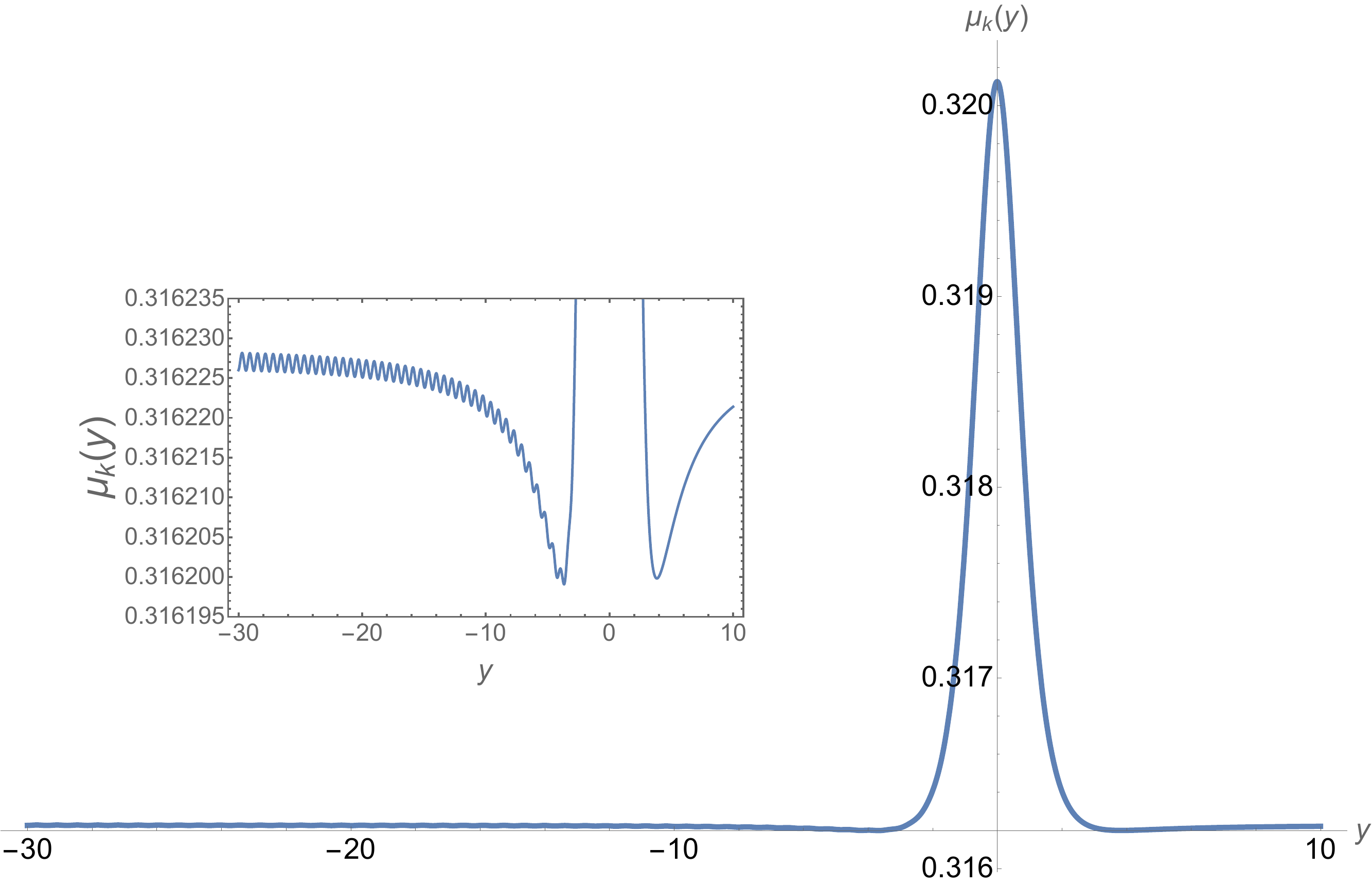}
\includegraphics[width=8.cm]{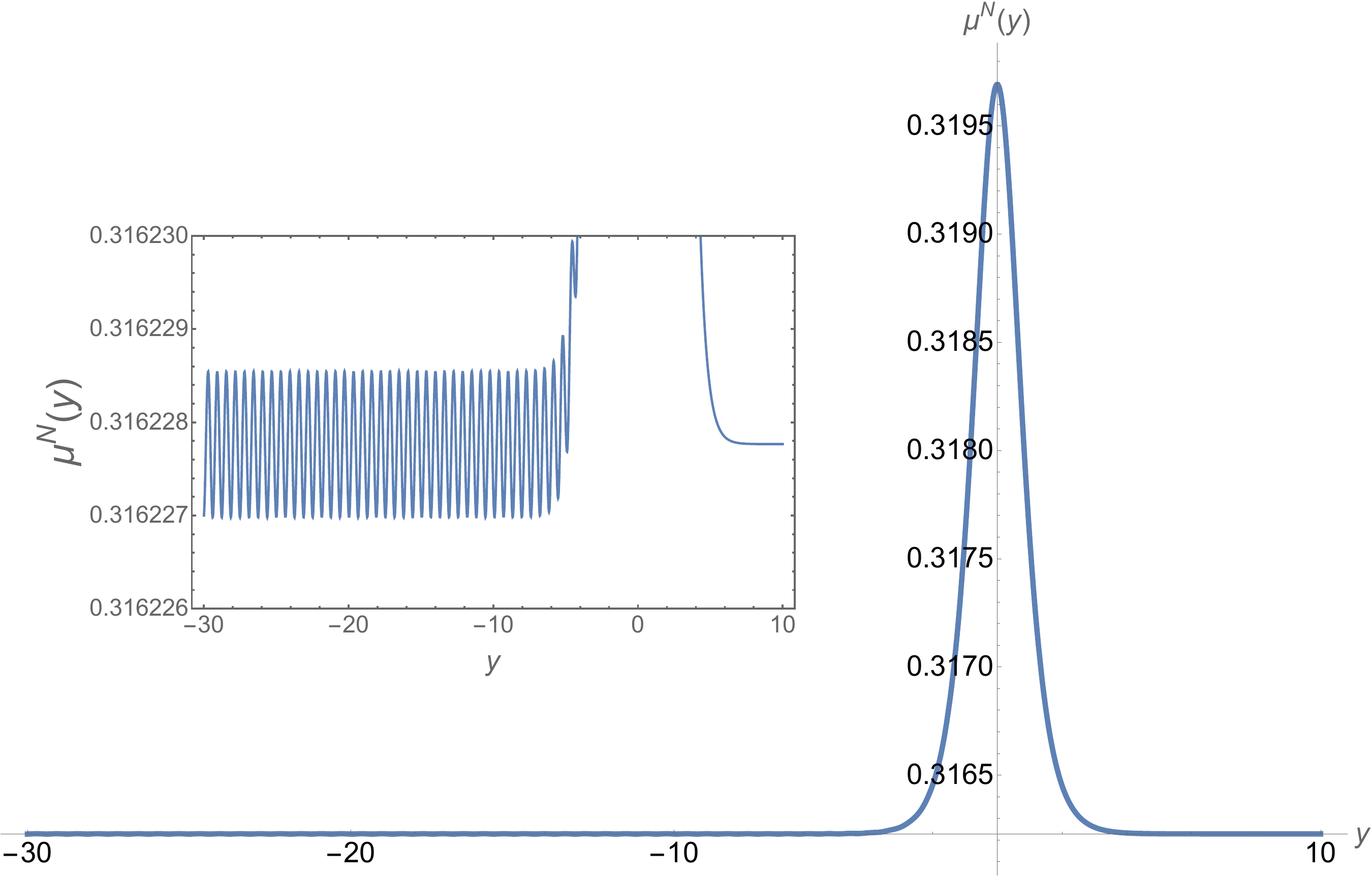}\\
\vspace{.5cm}
\includegraphics[width=8.cm]{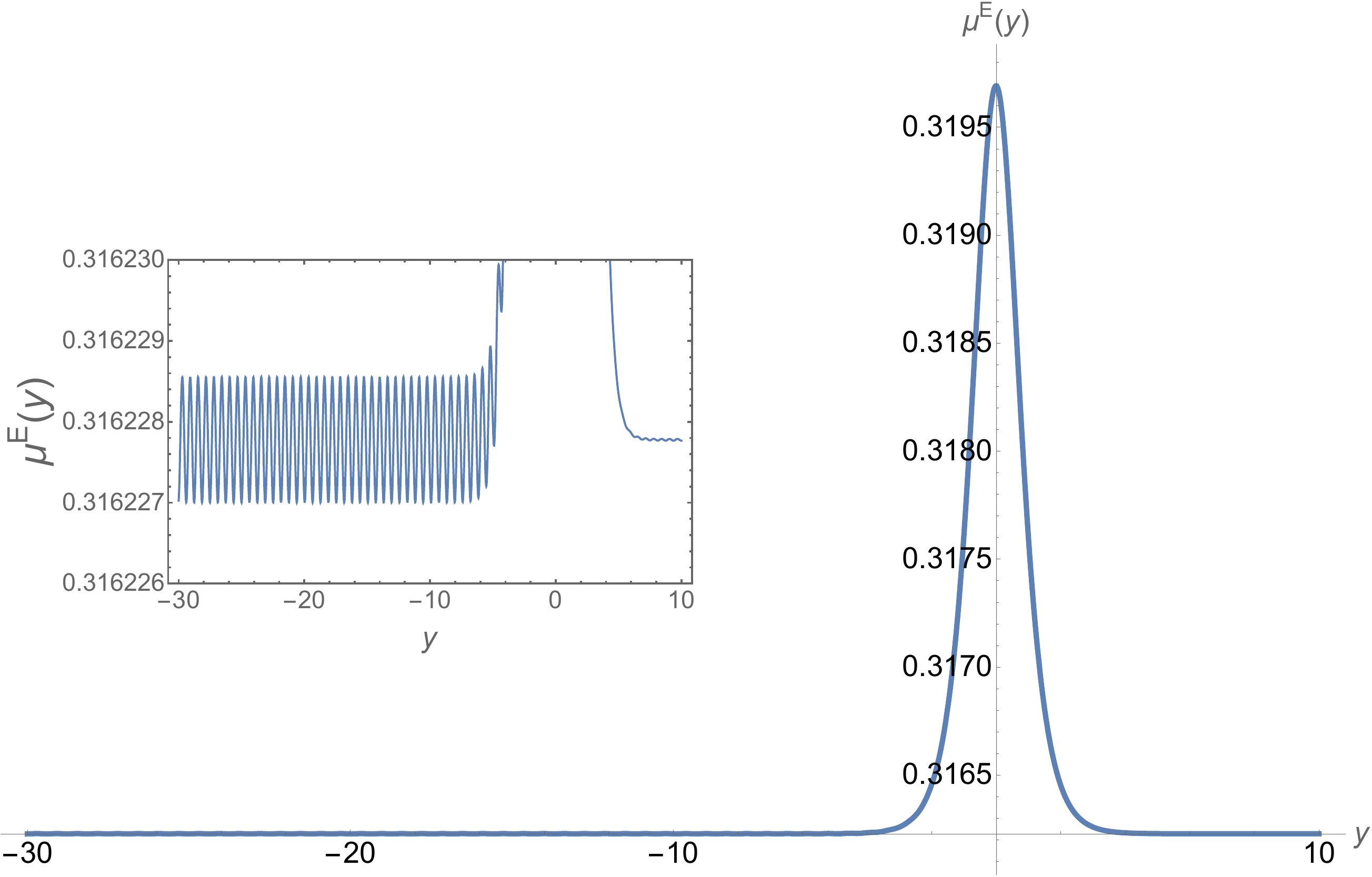}
\includegraphics[width=8.cm]{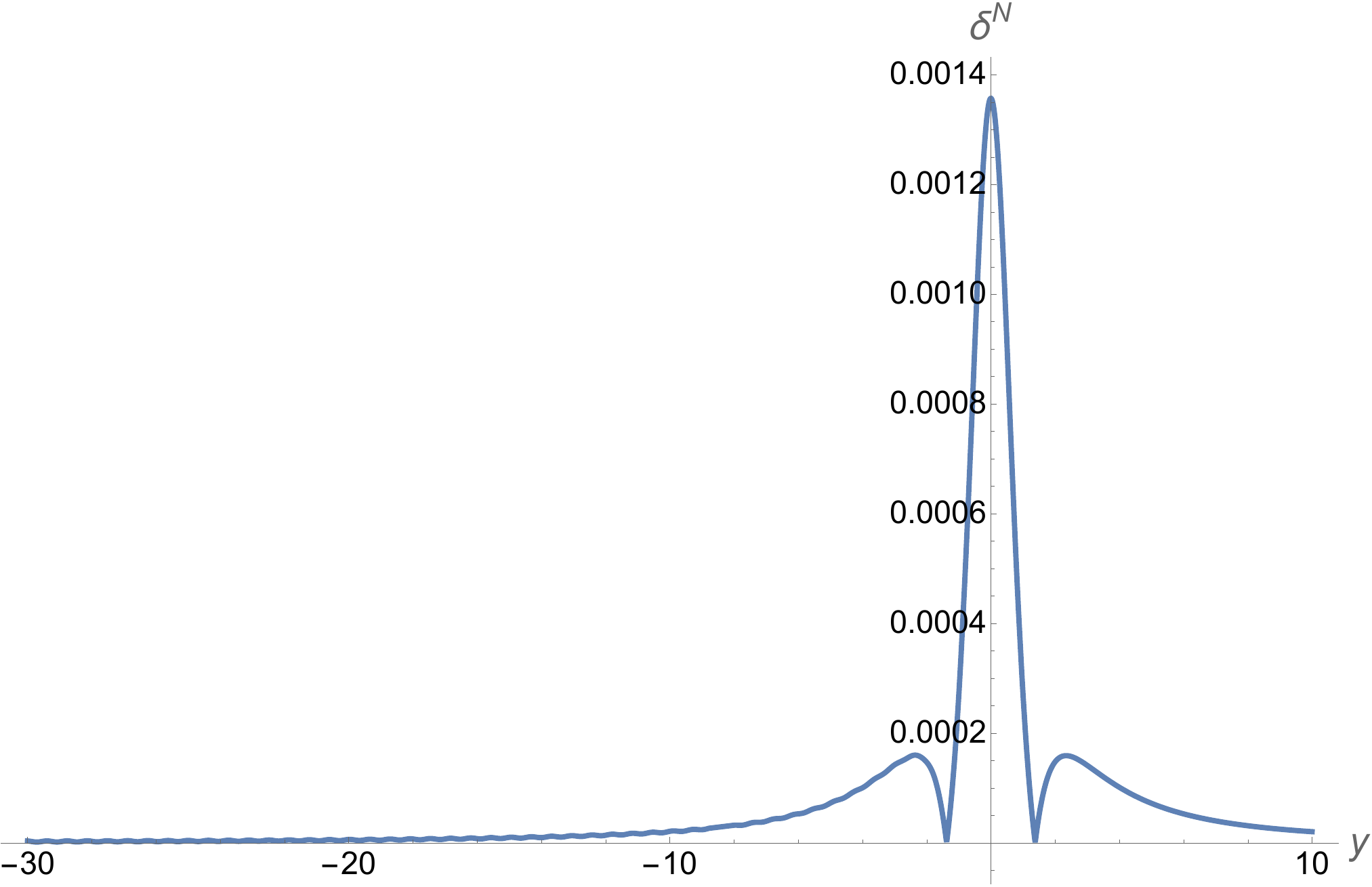}\\
\vspace{.5cm}
\includegraphics[width=8.cm]{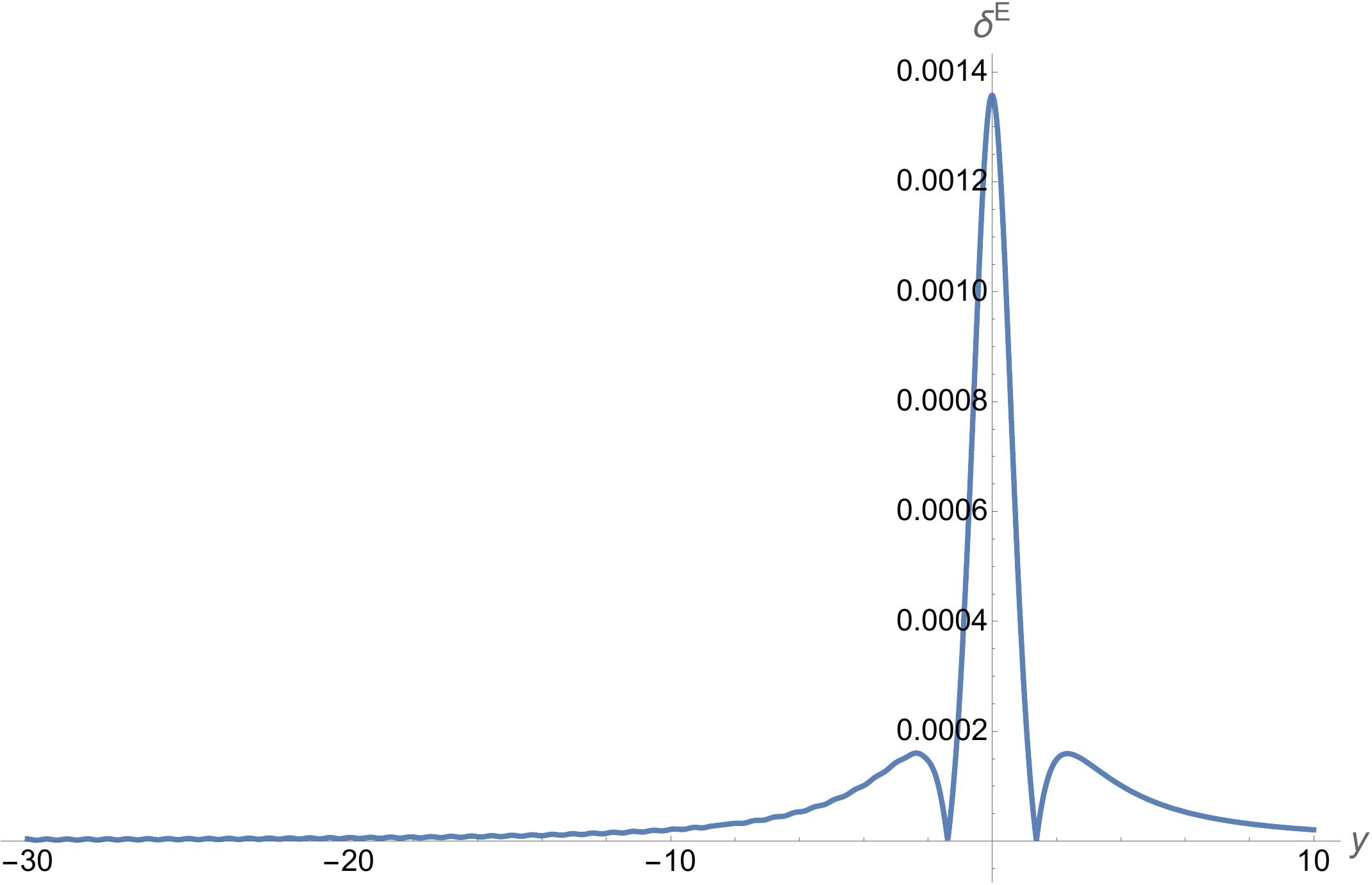}
\includegraphics[width=8.cm]{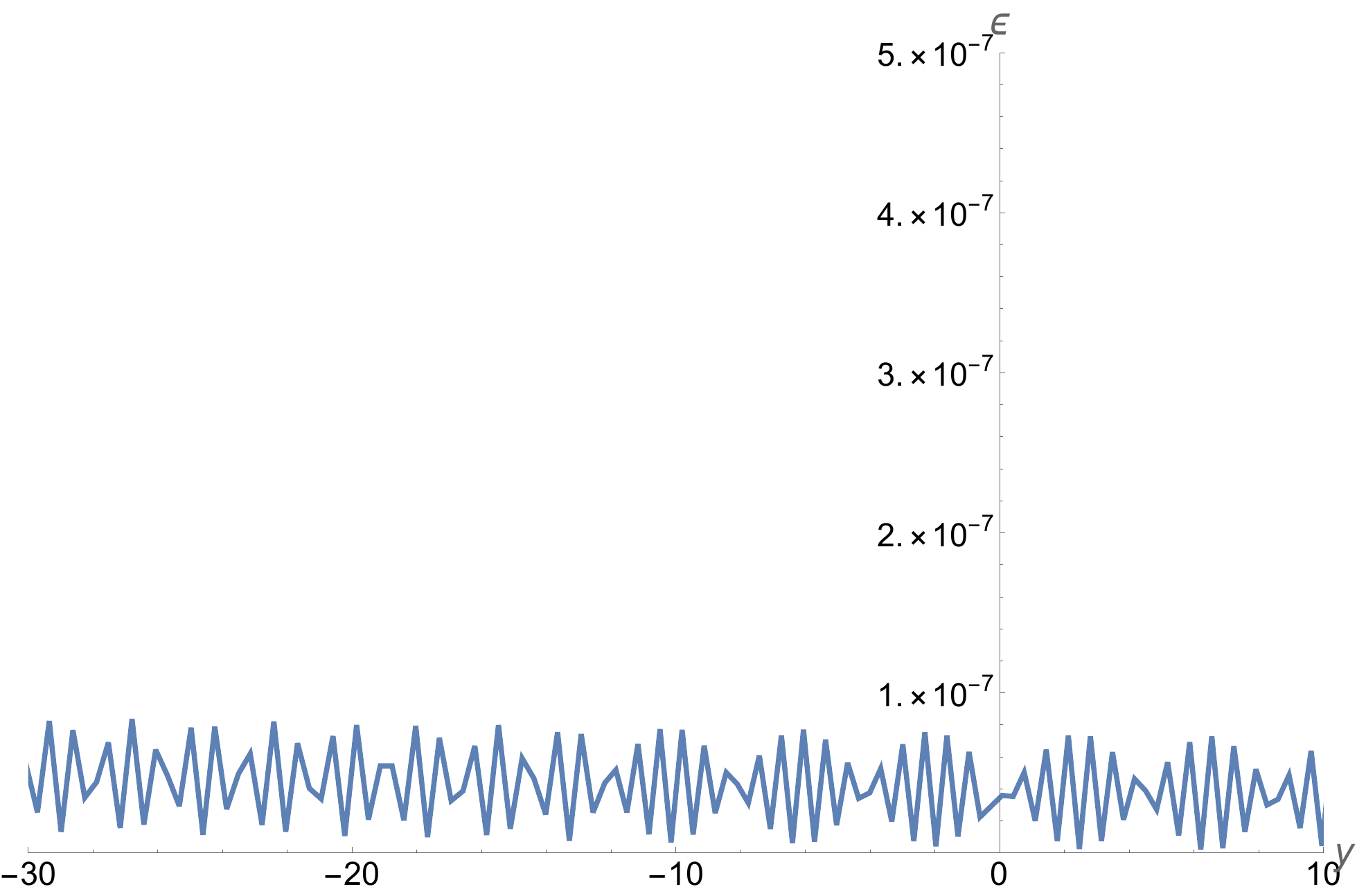}\\
\caption{Plots of the mode functions $\mu_k(y), \; \mu^N_k(y), \mu^E_k(y)$ and their
relative differences    $\delta^N(y), \; \delta^E(y)$ and $\epsilon(y)$  for
$k = 5.0$,  $\beta = 1.0$ and $q_0=1/\sqrt{24}$, for which we have $y_2 = y_1^* =1.36944i$.}
\label{k5b1_2}
\end{figure}

\end{widetext}

In Fig. \ref{k5b1_2}, we plot the mode functions $\mu_k(y)$, $\mu^{\text{N}}_k(y)$, $\mu^E_k(y)$, and the relative difference  $\delta^{\text{A}}(y)$ defined by 
\bqn
\lb{eq3.25}
\delta^{\text{A}} (y) &\equiv& \left|\frac{\left|\mu_k(y)\right| - \left|\mu^{\text{A}}_k(y)\right|}{\mu^{\text{A}}_{k}(y)}\right|,
\eqn
where $A = (N, E)$, $\mu_k(y)$ denotes the mode function obtained by the UAA method given by Eq.(\ref{sol1}), $\mu^{\text{N}}_k(y)$ is the numerical solution obtained by integrating Eq.(\ref{eq2.1}) directly with the same initial conditions, while $\mu^{\text{E}}_k(y)$ is the exact solution given by
Eq.(\ref{sol_PT}). From these figures we can see that the maximal errors occur in the region near $y = 0$, but the upper bound is no larger than $0.15\%$ at any given $y$, including the region near $y \simeq 0$. 

It is interesting to note that this analytical approximate solution is only up to the first-order approximation of the UAA method. With higher order approximations, the relative errors are even smaller. 

To check our numerical solutions, in Fig. \ref{k5b1_2}, we also plot the   relative differences  $\epsilon(y)$  between $\mu^{\text{N}}_k(y)$ 
and  $\mu^{\text{E}}_k(y)$, defined by
\bqn
\lb{eq3.25b}
\epsilon (y) &\equiv& \left|\frac{\left|\mu^{N}_k(y)\right| - \left|\mu^{\text{E}}_k(y)\right|}{\mu^{\text{E}}_{k}(y)}\right|.
\eqn
From these figures it can be seen that  $\epsilon(y)$ is no larger than $10^{-7}$, and our numerical code is well tested and  justified.

 {   {It is also interesting to note  that the mode functions are oscillating for $y \lesssim -10$, and these fine features are captured in all three
mode functions, although there are some differences in the details. Again, as shown by their relative variations, these differences are very small.}}
 In addition, we also consider other choices of $\beta$ and $k$, and find that they all have similar properties, as long as the condition $k^2 \gg \beta^2$.

\begin{widetext}
 
\begin{figure}
\includegraphics[width=8.5cm]{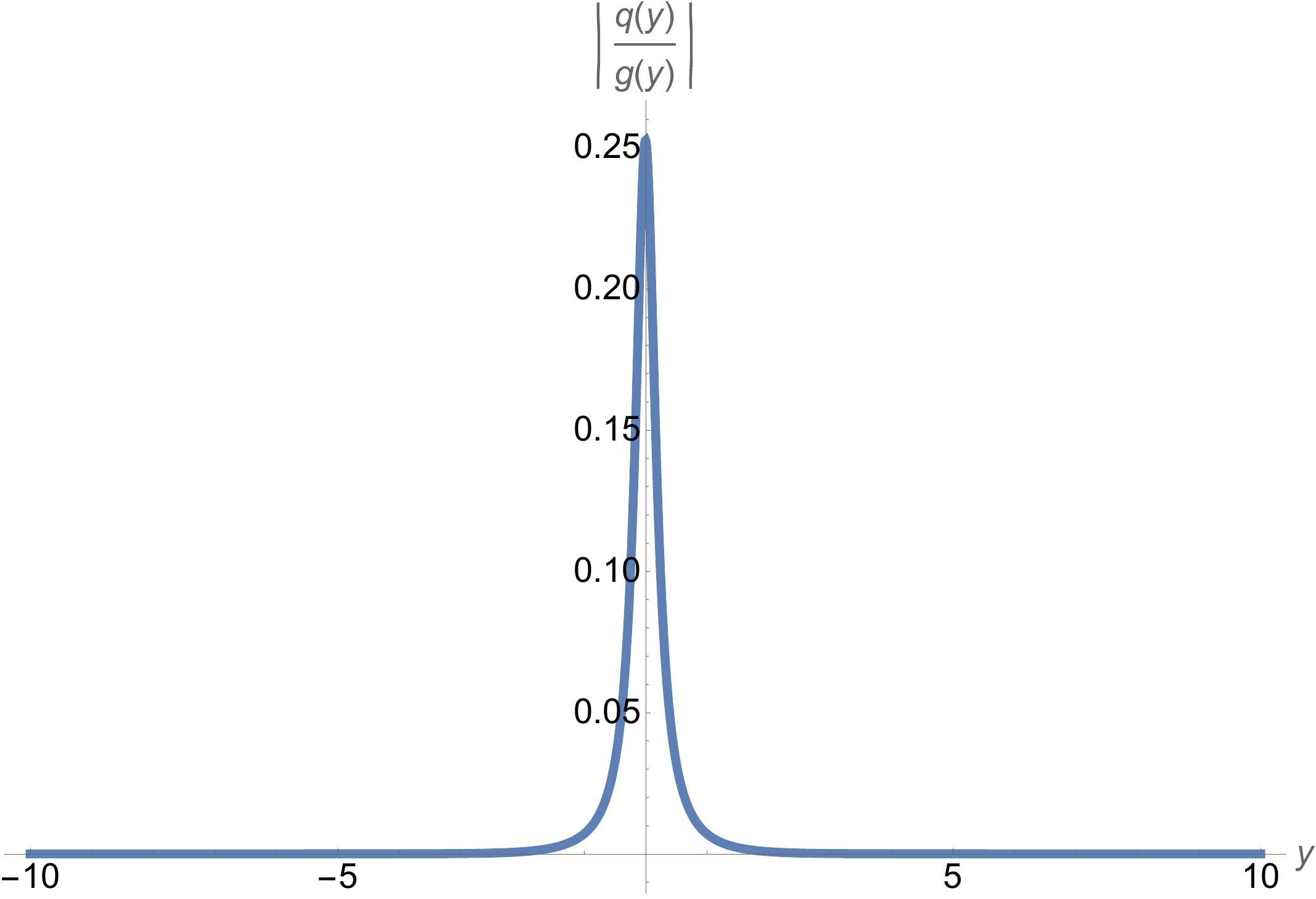}
\includegraphics[width=8.5cm]{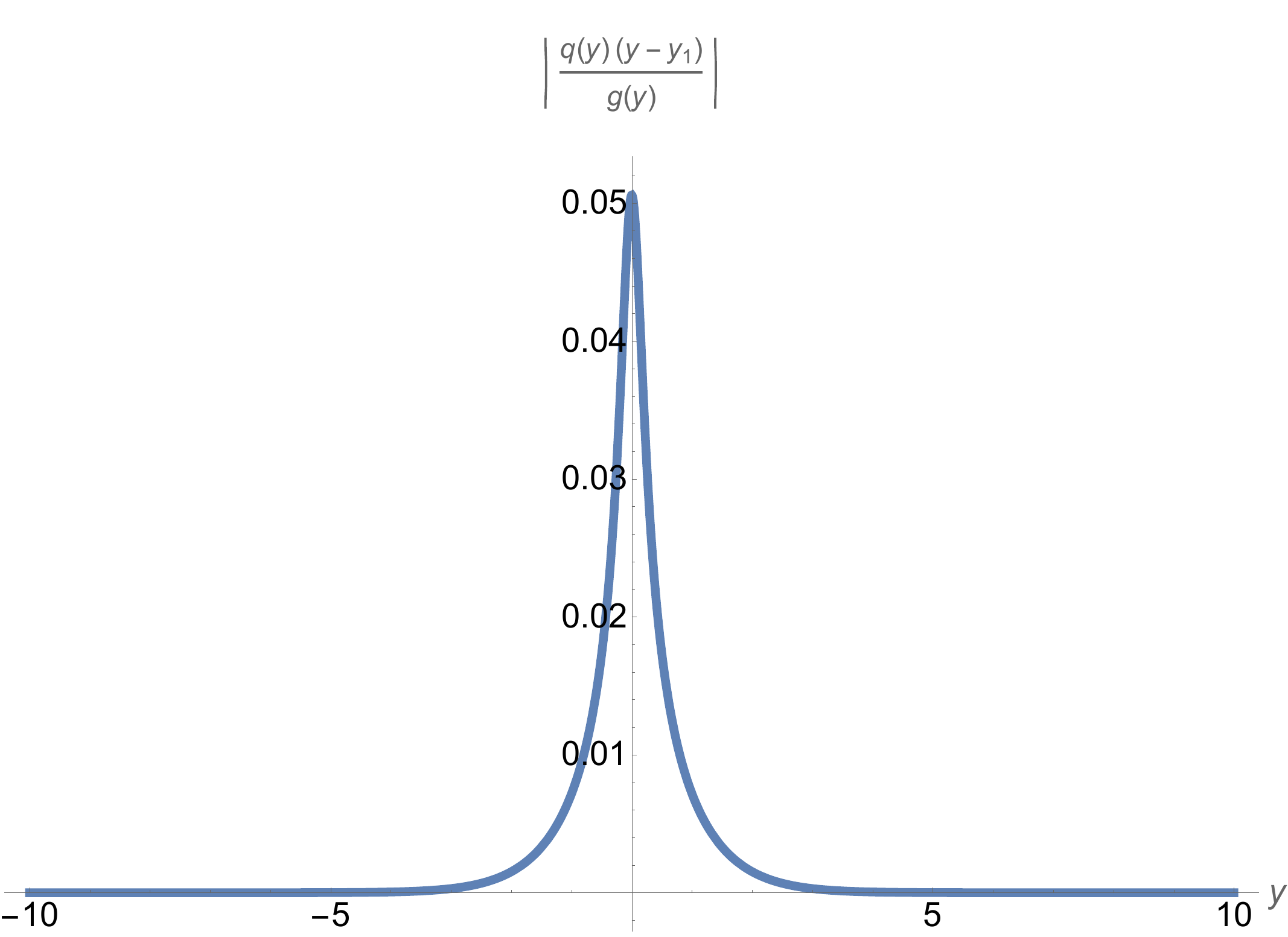}\\
\vspace{.5cm}
\includegraphics[width=8.5cm]{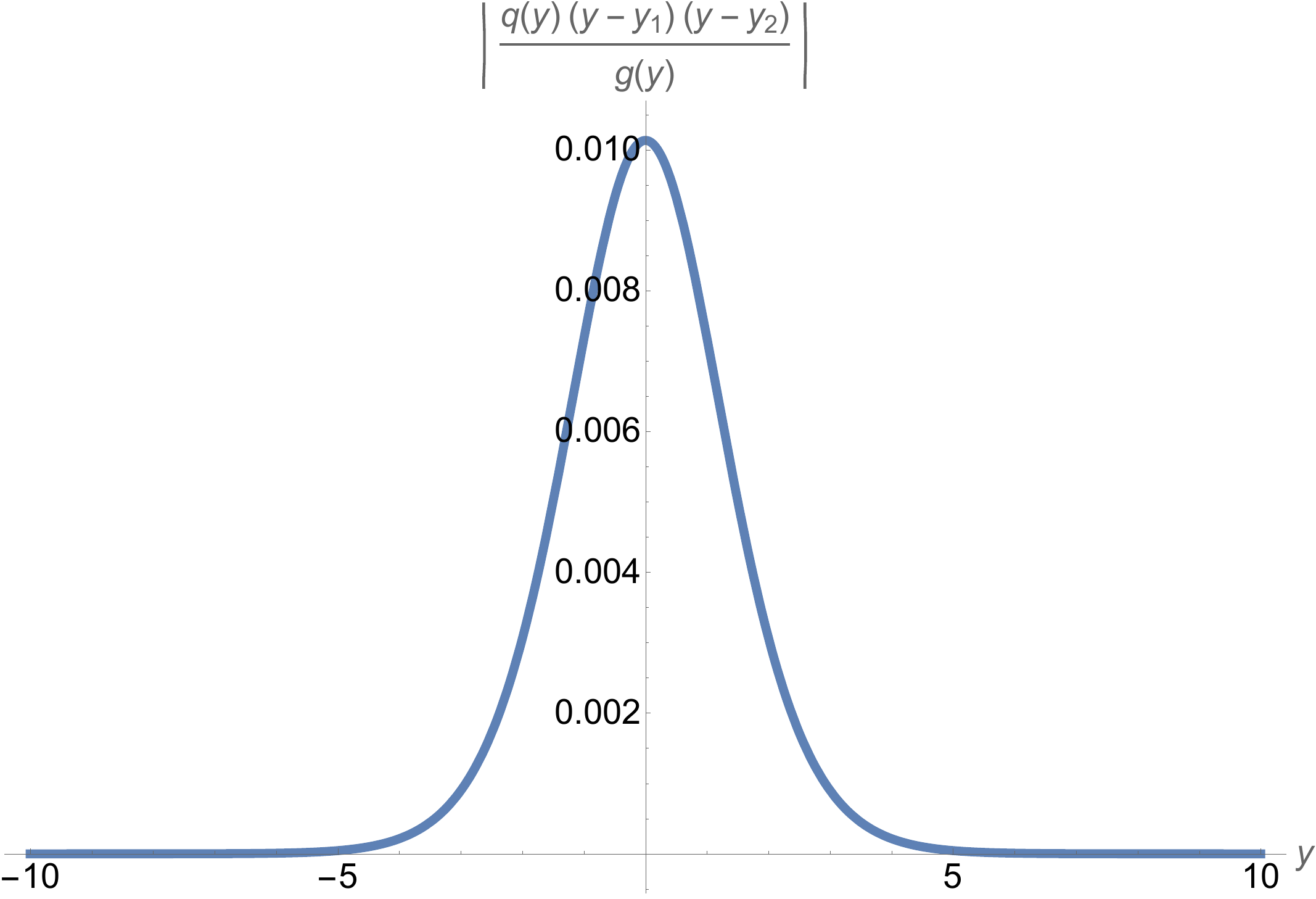}
\includegraphics[width=8.5cm]{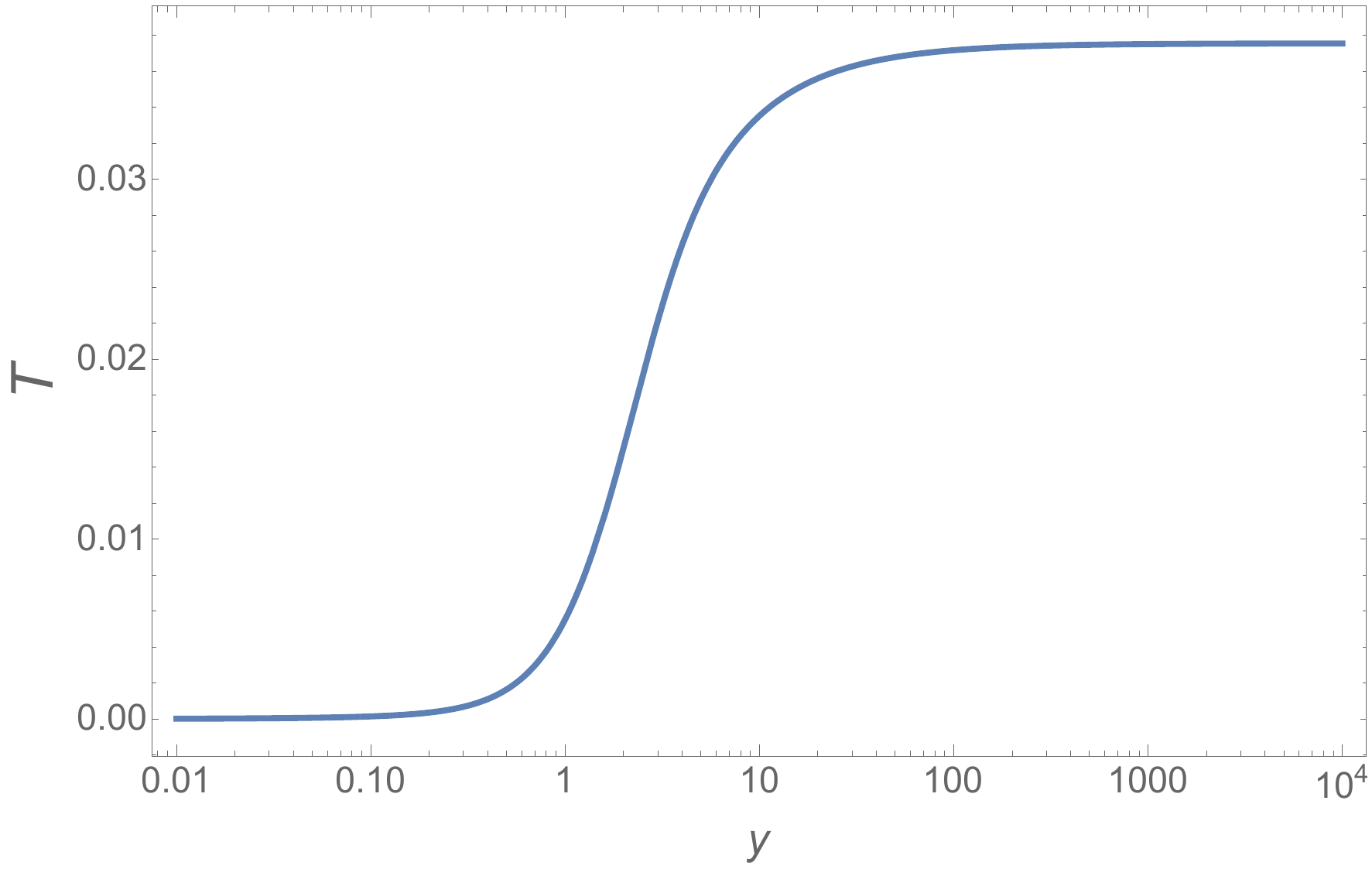}\\
\caption{Plots of the quantities $\left|{q}/{g}\right|$,  $\left|{q(y-y_1)}/{g}\right|$, $\left|{q(y-y_1)(y-y_2)}/{{  {g}}}\right|$,
 and the error control function ${\cal{T}}$  for
$k = 5.0$,  $\beta = 4.9$ and $q_0=1/2$, for which we have $y_1 = y_2^* =0.200335i$.} 
\label{k5b49}
\end{figure}

\begin{figure}
\includegraphics[width=8.cm]{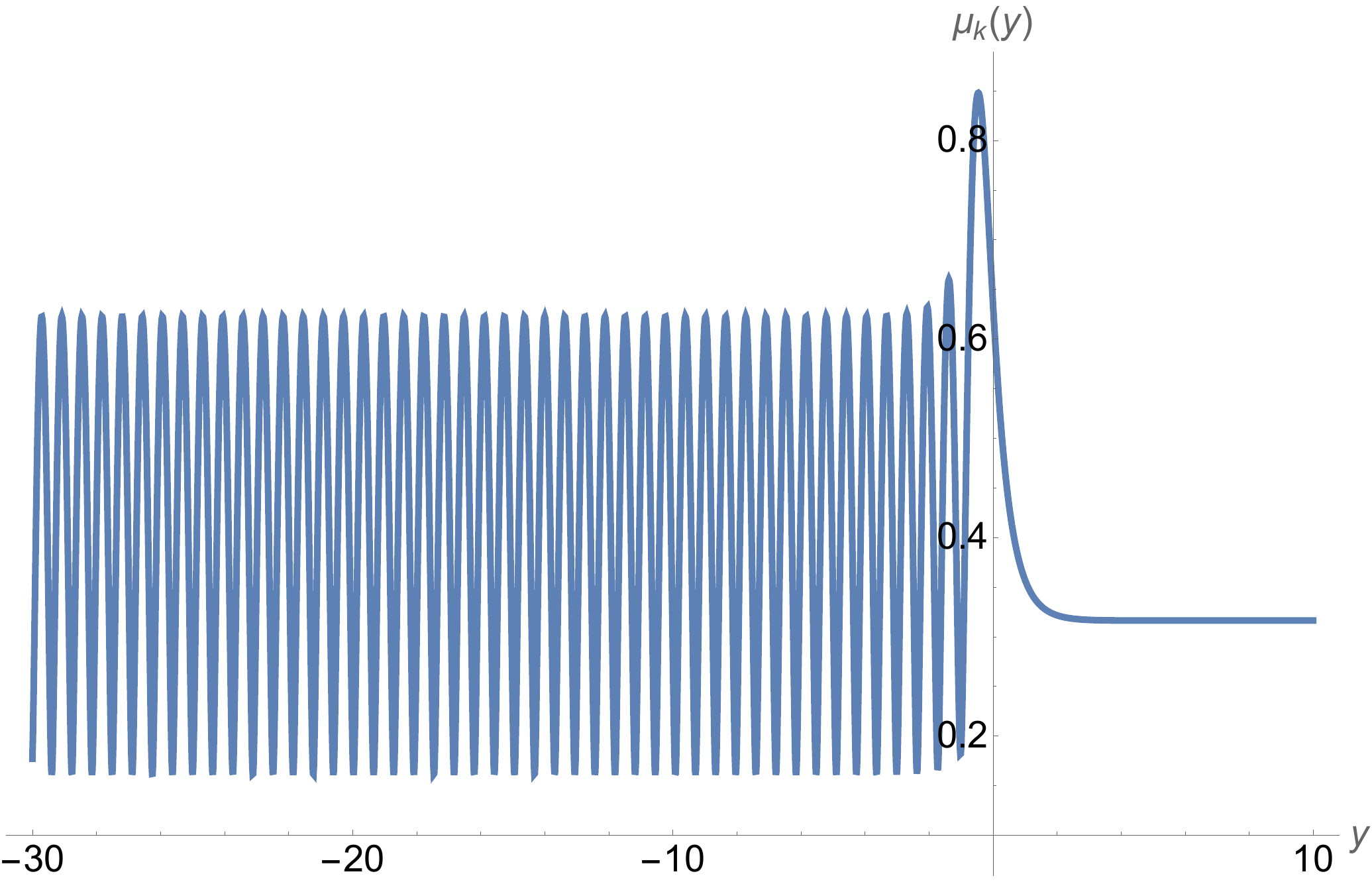}
\includegraphics[width=8.cm]{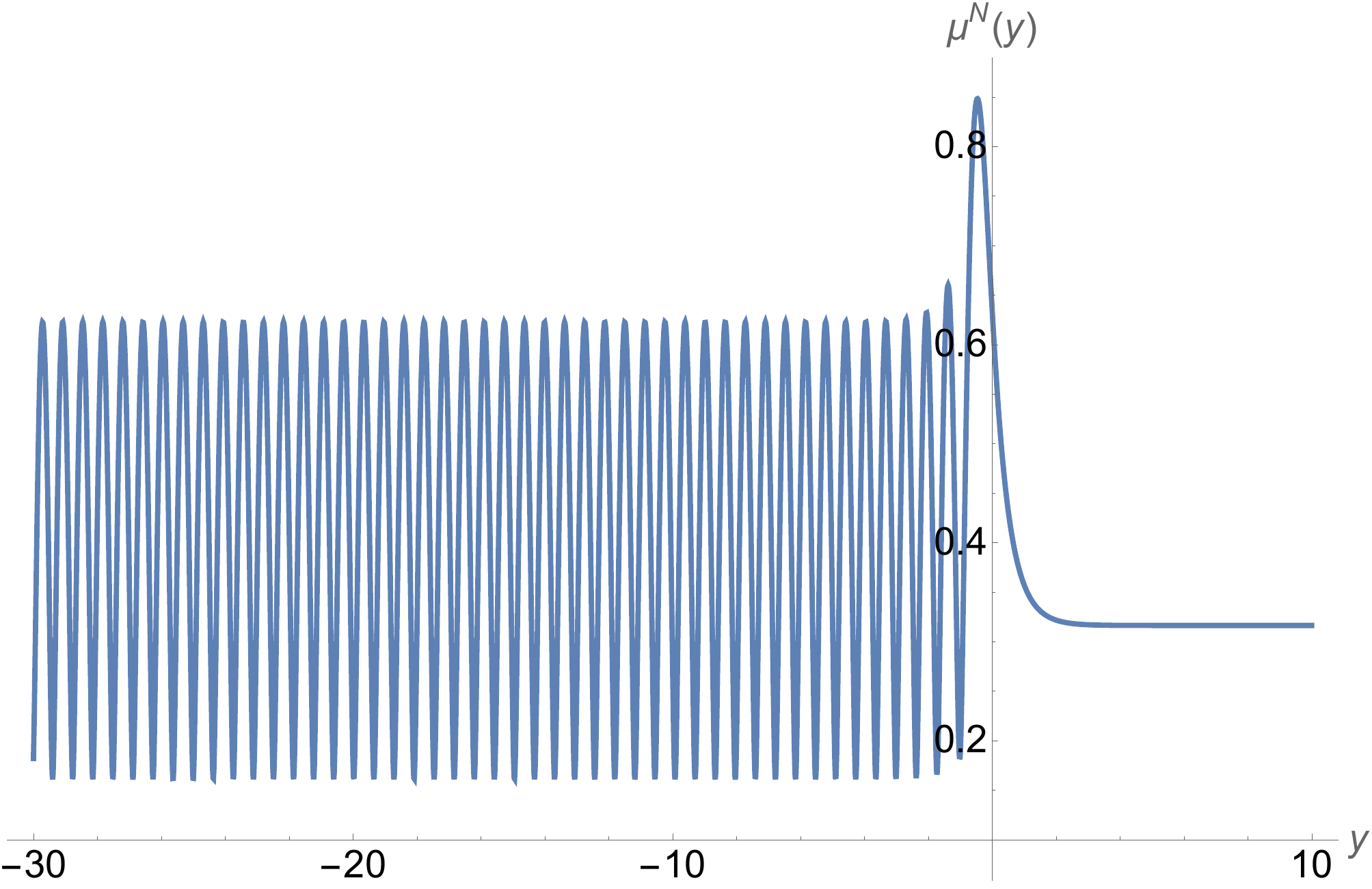}\\
\vspace{.5cm}
\includegraphics[width=8.cm]{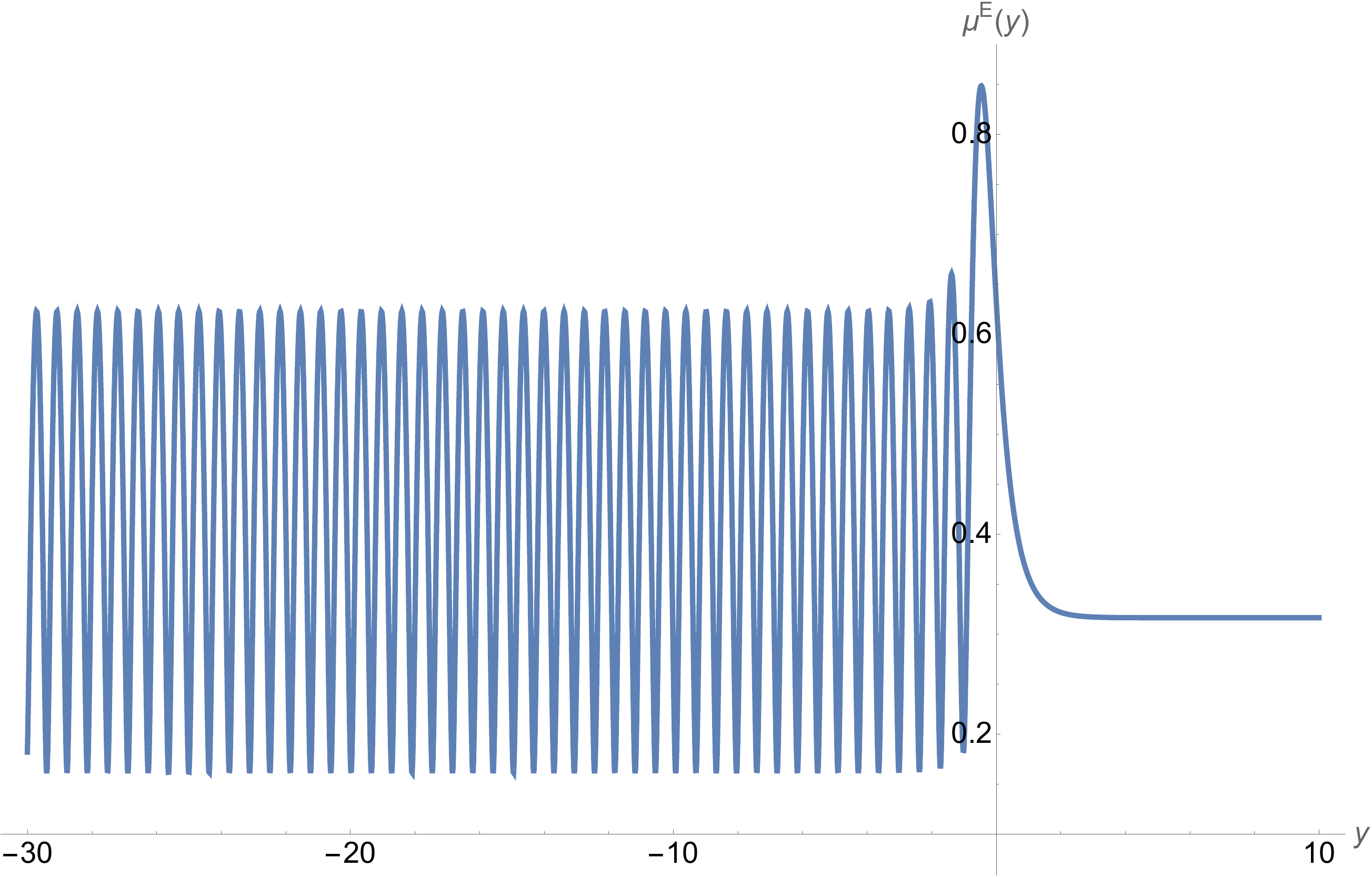}
\includegraphics[width=8.cm]{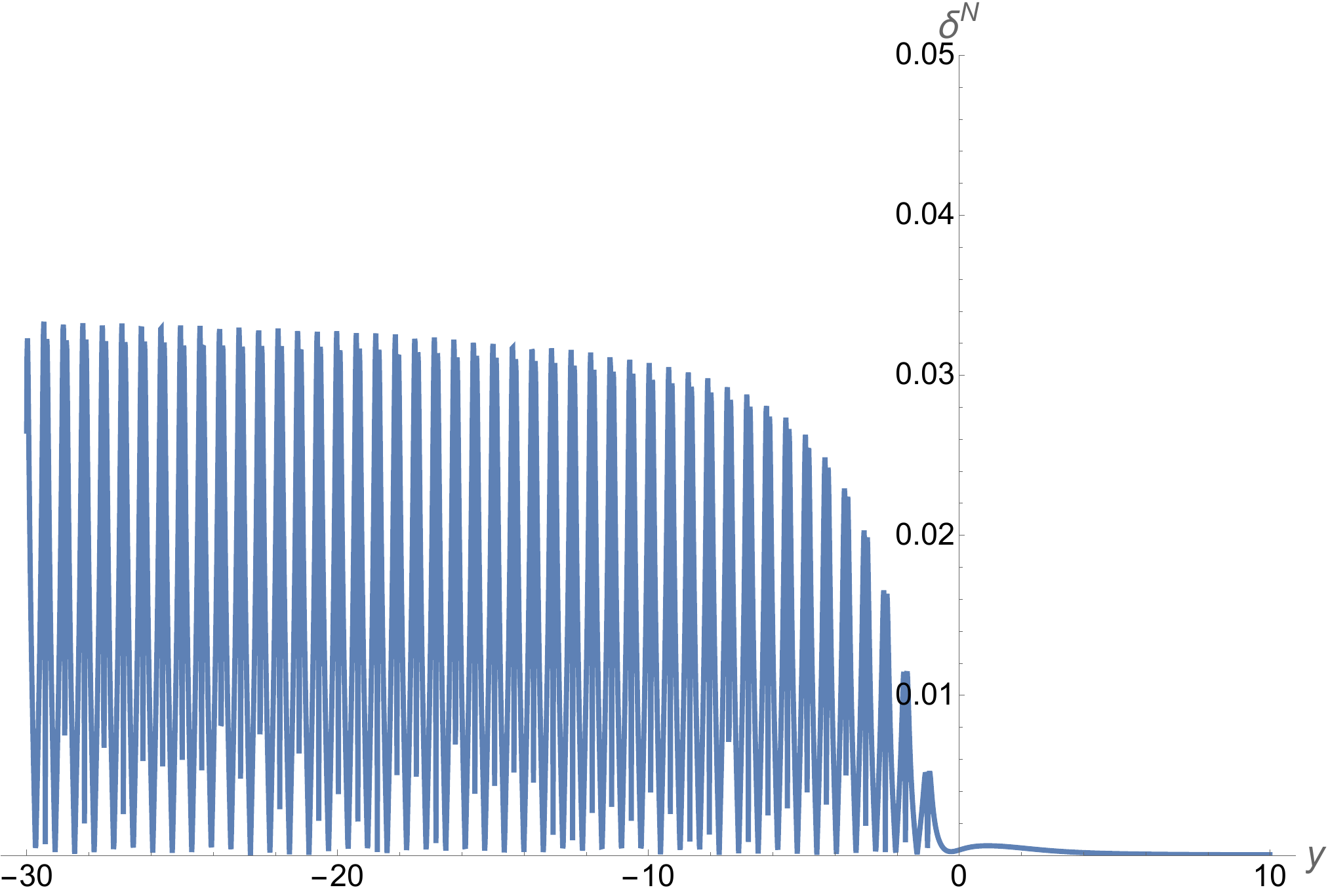}\\
\vspace{.5cm}
\includegraphics[width=8.cm]{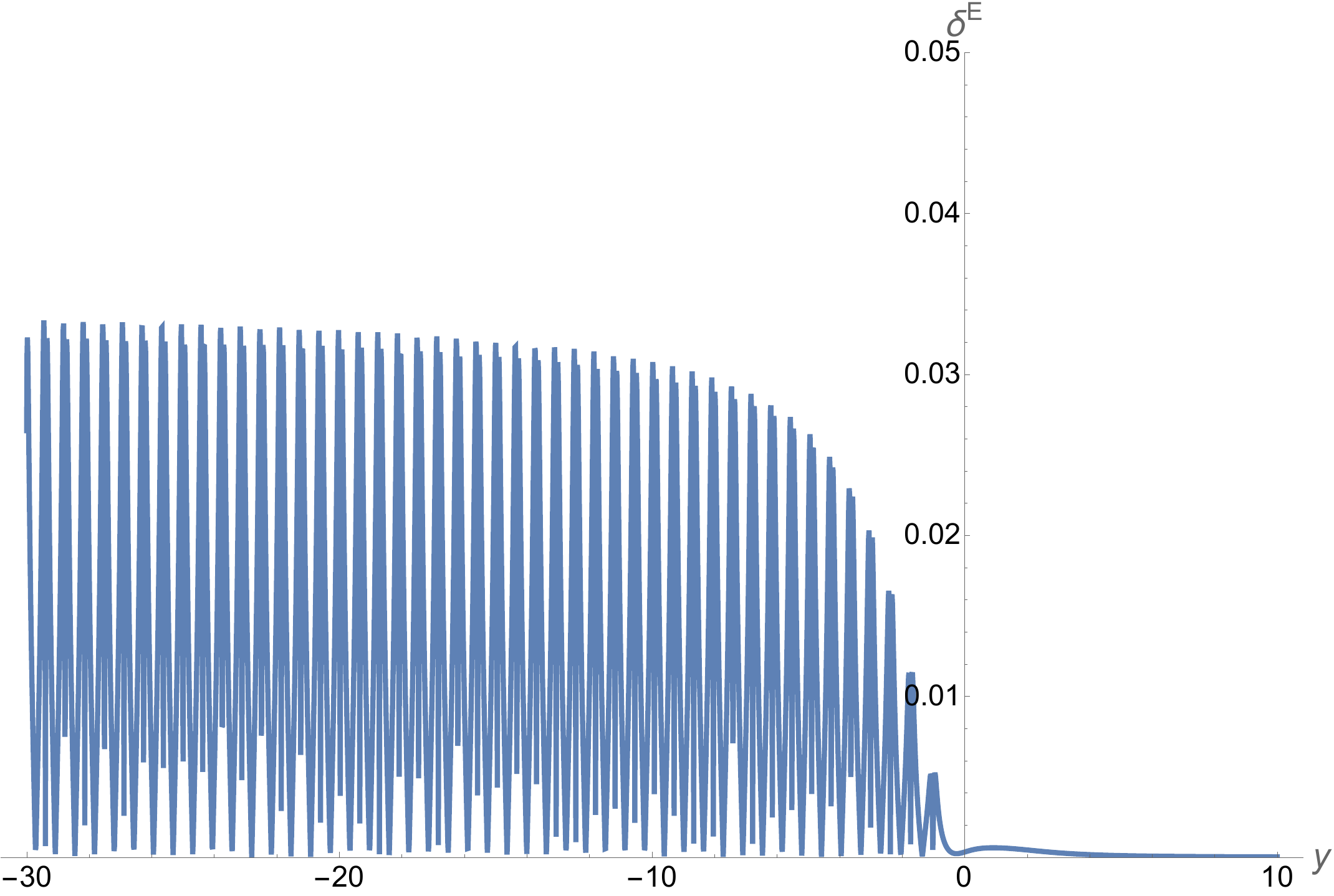}
\includegraphics[width=8.cm]{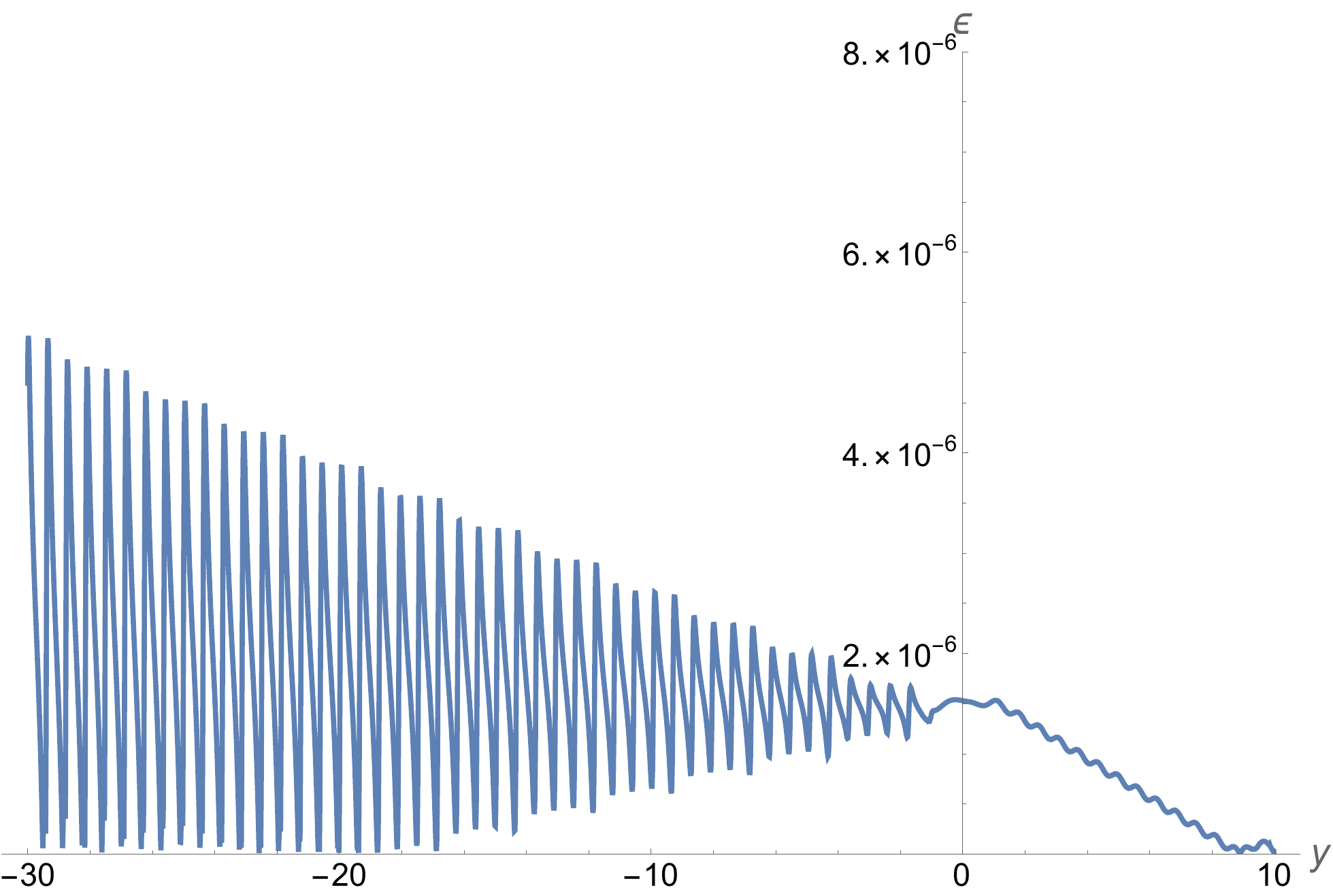}\\
 \caption{Plots of the mode functions $\mu_k(y), \; \mu^N_k(y), \mu^E_k(y)$ and their
relative differences    $\delta^N(y), \; \delta^E(y)$ and $\epsilon(y)$  for
$k = 5$,  $\beta = 4.9$ and $q_0=1/2$.
}
\label{k5b49_2}
\end{figure}

\end{widetext}

\subsection{$\beta^2 \simeq k^2$}  

In this case, depending on $k \gtrsim \beta$ or $k \lesssim \beta$, the function $g(y)$ has different properties, as shown in Fig. \ref{fig1}. Therefore, in the following {   {subsections}} let us consider them separately. 

\subsubsection{$k \gtrsim \beta$}

 When $k \gtrsim \beta$
the function $g(y)$ is always non-positive for $y \in (-\infty, \infty)$. Then, from Eqs.(\ref{C.3}) and (\ref{definition1}) we find that
\bqn
\lb{eq3.27}
\mathscr{T}(y) \simeq \frac{q_0^2-1/4}{2\beta}\ln\left(\frac{2}{\epsilon}\right) + \frac{9}{48k} + {\cal{O}}\left(\epsilon\right),
\eqn
as $y \rightarrow \infty$, but now with $\epsilon \equiv (k-\beta)/k$. Thus, to have the error control function be finite at $y = \infty$, now we must set
\bq
\lb{q0B}
q_0^2 = \frac{1}{4},
\eq
instead of the value given by Eq.(\ref{q0}) for the case $k \gg \beta^2$. 
In Fig. \ref{k5b49}, we plot the quantities $\left|{q}/{g}\right|$,  $\left|{q(y-y_1)}/{g}\right|$, $\left|{q(y-y_1)(y-y_2)}/{{  {g}}}\right|$,
 and the error control function ${\cal{T}}$  for
$k = 5.0$,  $\beta = 4.9$ and $q_0=1/2$, for which we have $y_1 = y_2^* =0.200335i$. From these figures we can see clearly that the conditions (\ref{CD1}) - (\ref{CD3}) are well satisfied, and the error control function remains small all the time. Then, the corresponding quantities
$\mu_k(y)$, $\mu^{\text{N}}_k(y)$, $\mu^E_k(y)$,   $\delta^{\text{A}}(y)$ and  $\epsilon(y)$   are plotted in Fig. \ref{k5b49_2}. From the curves of $\delta^{\text{N}}(y)$ and 
$\delta^{\text{E}}(y)$ we can see that now the errors of the first-order UAA solution are $\le 4\%$, which are larger than those of the last subcase. This is  {mainly} because of the fast oscillations of the solution in the region $y < 0$. Therefore, in order to obtain solutions with high {  {precision}}, high-order approximations for this case are needed. However, we do like to note that our numerical solution still matches to the exact one very well, as shown by the {  {curve}} of $\epsilon(y)$, which is no larger than $6.0 \times 10^{-6}$.

\begin{widetext}
 
\begin{figure}
\includegraphics[width=8.5cm]{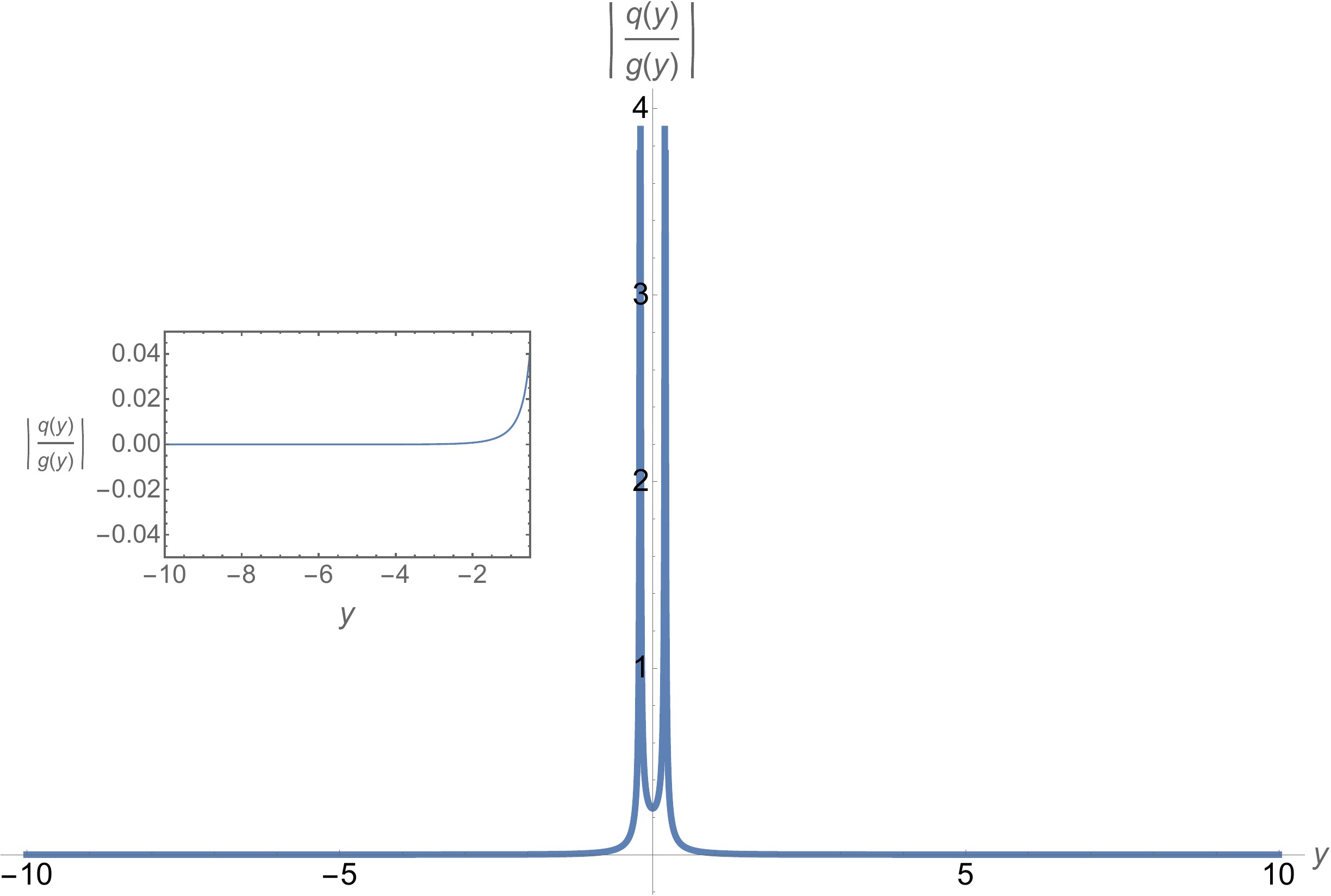}
\includegraphics[width=8.5cm]{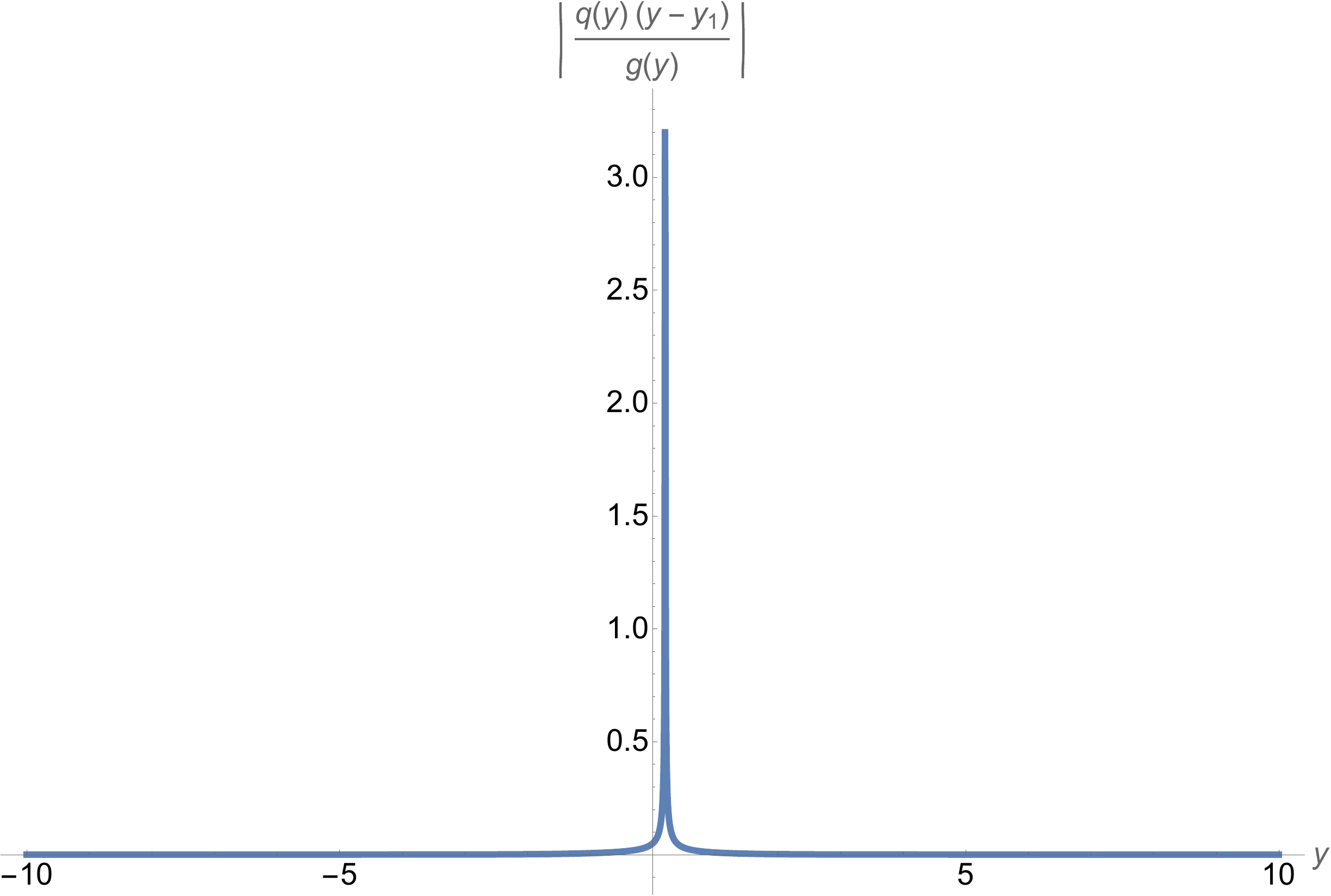}\\
\vspace{.5cm}
\includegraphics[width=8.5cm]{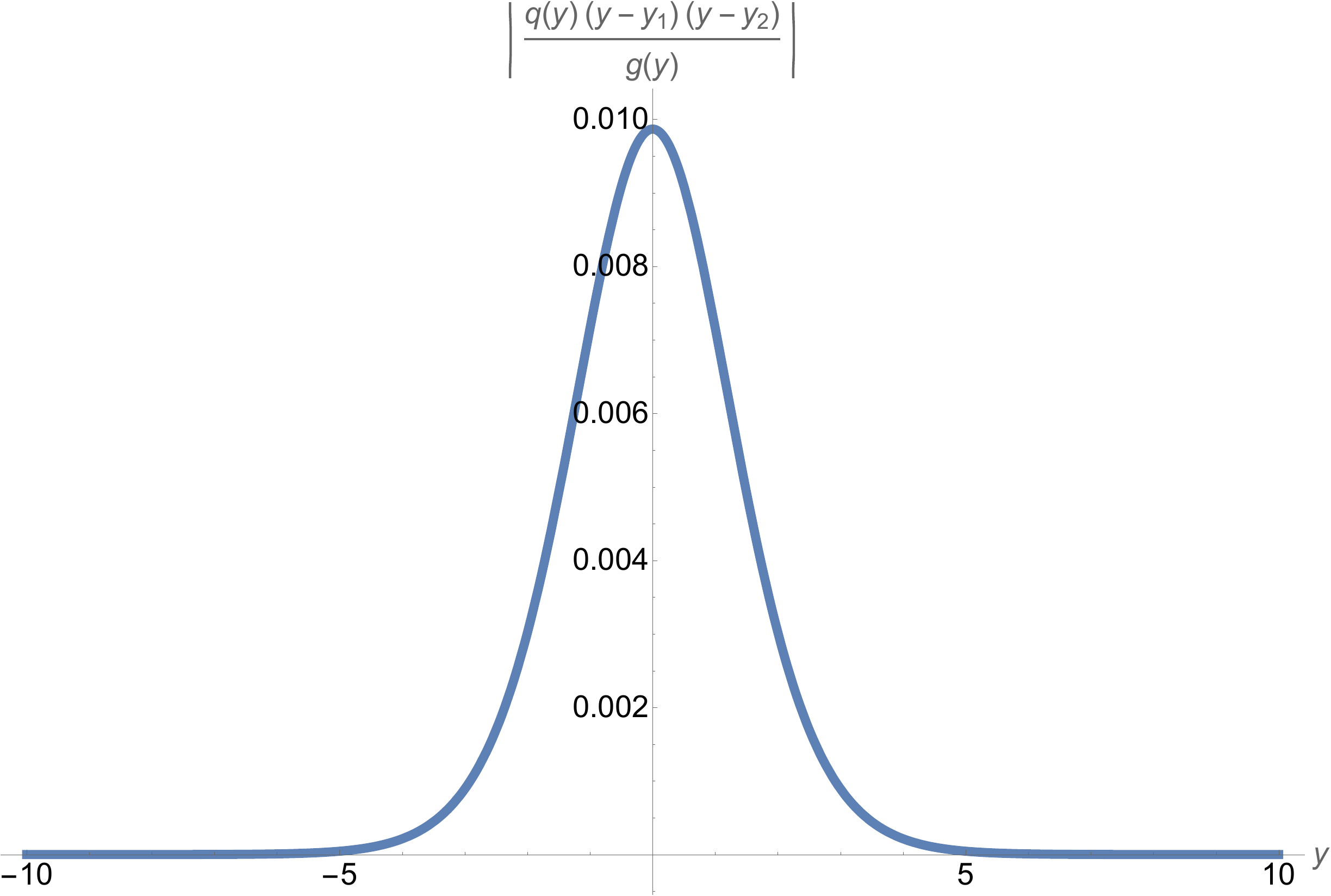}
\includegraphics[width=8.5cm]{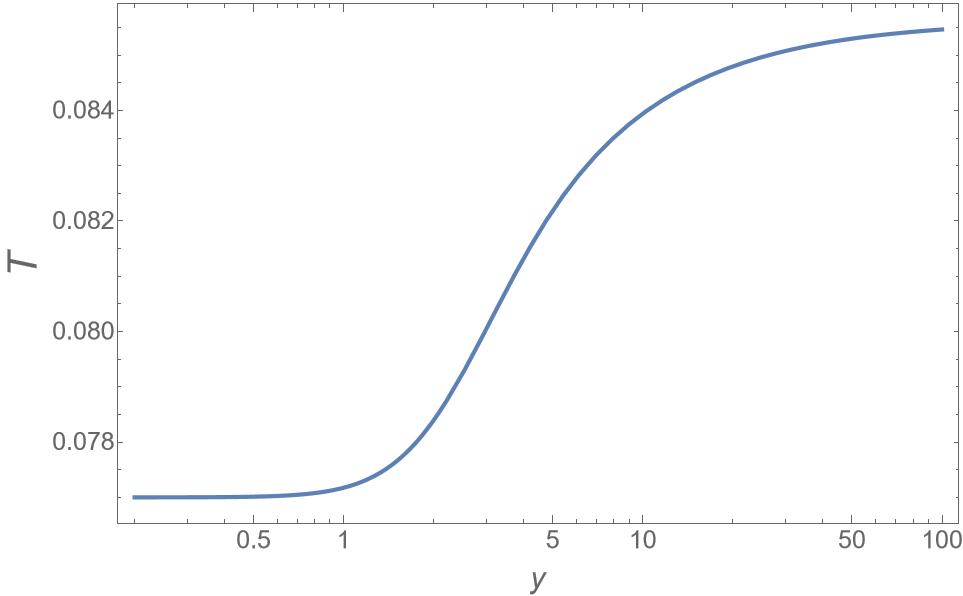}\\
\caption{Plots of the quantities $\left|{q}/{g}\right|$,  $\left|{q(y-y_1)}/{g}\right|$, $\left|{q(y-y_1)(y-y_2)}/{{  {g}}}\right|$,
 and the error control function ${\cal{T}}$  for
$k = 5.0$,  $\beta = 5.1$ and $q_0=1/2$, for which we have $y_1 = -y_2 =-0.199668$.} 
\label{k5b51}
\end{figure}
\end{widetext}

\subsubsection{$k \lesssim \beta$}

In this case, we find that  
\bq
\lb{eq3.26b}
\zeta_0^2 = \frac{2}{\pi}\left|\int_{y_1}^{y_2}{\sqrt{g(y)}\; dy}\right| = 2\left|k-\beta\right|.
\eq
On the other hand, from Eqs.(\ref{ECFb}), (\ref{C.3}) and (\ref{C.6}) we find that
\bqn
\lb{eq3.27b}
\mathscr{T}(y) \simeq \begin{cases}
\frac{\zeta(0)\left(6\zeta^2_0-\zeta^2(0)\right)}{12\zeta^2_0\left(\zeta^2_0-\zeta^2(0)\right)^{3/2}}, & y \rightarrow 0, \cr
\frac{\pi\left(q_0^2-1/4\right)}{2\beta}, & y \rightarrow y_2, \cr
\end{cases}
\eqn
 where $\zeta (0) \equiv \left.\zeta(y)\right|_{y= 0} < \zeta_0$. Note that in calculating the error control function near the turning point $y \simeq y_2$, we  {have} used the relation
 \bq
 \lb{eq3.27c}
 \frac{\beta}{k^2\sqrt{\beta^2 - k^2}}\left(\beta^2x^2 - k^2\right)^{3/2} \simeq \frac{1}{\zeta_0}(\zeta_0^2 - \zeta^2)^{3/2},
 \eq
 so that the divergence of the second term of $\mathscr{T}_2$ cancels exactly with that of $\mathscr{T}_3$. Eq.(\ref{eq3.27c}) can be obtained directly from the relation $\sqrt{g} dy = \sqrt{\zeta_0^2-\zeta^2}d\zeta$ for the case $g \ge 0$.
 Similarly, it can be shown that 
 \bqn
\lb{eq3.27d}
\mathscr{T}(y) \simeq 
\frac{q_0^2-1/4}{2\beta} \ln\left(\frac{2}{\epsilon}\right),  \;\;\;\;  y \rightarrow \infty.
\eqn
 It is clear that to minimize the errors, in the present case $q_0^2$ must be also chosen to be
 \bq
\lb{q0C}
q_0^2 = \frac{1}{4},
\eq
as that given by Eq.(\ref{q0B}).  %}
 In Fig. \ref{k5b51}, we plot the quantities $\left|{q}/{g}\right|$,  $\left|{q(y-y_1)}/{g}\right|$, $\left|{q(y-y_1)(y-y_2)}/{{  {g}}}\right|$,
 and the error control function ${\cal{T}}$  for
$k = 5.0$,  $\beta = 5.1$ and $q_0=1/2$, for which we have $y_1 = -y_2 =-0.199668$. It is clear that in this case the two turning points are very {  {close}}, and the conditions 
 $\left|{q}/{g}\right| \ll 1$ and $\left|{q(y-y_1)}/{g}\right| \ll 1$ are violated near these points. But, the condition  $\left|{q(y-y_1)(y-y_2)}/{{  {g}}}\right| \ll 1$ holds near them. So, 
the conditions (\ref{CD1}) - (\ref{CD3}) are also satisfied, and the error control function remains small all the time.

Then, the corresponding quantities
$\mu_k(y)$, $\mu^{\text{N}}_k(y)$, $\mu^E_k(y)$,   $\delta^{\text{A}}(y)$ and  $\epsilon(y)$   are plotted in Fig. \ref{k5b51_2}. From the curves of $\delta^{\text{N}}(y)$ and 
$\delta^{\text{E}}(y)$ we can see that now the errors of the first-order UAA solution are $\lesssim 10\%$. Similar to the last subcase,  this is  {mainly} because of the fast oscillations of the solution in the region $y < 0$. Therefore, in order to obtain high precision, high-order approximations for this case are needed, too. In addition, our numerical solution still matches well to the exact one, as shown by the  {curve} of $\epsilon(y)$, which is no larger than $2.0 \times 10^{-6}$.

\begin{widetext}

\begin{figure}
\includegraphics[width=8.cm]{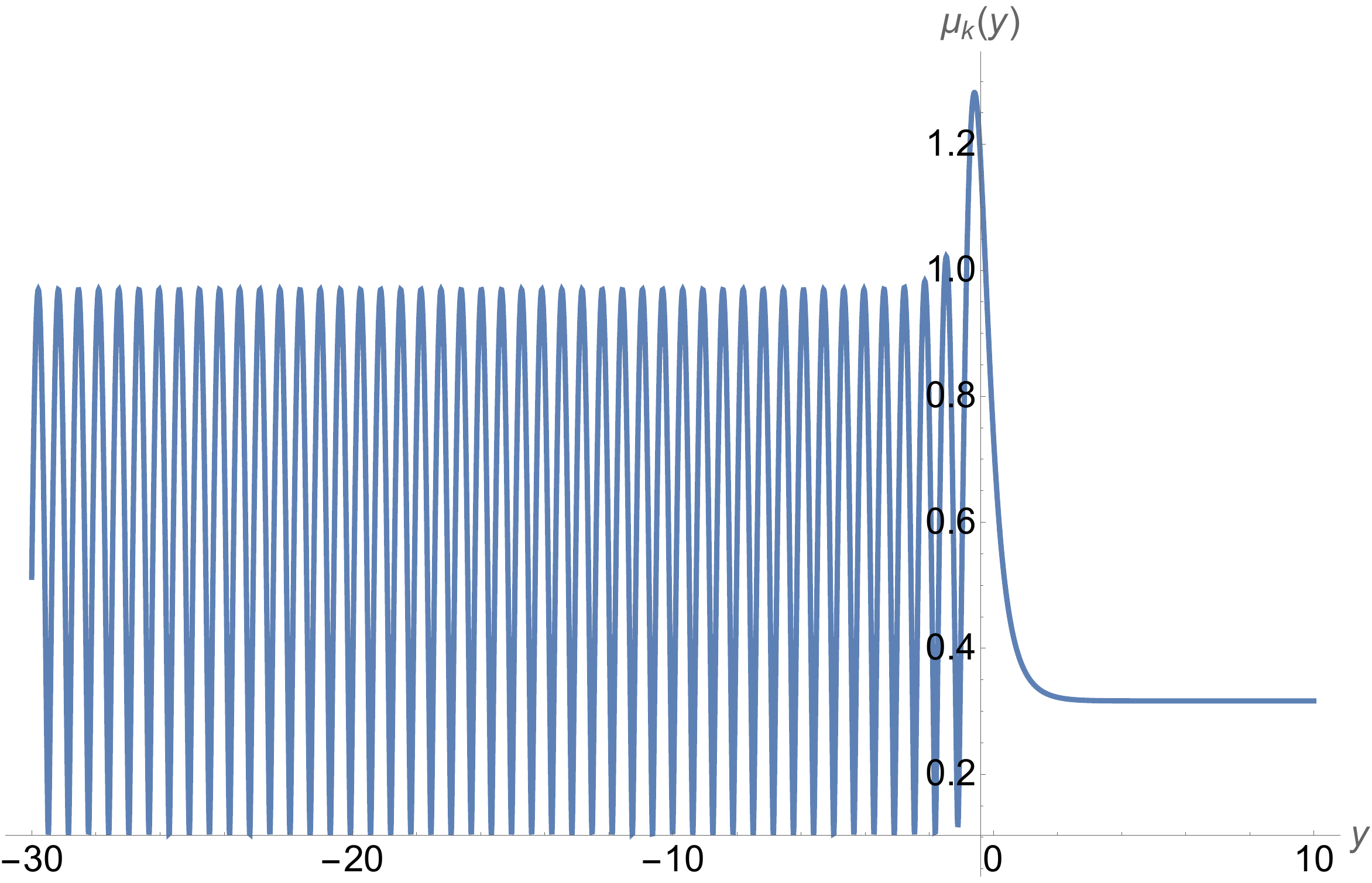}
\includegraphics[width=8.cm]{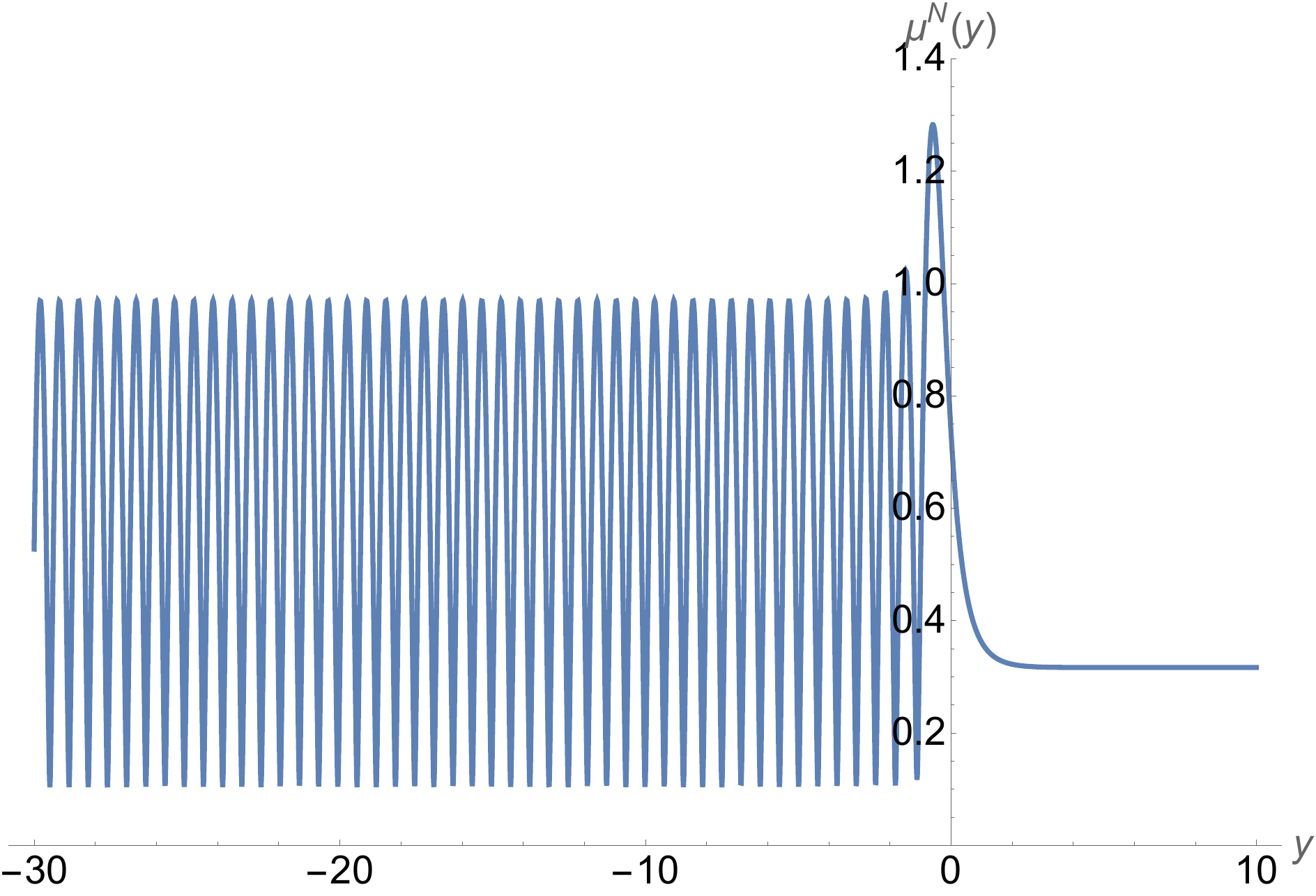}\\
\vspace{.5cm}
\includegraphics[width=8.cm]{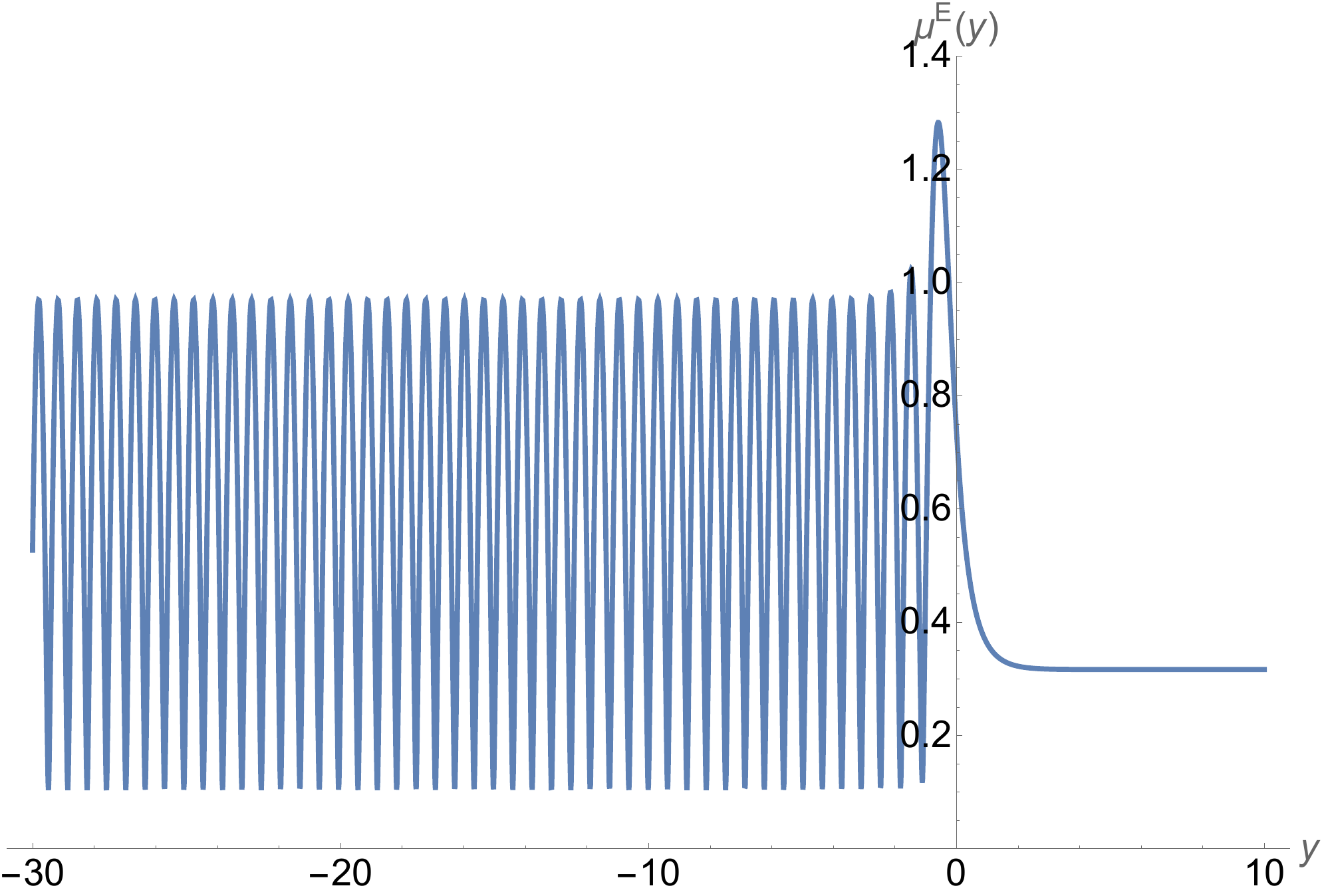}
\includegraphics[width=8.cm]{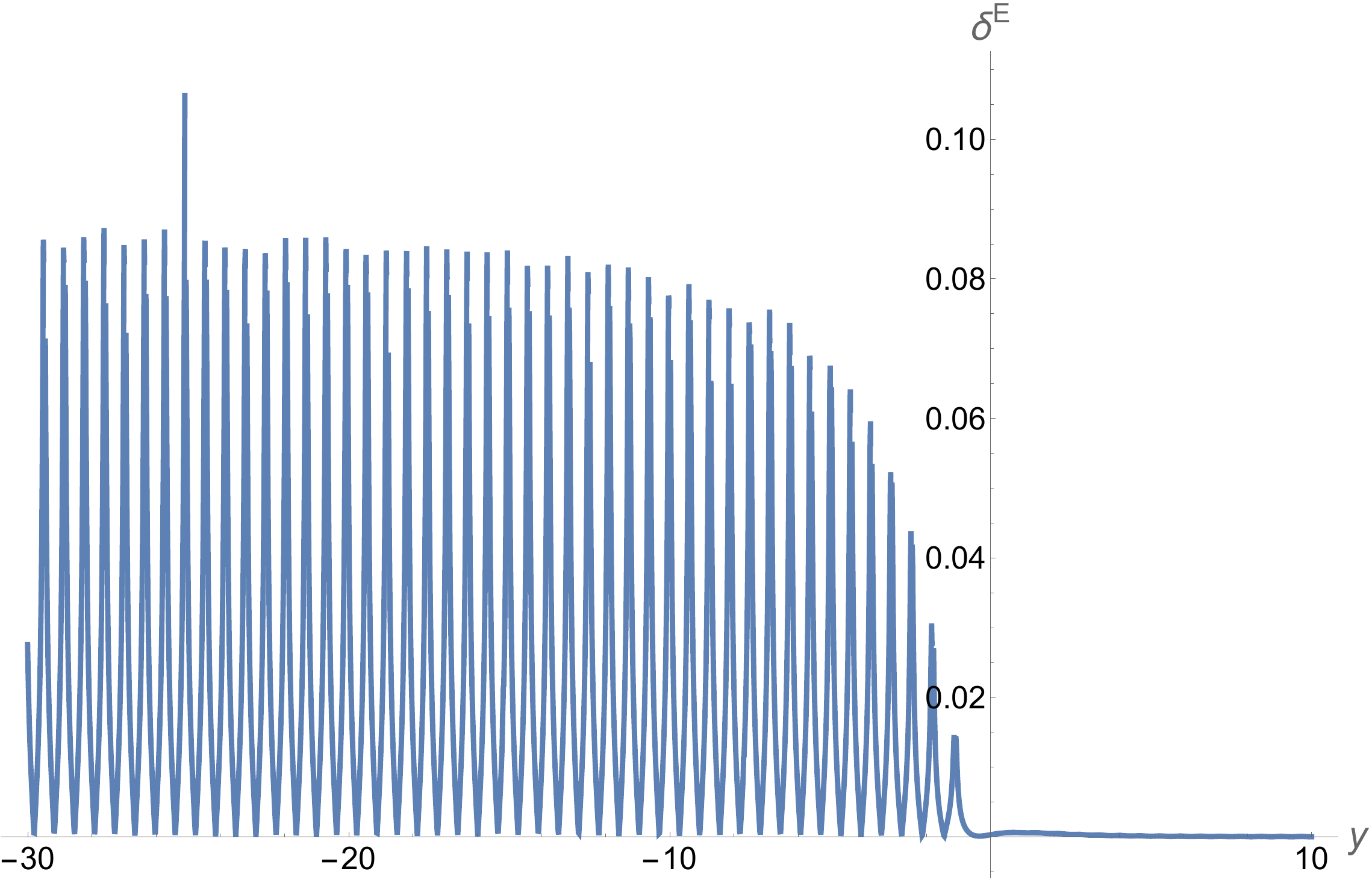}\\
\vspace{.5cm}
\includegraphics[width=8.cm]{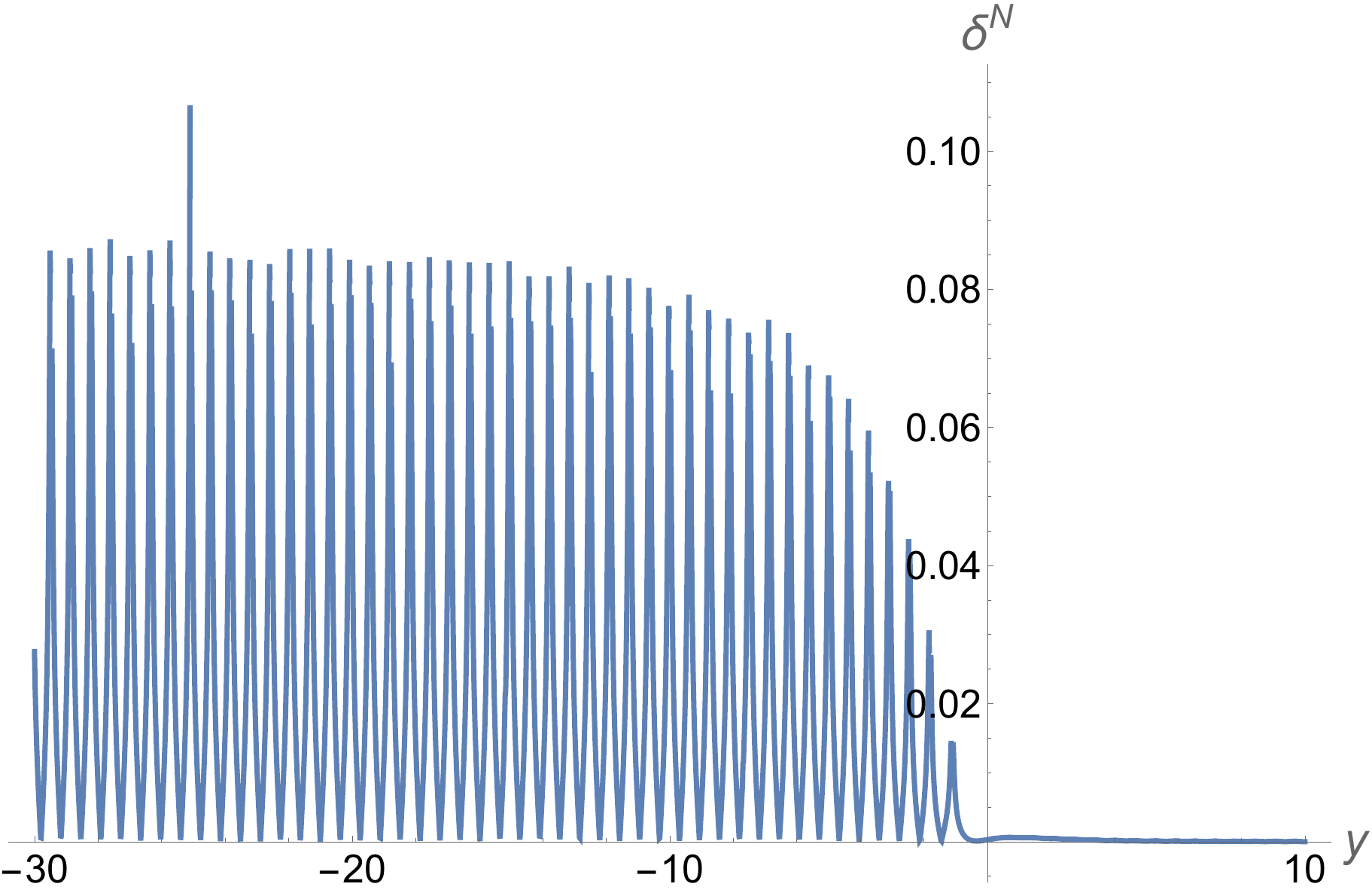}
\includegraphics[width=8.cm]{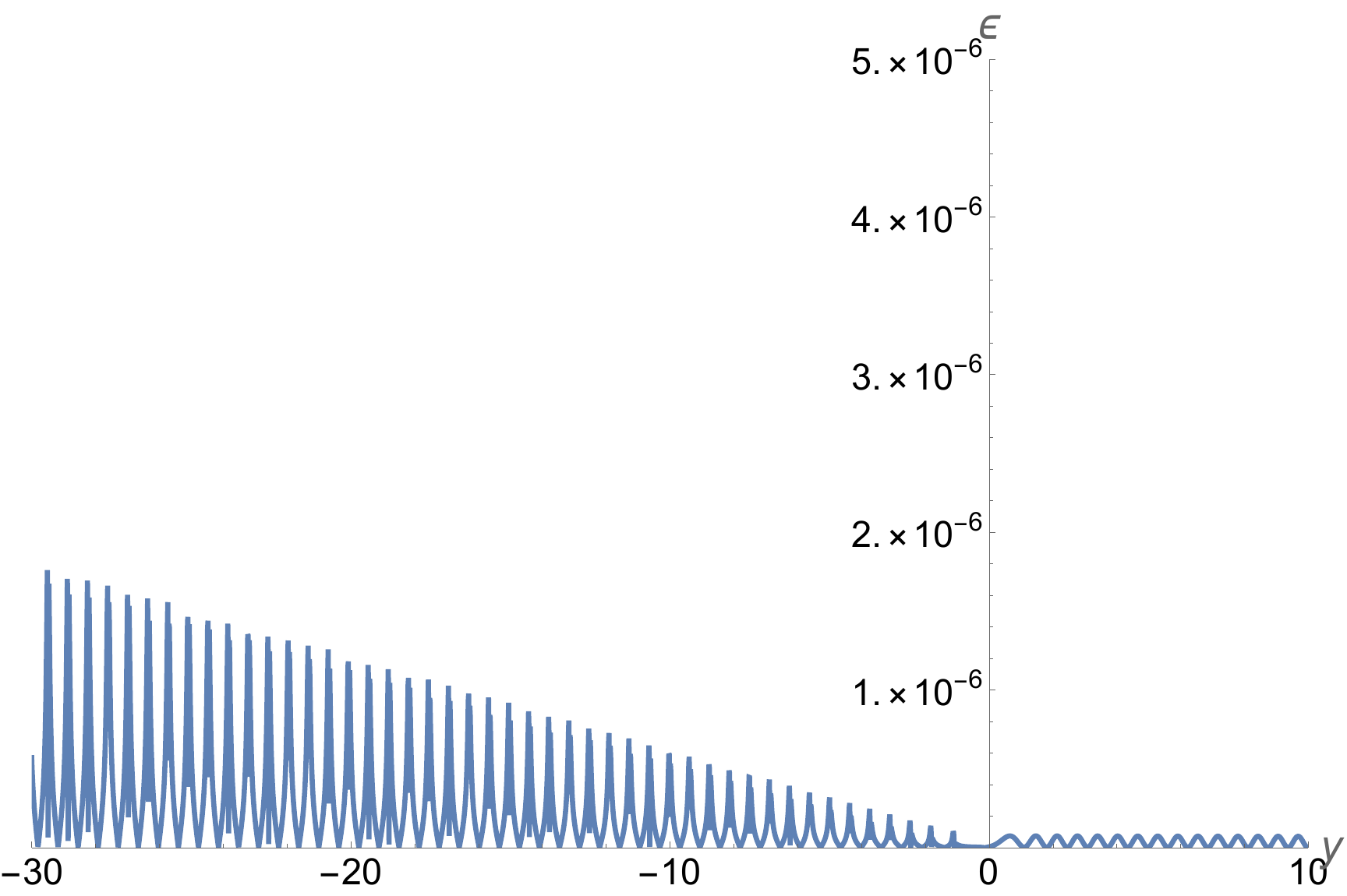}\\
 \caption{Plots of the mode functions $\mu_k(y), \; \mu^N_k(y), \mu^E_k(y)$ and their
relative differences    $\delta^N(y), \; \delta^E(y)$ and $\epsilon(y)$  for
$k = 5$,  $\beta = 5.1$ and $q_0=1/2$.
}
\label{k5b51_2}
\end{figure}

\begin{figure}
\includegraphics[width=8.5cm]{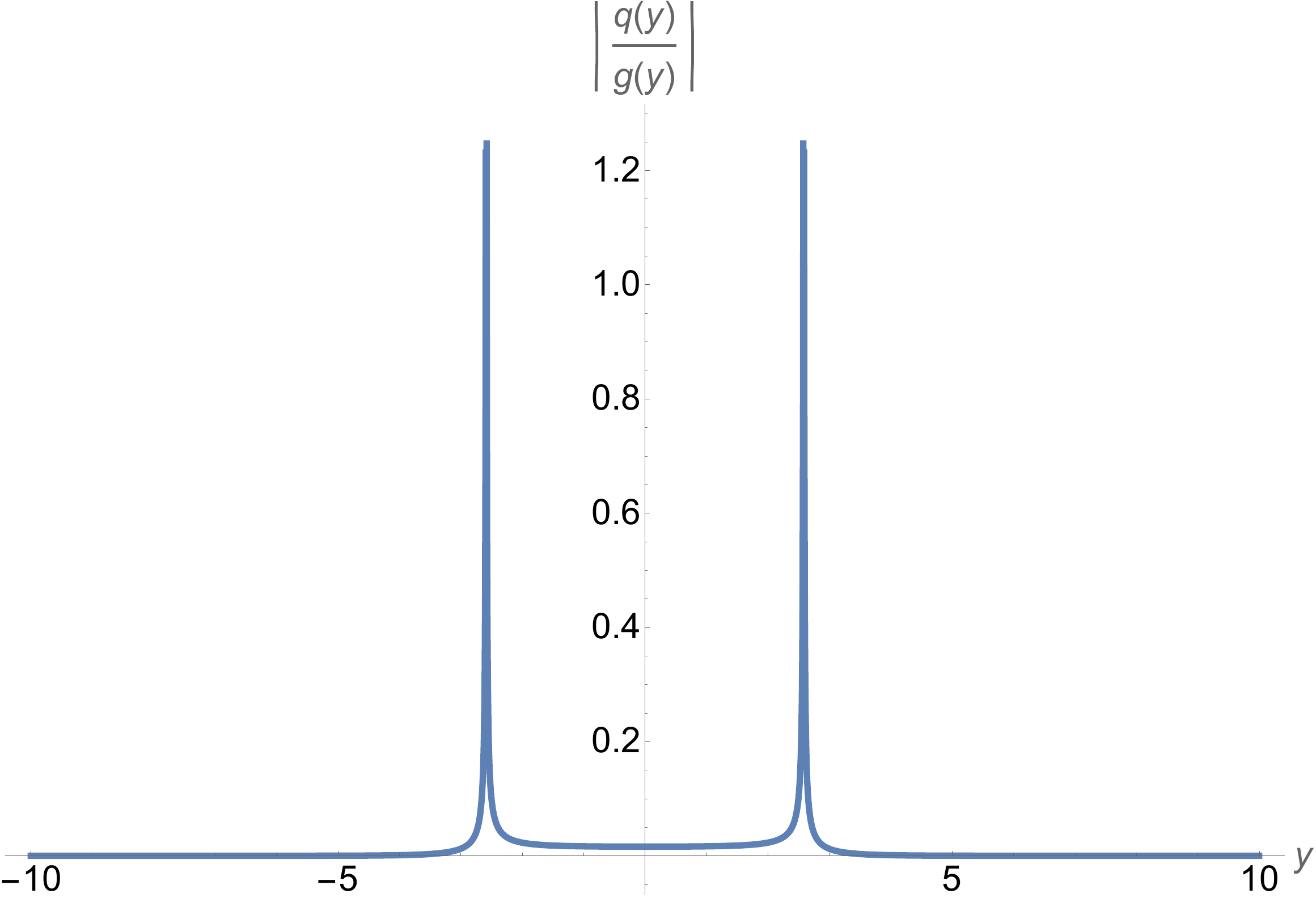}
\includegraphics[width=8.5cm]{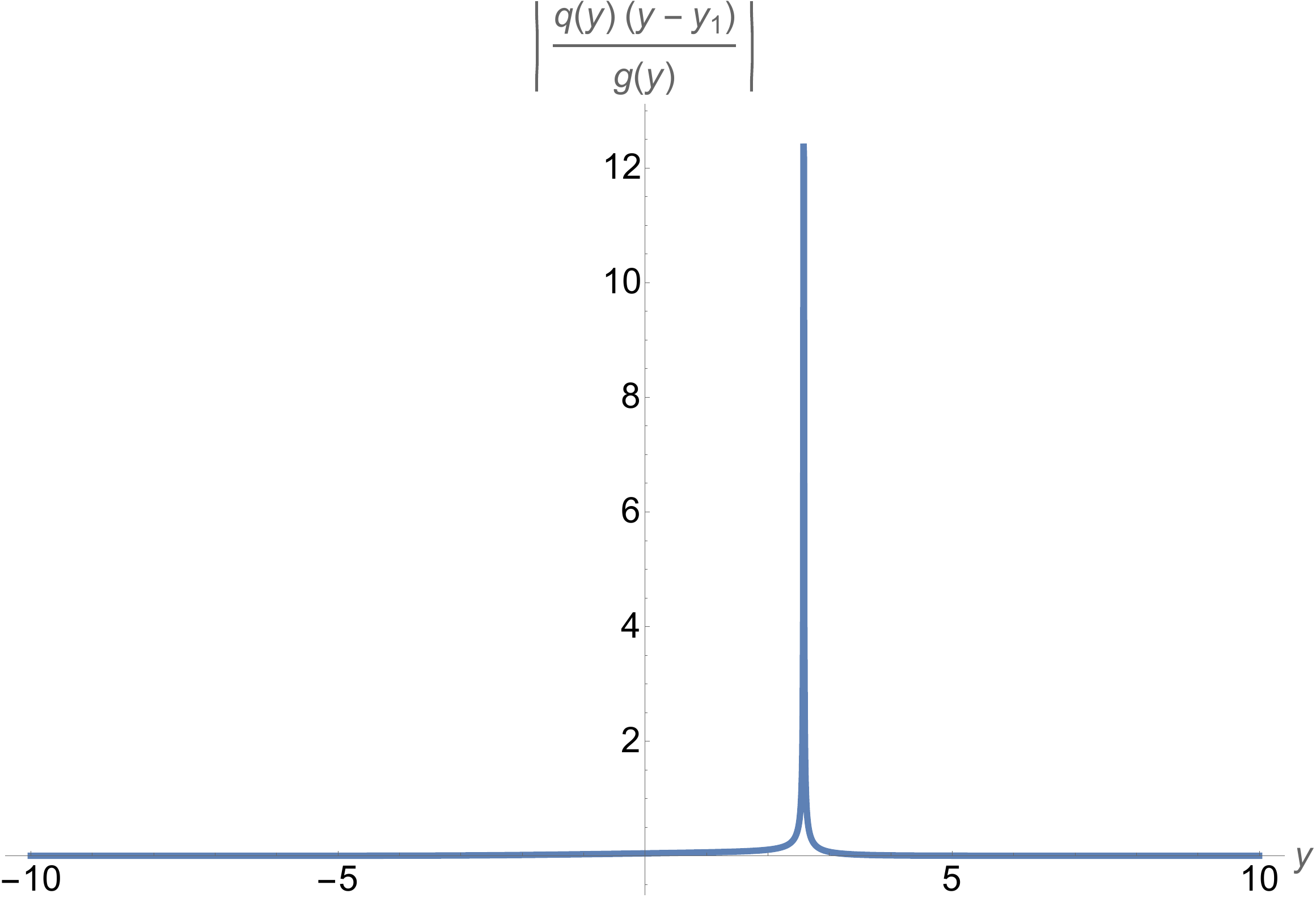}\\
\vspace{.5cm}
\includegraphics[width=8.5cm]{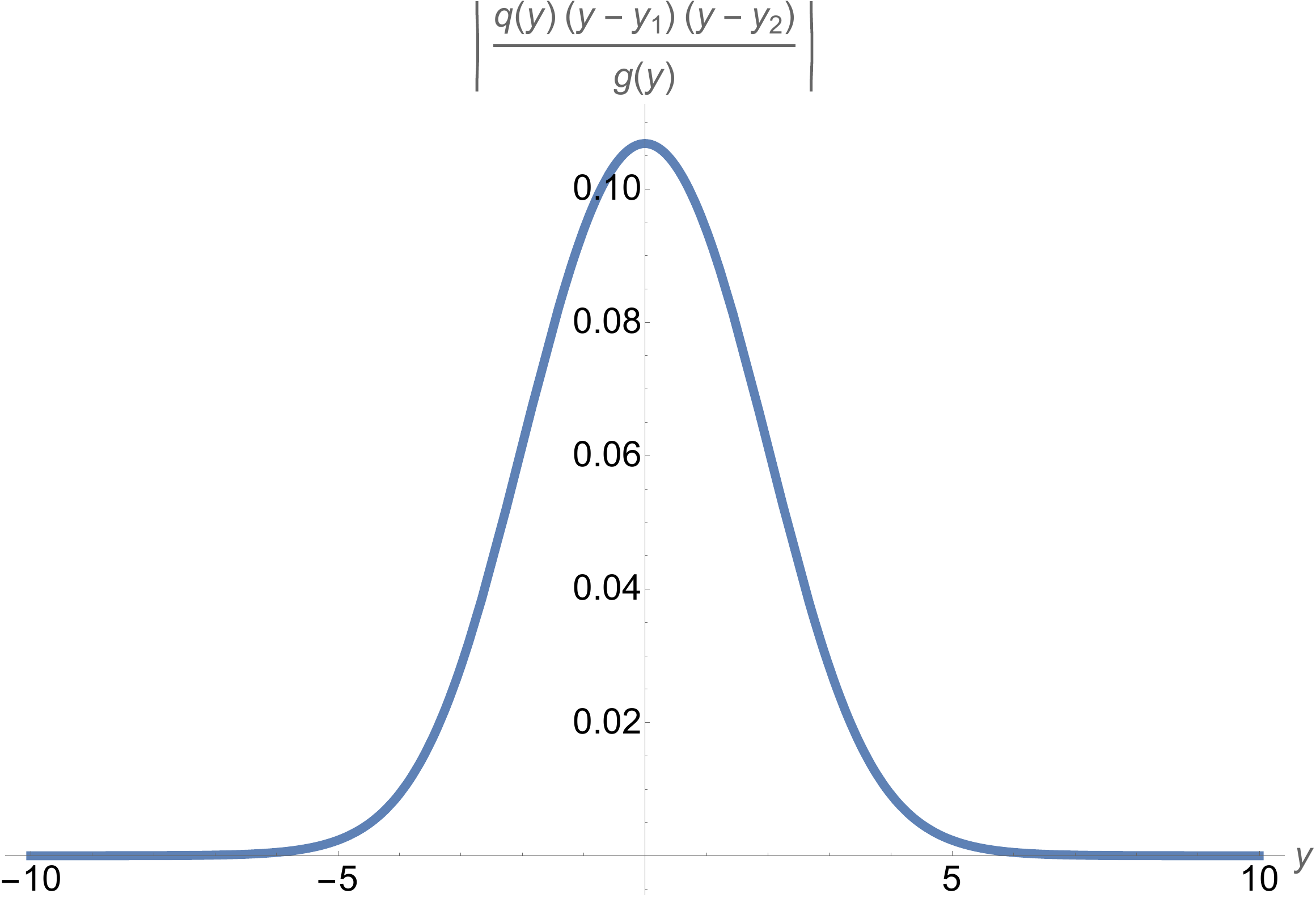}
\includegraphics[width=8.5cm]{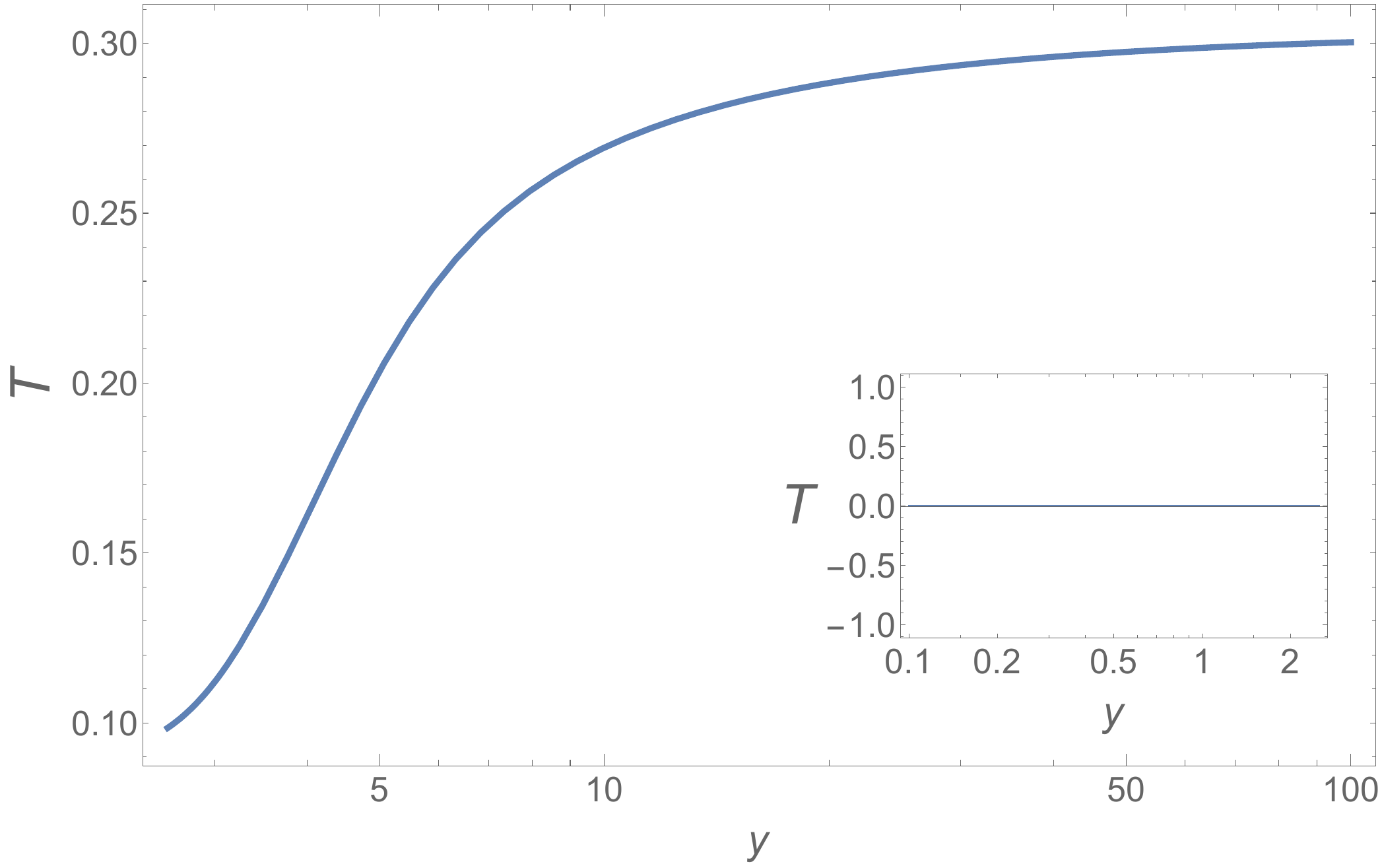}\\
\caption{Plots of the quantities $\left|{q}/{g}\right|$,  $\left|{q(y-y_1)}/{g}\right|$, $\left|{q(y-y_1)(y-y_2)}/{{  {g}}}\right|$,
 and the error control function ${\cal{T}}$  for
$k = 0.6$,  $\beta = 4.0$ and $q_0=1/2$, for which we have $y_1 = -y_2 \simeq -2.58459$.} 
\label{k06b4}
\end{figure}

\begin{figure*}[htbp]
\includegraphics[width=8.cm]{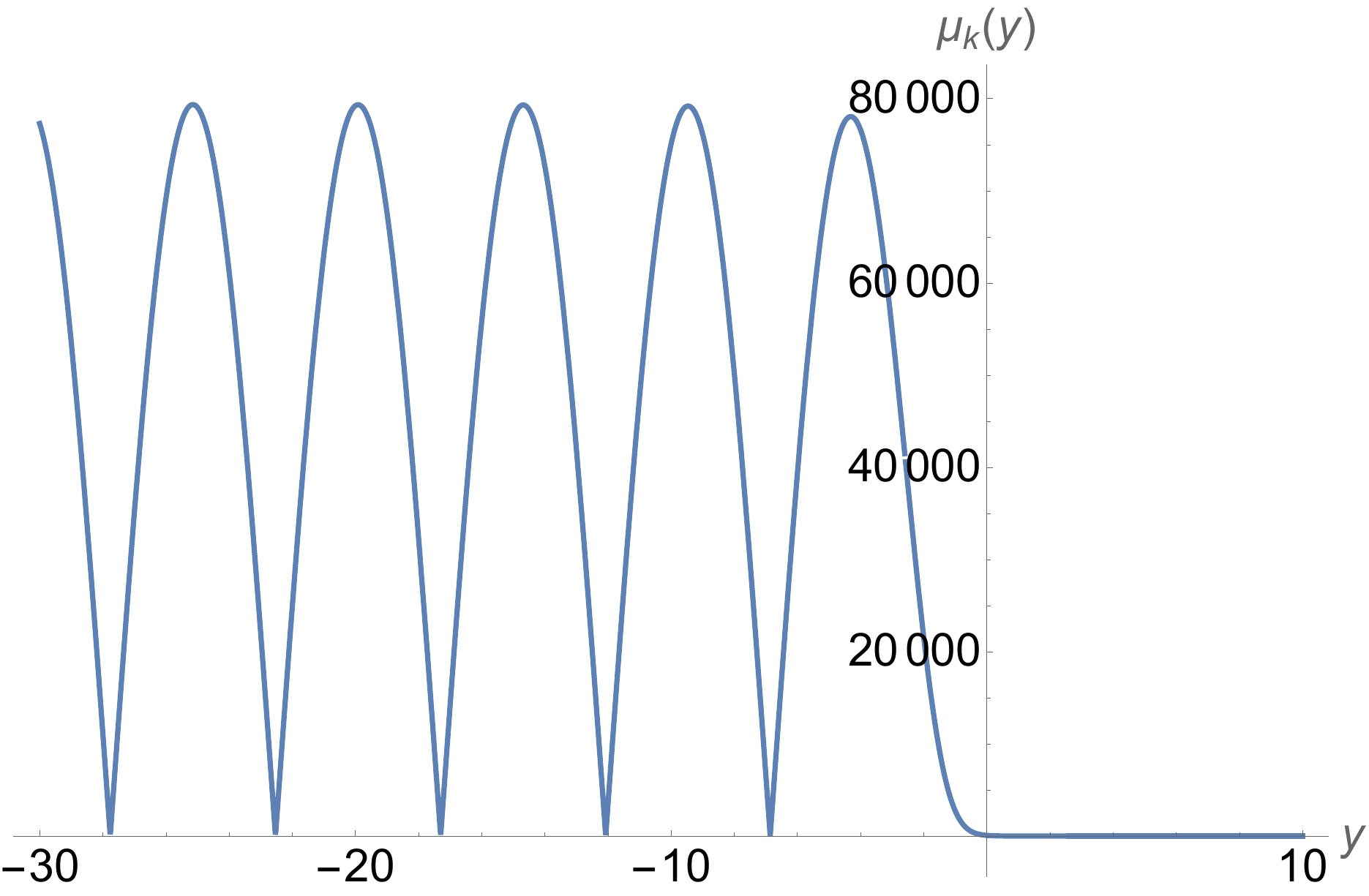}
\includegraphics[width=8.cm]{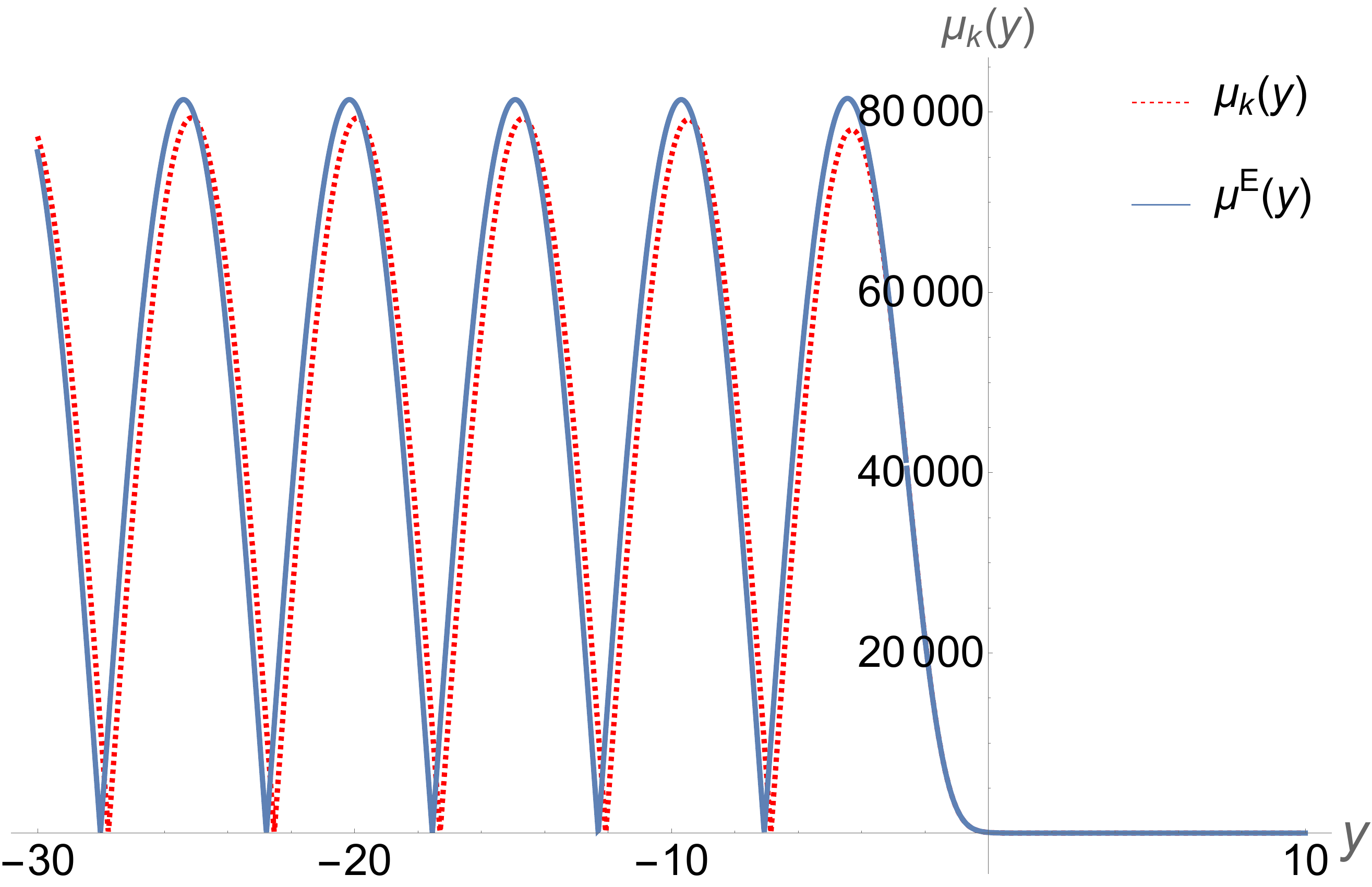}\\
\vspace{.5cm}
(a) $~~~~~~~~~~~~~~~~~~~~~~~~~~~~~~~~~~~~~~~~~~~~~~~~~~~~~~~~~~~~~~~~~~~~~~~~~~$ (b)\\
\vspace{.5cm}
\includegraphics[width=8.cm]{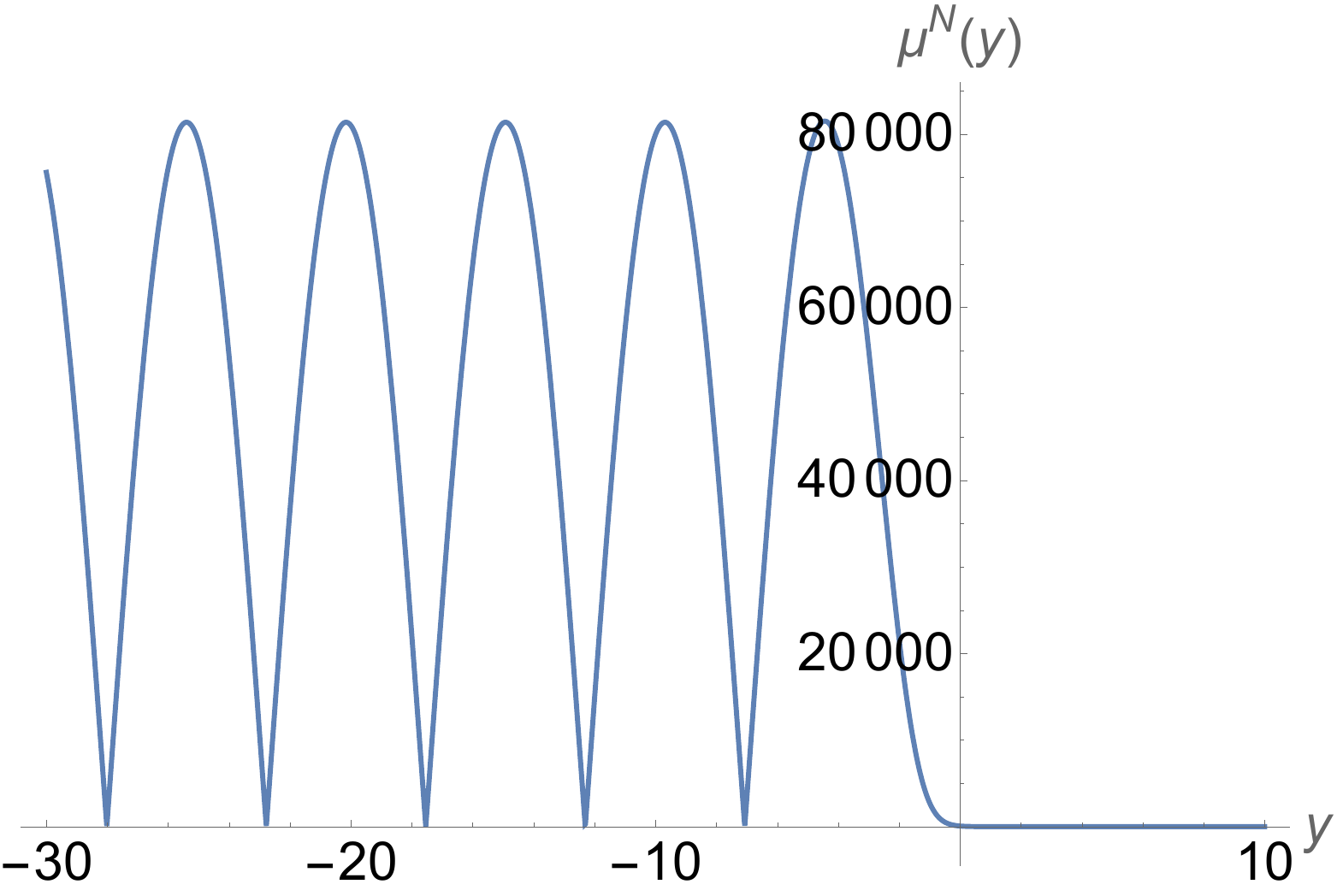}
\includegraphics[width=8.cm]{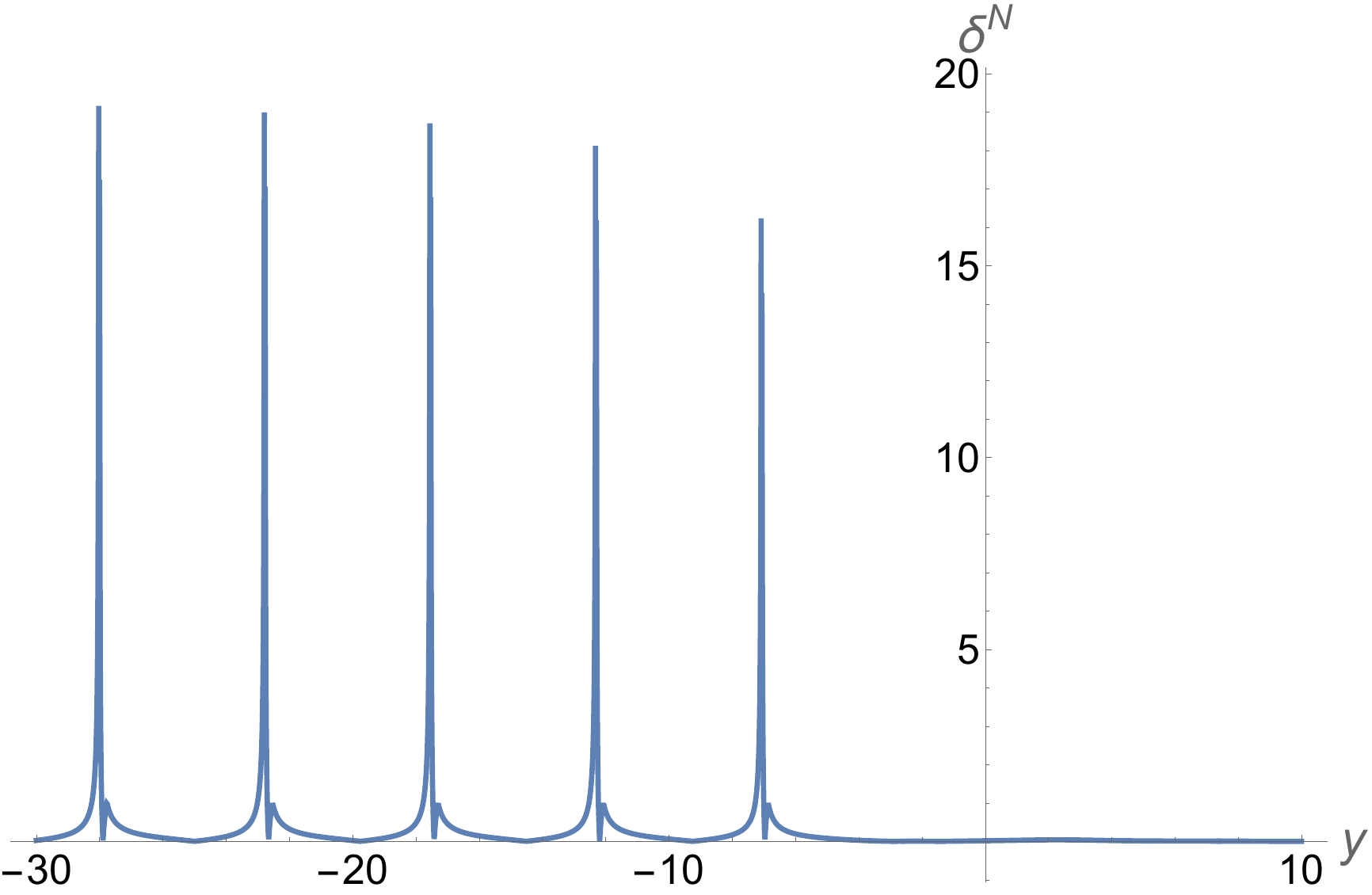}\\
\vspace{.5cm}
(c) $~~~~~~~~~~~~~~~~~~~~~~~~~~~~~~~~~~~~~~~~~~~~~~~~~~~~~~~~~~~~~~~~~~~~~~~~~~$ (d)\\
\vspace{.5cm}
\includegraphics[width=8.cm]{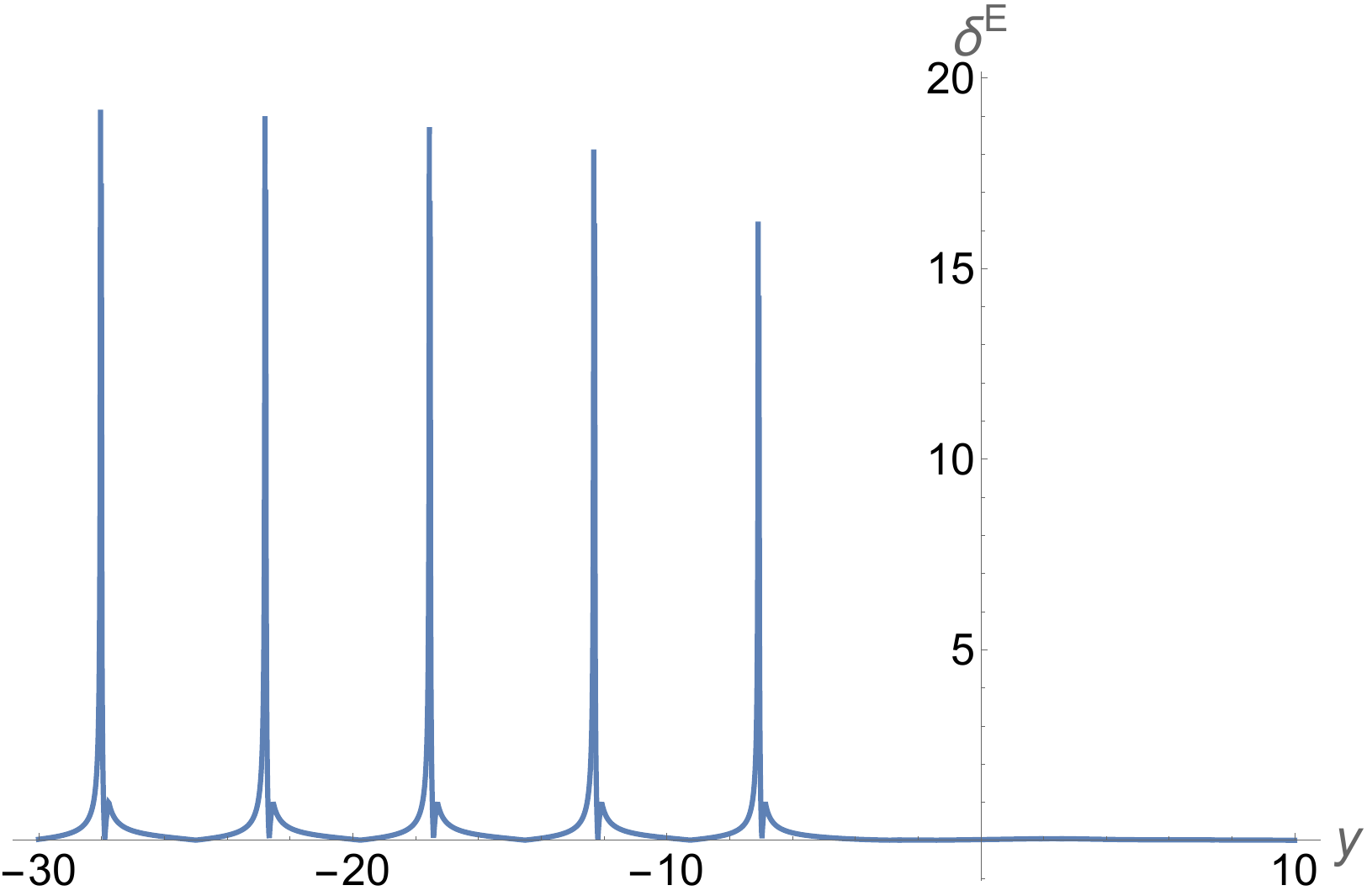}
\includegraphics[width=8.cm]{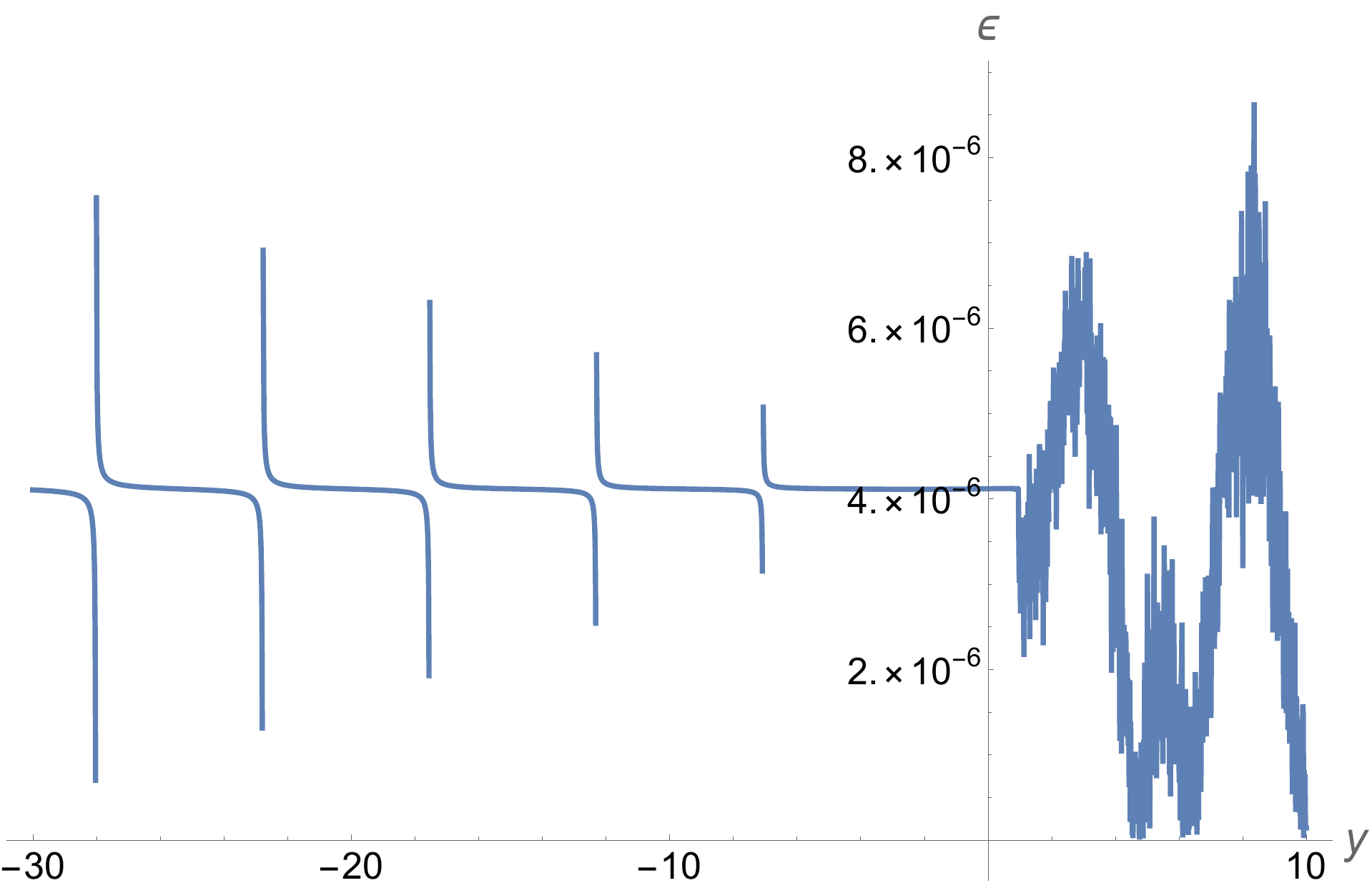}\\
\vspace{.5cm}
(e) $~~~~~~~~~~~~~~~~~~~~~~~~~~~~~~~~~~~~~~~~~~~~~~~~~~~~~~~~~~~~~~~~~~~~~~~~~~$ (f)\\
\vspace{.5cm}
 \caption{Plots of the mode functions $\mu_k(y), \; \mu^N_k(y), \mu^E_k(y)$ and their
relative differences    $\delta^N(y), \; \delta^E(y)$ and $\epsilon(y)$  for
$k = 0.6$,  $\beta = 4.0$ and $q_0=1/2$, for which we have $y_1 = -y_2 \simeq -2.58459$.
}
\label{k06b4_2}
\end{figure*}

\end{widetext}

\subsection{$\beta^2 \gg k^2$}  

In this case,  two real turning points appear,  given, respectively by  
\bqn
\lb{2b1}
 y_1 = - y_2 = -  \cosh^{-1}  \left(\frac{\beta^2}{k^2}\right).
  \eqn
 Then, we find that Eqs.(\ref{eq3.26b}) and (\ref{eq3.27b}) still hold in the current case,
  while Eq.(\ref{eq3.27c}) is replaced by
  \bqn
\lb{eq3.27d}
\mathscr{T}(y) \rightarrow
\frac{q_0^2-1/4}{2\beta} \ln\left(\frac{1+ \epsilon}{1-\epsilon}\right) + \frac{4 + \epsilon - 5 \epsilon^2}{24k(1- \epsilon^2)}, ~~~
\eqn
as $y \rightarrow \infty$, but now with $\epsilon \equiv k/\beta$. Combining  Eqs.(\ref{eq3.26b}), (\ref{eq3.27b}) and (\ref{eq3.27d}), 
we find that currently the proper choice of $q_0$ is still that given by $q_0 = 1/2$, as those given in the last two subcases.

  In Fig. \ref{k06b4}, we plot the quantities $\left|{q}/{g}\right|$,  $\left|{q(y-y_1)}/{g}\right|$, $\left|{q(y-y_1)(y-y_2)}/{{  {g}}}\right|$,
 and the error control function ${\cal{T}}$  for
$k = 0.6$,  $\beta = 4.0$ and $q_0=1/2$, for which we have $y_1 = -y_2 \simeq - 2.58459$. From this figure we can see that the preconditions (\ref{CD1})-(\ref{CD3}) are well satisfied. 
Then, to the first-order approximation of the UAA method, the solution can be approximated by Eq.(\ref{solutionW}), where $\zeta_0^2$ is given by Eq.(\ref{eq3.26b}), 
$\alpha_k$ and $\beta_k$ are two integration constants, and $\epsilon_1$ and $\epsilon_2$   the errors of the corresponding approximate solutions, whose upper bounds are given by
Eqs.(\ref{errors}) and (\ref{ecf}) in Appendix A.

 In Fig. \ref{k06b4_2} (a), we plot out our first-order approximate solution, while Fig. \ref{k06b4_2} (b) to compare the approximate solution with the exact one, we plot both of them. 
 In particular, the solid line represents the exact solution, while the red dotted line 
 the approximate solution. From this figure it can be seen that except the minimal points, the two solutions match well. However, at these extreme minimal points, they deviate 
 significantly from each. The causes of such errors are not clear, and we hope to come back to this issue in another occasion. 
 
 Finally, similar to all other cases,   our numerical solution still matches well to the exact one, as shown by the  {curve} of $\epsilon(y)$, which is no larger than $8.0 \times 10^{-6}$.

\section{Conclusions}
\renewcommand{\theequation}{4.\arabic{equation}}
\setcounter{equation}{0}

In this paper, we have applied the UAA method to the mode function $\mu_k$ with a PT potential, for which it satisfies the second-order differential equation 
\bqn
\lb{eq4.1}
\frac{d^2\mu_k(y)}{dy^2} + \left(k^2 - \frac{\beta_0^2}{\cosh y}\right)\mu_k(y) = 0,
\eqn
where $k$ and $\beta_0$ are real constants. In this case, the exact solution is known and given by Eq.(\ref{sol_PT}). The implementation of the UAA method  {includes} the introduction of an auxiliary function
$q(y)$, which is taken as 
\bqn
\lb{eq4.2}
q(y) =   \frac{q_0^2}{\cosh y},
\eqn
where $q_0$ is a free parameter. Then, we carry out the integration of the error control function, defined by 
\bqn
\lb{eq4.3}
\mathscr{T}(\zeta)   &\equiv& - \int\frac{\psi(\zeta)}{{\left|\dot y^2 g\right|^{1/2}} } d\zeta, 
\eqn
where
\bqn
\lb{eq4.4}
\psi(\zeta) &\equiv& \dot y^2 q + \dot y^{1/2}\frac{d^2}{d\zeta^2}\left(\dot y^{-1/2}\right), \nb\\
\dot y^2 g &=& \zeta_0^2 - \zeta^2. 
 \eqn
 Clearly, the error control function  $\mathscr{T}(\zeta)$ will depend on $q_0$. After working out the details, we find that it is convenient to distinguish the three
 cases: A) $k^2 \gg \beta^2$, B) $k^2 \simeq \beta^2$, and C) $k^2 \ll \beta^2$, where $\beta^2 \equiv \beta_0^2 - q_0^2 > 0$. In particular, in Case A), a proper choice
 of $q_0$ is $q_0 = 1/\sqrt{24}$, while in Cases B) and C), it is  $q_0 = 1/2$. 
 
 Once $q_0$ is fixed, the analytical approximate solutions are uniquely determined by the linear combination of the two parabolic cylindrical functions $W(\zeta_0^2/2, \pm \sqrt{2} \zeta)$, 
 as shown by Eq.(\ref{solutionW}).   In particular, in Case A) the upper bounds of errors are $\lesssim  0.15\%$, as shown in Fig. \ref{k5b1_2}.
 In Case B), two subcases are considered, one with $k \gtrsim \beta$, and the other with $k \lesssim \beta$. In the first case, the  upper bounds of errors are $\lesssim  4\%$, while in the second case 
 they are  $\lesssim  10\%$, as shown, respectively, by Figs. \ref{k5b49_2} and \ref{k5b51_2}. In Case C), the approximate solutions trace also very well to the exact one, except the minimal points,
 as shown in Fig. \ref{k06b4_2}. {   {This might be caused by the fact that at these points the mode function $\mu_k$ is almost zero}}, and very small non-zero values will cause significantly deviations. 
 We are still working on this case, and hope to come back to this point in another occasion. 
 
 As mentioned in the Introduction, the potentials of the mode functions in both dressed metric and hybrid approaches can be well modeled by  PT potentials. Therefore, the current analysis on the choice
 of the function $q(y)$ and the minimization of the error control function shall shed great light on how to carry out similar analyses in order to obtain more accurate approximate solutions in these models. 
 We have been working on it recently, and wish to report our results soon in another occasion.
 
In addition,  the differential equations for the quasi-normal modes of black holes usually  also take the form of Eq.(\ref{eq2.1}) with potentials that have no singularities \footnote{Recall the inner boundaries
of black hole perturbations are the horizons, at which the potentials are usually finite and non-singular.},  but normally do have turning points \cite{BCS09,KZ11}. For example, the effective potential for the axial perturbations of the Schwarzschild black hole is given by 
\bqn
\lb{eq4.5}
\mathscr{V}(r) = \frac{r - 2m}{r^4} \Big\{l(l+1)r - 6m\Big\},
 \eqn
where $\omega$ denotes the qusinormal mode. Clearly, for $r \ge 2m$, this potential also has no poles, but in general $f(r) \equiv \mathscr{V}(r) - \omega^2$ have   two turning points.
From \cite{BCS09,KZ11}, it can be seen that  the properties of this potential are shared by many other cases, including those from modified theories of gravity. Thus, one can equally apply the analysis presented here 
to the studies of quasi-normal modes of black holes.

\section*{Acknowledgements}

RP and AW  were partially supported  by the  US Natural Science Foundation (NSF) with the Grant No.  PHY2308845, and JJM and JS were supported through the Baylor Physics graduate program. 
BFL and TZ were  supported in part by the National Key Research and Development Program of China under Grant No. 2020YFC2201503, the National Natural Science Foundation of China under Grant Nos. 11975203, 11675143, 12205254,  12275238, and 12005186, the Zhejiang Provincial Natural Science Foundation of China under Grant Nos. LR21A050001 and LY20A050002, and the Fundamental Research Funds for the Provincial Universities of Zhejiang in China under Grant No. RF-A2019015.

\newpage

 \appendix
\section*{Appendix A. Upper Bounds of Errors}
\lb{ApendixA}
\renewcommand{\theequation}{A.\arabic{equation}} \setcounter{equation}{0}

The upper bounds of the errors $\epsilon_1$ and $\epsilon_2$ appearing in Eq.(\ref{Uf}) are given by
\begin{widetext}
\bqn
\lb{errors}
&&\frac{|\epsilon_1|}{M\left(\frac{1}{2}\lambda \zeta_0^2,\sqrt{2\lambda }\zeta\right)},\;\frac{|\partial \epsilon_1/\partial \zeta|}{\sqrt{2} N\left(\frac{1}{2}\lambda \zeta_0^2,\sqrt{2\lambda }\zeta\right)}
\leq \frac{\kappa}{\lambda_0 E\left(\frac{1}{2}\lambda \zeta_0^2,\sqrt{2\lambda }\zeta\right)} \Bigg\{\exp{\Big(\lambda \mathscr{V}_{\zeta, \zeta_2}(\mathscr{T})\Big)}-1\Bigg\}. ~~~~~~\nb\\
&&\frac{|\epsilon_2|}{M\left(\frac{1}{2}\lambda \zeta_0^2,\sqrt{2\lambda }\zeta\right)},\;\frac{|\partial \epsilon_2/\partial \zeta|}{\sqrt{2} N\left(\frac{1}{2}\lambda \zeta_0^2,\sqrt{2\lambda }\zeta\right)}
\leq \frac{\kappa E\left(\frac{1}{2}\lambda \zeta_0^2,\sqrt{2\lambda }\zeta\right)}{\lambda } \Bigg\{\exp{\Big(\lambda_0 \mathscr{V}_{0,\zeta}(\mathscr{T})\Big)}-1\Bigg\}, ~~~~
\eqn
\end{widetext}
where $M\left(\frac{1}{2}\lambda \zeta_0^2,\sqrt{2\lambda }\zeta\right)$, $N\left(\frac{1}{2}\lambda \zeta_0^2,\sqrt{2\lambda }\zeta\right)$, and $E\left(\frac{1}{2}\lambda \zeta_0^2,\sqrt{2\lambda }\zeta\right)$ are auxiliary functions of the parabolic cylinder functions defined explicitly in \cite{Olver75}, and \footnote{This corresponds to
choosing the function $\Omega(x)$ introduced by Olver in \cite{Olver75} as  $\Omega(x) = \sqrt{\left|x^2 - \zeta_0^2\right|}$, which satisfies the requirement
$\Omega(x) = {\cal{O}}(x)$, as $x \rightarrow \pm \infty$. For more details, see \cite{Olver75}.}
\bqn
\lb{ecf}
 \mathscr{V}_{\zeta_1,\zeta_2} \equiv  \int_{\zeta_1}^{\zeta_2}{ \frac{\left|\psi(\zeta)\right|}{\sqrt{|\zeta^2-\zeta_0^2|}}d\zeta},
\eqn
is {\em  the associated error control function}.

\appendix
\section*{Appendix B. Exact Solutions with the P\"oschl-Teller Potential}
\lb{ApendixB}
\renewcommand{\theequation}{B.\arabic{equation}} \setcounter{equation}{0}

Let us consider the case with the P\"oschl-Teller Potential given by
\bq
\lb{eqB.1}
\left(\lambda^2  g+q\right)=-\left(k^2-\frac{\beta_0^2}{\cosh^2 y}\right).
\eq
 Then, introducing the two new variables $x$ and ${\cal{Y}}$ via the relations
\bqn
\lb{eq3.2}
x = \frac{1}{1+ e^{-2y}}, \quad
{\cal{Y}}(x) = [x(1-x)]^{ik/2} \mu_k,
\eqn
we find that  Eq.(\ref{eq2.1}) with the above PT potential reads
\bqn
\lb{eq3.3}
x(1-x) \frac{d^2{\cal{Y}}}{dx^2} + \left[a_3 - \left(a_1 + a_2+1\right)x\right] \frac{d{\cal{Y}}}{dx} - a_1 a_2 {\cal{Y}}    = 0,\nb\\
\eqn
where
\bqn
\lb{eq3.4}
a_1 &=& \frac12(1+\sqrt{1-4\beta_0^2})-ik, \nb\\
a_2 &=& \frac12(1-\sqrt{1-4\beta_0^2})-ik, \nb\\
a_3 &=& 1-ik.
\eqn
Eq.(\ref{eq3.3}) is the standard hypergeometric equation, and has the general solution \cite{ZWCKS17}
\bqn\lb{sol_PT}
\mu^{\text{E}}_k(\eta) &=& a_k \left(\frac{x}{1-x}\right)^{ik/2}\nb\\
&& \times \; _2F_1(a_1-a_3+1,a_2-a_3+1,2-a_3,x)\nb\\
&& +\frac{b_k}{[x (1-x)]^{ik /2}} \;_2F_1(a_1,a_2,a_3,x).\nb\\
\eqn
Here $_2F_1(a_1,a_2,a_3,x)$ denotes the hypergeometric function, and $a_k$ and $b_k$ are two independent integration constants, and   are uniquely determined by the
initial conditions.

\appendix
\section*{Appendix C. Computing the error control function}
\renewcommand{\theequation}{C.\arabic{equation}} \setcounter{equation}{0}
In this appendix, we collect some useful formulae for working out the error control function explicitly.  Considering the particular form of the PT potential, it is easier to compute the error control function by using the new variable $x=\mathrm{sech}(y)$, thus 
\bq
\lb{C.1}
dy=-\frac{\epsilon_y~dx}{x\sqrt{1-x^2}},
\eq
where $\epsilon_y$ denotes the sign of $y$.  In terms of the new variable, 
\bqn
\lb{C.2}
q=q^2_0 x^2,\quad \quad  g =\beta^2 x^2-k^2.
\eqn

To calculate the error control function explicitly,  let us consider the cases $g < 0$ and $g > 0$ separately.

\subsection{$g < 0$}

In this case, the  error control function is defined by Eq.(\ref{ECFa}), which can be written as
\bq
\lb{C.3}
\mathscr{T}(\zeta) = \mathscr{T}_1(\zeta) + \mathscr{T}_2(\zeta) + \mathscr{T}_3(\zeta),
\eq
where
\begin{widetext}
\bqn
\lb{C.4}
\mathscr{T}_1&\equiv&\int \frac{q}{\sqrt{- g}}dy = - q^2_0  \epsilon_y \int \frac{x dx}{\sqrt{1-x^2}\sqrt{k^2-\beta^2x^2}}
=\frac{q^2_0\epsilon_y}{\beta}\ln \left( \frac{\sqrt{1-x^2}\beta+\sqrt{k^2-\beta^2 x^2}}{\sqrt{| k^2-\beta^2|}} \right),\nb\\
\mathscr{T}_2&\equiv&\int \left(\frac{5 g'^2}{16 g^3}-\frac{ g^{''}}{4 g^2}\right)\sqrt{- g}dy
= \epsilon_y \int dx \left( \frac{5\beta^4(x^3-x^5)}{4\sqrt{1-x^2}(k^2-\beta^2x^2)^{5/2}}+\frac{\beta^2(2x-3x^3)}{2\sqrt{1-x^2}(k^2-\beta^2x^2)^{3/2}}\right)\nb\\
&=&\epsilon_y \Bigg\{-\frac{1}{4\beta}\ln\left (\frac{\sqrt{1-x^2}\beta+\sqrt{k^2-x^2\beta^2}}{\sqrt{|k^2-\beta^2|}}\right ) +\frac{\sqrt{1-x^2}A}{12(k^2-\beta^2)(k^2-\beta^2 x^2)^{3/2}}\Bigg\},\nb\\
\mathscr{T}_3&\equiv& \int^{\zeta} \left\{\frac{5\zeta_0^2}{4(v^2-\zeta_0^2)^{5/2}}+\frac{3}{4(v^2-\zeta_0^2)^{3/2}}\right\}dv =  \frac{\zeta ^3-6\zeta  \zeta^2_0}{12\zeta^2_0(\zeta ^2-\zeta^2_0)^{3/2}},
\eqn
\end{widetext}
where 
\bq
\lb{C.5}
A(x) \equiv 3k^4+2k^2\beta^2\left(x^2- 1\right) -3x^2\beta^4.
\eq

\subsection{$g > 0$}

In this case, the  error control function is defined by Eq.(\ref{ECFb}), which can be also written as Eq.(\ref{C.3}), but now with
\begin{widetext}
\bqn
\lb{C.6}
\mathscr{T}_1&\equiv& \int \frac{q}{\sqrt{ g}}dy
= \epsilon_y\frac{q^2_0}{\beta}\arcsin \left( \frac{\beta \sqrt{1-x^2}}{\sqrt{\beta^2-k^2}}\right),\nb\\
\mathscr{T}_2&\equiv&\int \left(-\frac{5 g'^2}{16 g^3}+\frac{ g^{''}}{4 g^2}\right)\sqrt{ g}dy =  \epsilon_y \int dx \left( \frac{5\beta^4(x^3-x^5)}{4\sqrt{1-x^2}(\beta^2x^2-k^2)^{5/2}} 
-\frac{\beta^2(2x-3x^3)}{2\sqrt{1-x^2}(\beta^2x^2-k^2)^{3/2}}\right)\nb\\
&=&\epsilon_y \Bigg\{-\frac{1}{4\beta}\arcsin\left(\frac{\sqrt{1-x^2}\beta}{\sqrt{\beta^2-k^2}}\right) +\frac{\sqrt{1-x^2}A}{12(\beta^2-k^2)(\beta^2 x^2-k^2)^{3/2}}\Bigg\},\nb\\
\mathscr{T}_3&\equiv& \int^{\zeta} \left\{\frac{5\zeta_0^2}{4(\zeta_0^2-v^2)^{5/2}}-\frac{3}{4(\zeta_0^2-v^2)^{3/2}}\right\}dv =  \frac{6\zeta \zeta^2_0-\zeta ^3}{12\zeta^2_0(\zeta^2_0-\zeta ^2)^{3/2}}. 
\eqn
\end{widetext}

\end{document}